\documentstyle[11pt,draft]{article}
\newfont{\blackb}{msbm10 scaled\magstep1}
\def\Bbb#1{\hbox{\blackb #1}}

\newtheorem{bbb}{Proposition}[section]
\newtheorem{ccc}{Lemma}[section]
\newtheorem{ddd}{Theorem}[section]
\newtheorem{eee}{Remark}[section]
\newtheorem{fff}{Corollary}[section]

\newtheorem{bbbb}{Proposition}[subsection]
\newtheorem{cccc}{Lemma}[subsection]

\newtheorem{eeee}{Remark}[subsection]
\newtheorem{ffff}{Corollary}[subsection]
\begin{document}
\baselineskip=12.5pt
\frenchspacing
\abovedisplayshortskip=4.25pt
\belowdisplayshortskip=4.25pt
\abovedisplayskip=4.25pt
\belowdisplayskip=4.25pt
\title{Asymptotics of Solutions to the Modified Nonlinear 
Schr\"{o}dinger Equation: Solitons on a Non-Vanishing 
Continuous Background}
\author{A. V. Kitaev\footnote{{\sf E-mail: kitaev@pdmi.ras.ru}. 
Supported by the Alexander von Humboldt Foundation.} \, \, and 
\, A. H. Vartanian\footnote{{\sf E-mail: arthur@pdmi.ras.ru}. 
Partially supported by the Russian Academy of Sciences.} \\
Steklov Mathematical Institute \\
Fontanka 27 \\
St.~Petersburg 191011 \\
Russia}
\date{December 27, 1997}
\maketitle
\begin{abstract}
\noindent
Using the matrix Riemann-Hilbert factorization approach 
for nonlinear evolution systems which take the form of 
Lax-pair isospectral deformations and whose corresponding 
Lax operators contain both discrete and continuous spectra, 
the leading-order asymptotics as $t \! \to \! \pm \infty$ 
of the solution to the Cauchy problem for the modified 
nonlinear Schr\"{o}dinger equation, $i \partial_{t} u \! + 
\! \frac{1}{2} \partial_{x}^{2} u \! + \! \vert u \vert^{2} 
u \! + \! i s \partial_{x} (\vert u \vert^{2} u) \! = \! 0$, 
$s \! \in \! \Bbb R_{> 0}$, which is a model for nonlinear 
pulse propagation in optical fibers in the subpicosecond 
time scale, are obtained: also derived are analogous results 
for two gauge-equivalent nonlinear evolution equations; in 
particular, the derivative nonlinear Schr\"{o}dinger equation, 
$i \partial_{t} q \! + \! \partial_{x}^{2} q \! - \! i 
\partial_{x}(\vert q \vert^{2} q) \! = \! 0$. As an application 
of these asymptotic results, explicit expressions for position 
and phase shifts of solitons in the presence of the continuous 
spectrum are calculated.

\vspace{1.35cm}

\vspace{0.55cm}

\vspace{0.55cm}
{\bf Abbreviated title.} Modified Nonlinear Schr\"{o}dinger 
Equation
\end{abstract}
\clearpage
\section{Introduction}
With the current emphasis on the utilization of optical fibers, 
capable of supporting solitons, as the communication channel 
in the practical realization and implementation of all-optical 
(lightwave), ultrahigh-bit-rate, long-distance communication 
systems using the return-to-zero (RZ) format for generating 
the optical bit stream, design issues requiring the consideration 
of several factors, e.g., soliton widths and inter-soliton 
spacings, are intimately related to the study of the fundamental 
dynamical processes associated with the propagation of high-power 
ultrashort pulses in optical fibers (at the present stage of 
technology, these systems can at best still only be called 
near-soliton(ic)-based) \cite{a1}. The standard, classical 
mathematical model for nonlinear pulse propagation in the 
picosecond time scale in the anomalous dispersion regime 
in an isotropic, homogeneous, lossless, non-amplifying, 
polarization-preserving single-mode optical fiber is the 
nonlinear Schr\"{o}dinger equation (NLSE) \cite{a1,a2}. 
However, experiments and theories on the propagation of 
high-power ultrashort pulses in the subpicosecond-femtosecond 
time scale in monomode optical fibers have shown that the NLSE 
is no longer a valid model, and that additional nonlinear 
terms (dispersive and dissipative) and higher-order linear 
dispersion must be taken into account: in this case, pulse-like 
propagation is described (in dimensionless and normalized form) 
by the following nonlinear evolution equation (NLEE) \cite{a1},
\begin{eqnarray}
i \partial_{z} u + \frac{1}{2} \partial^{2}_{\tau} u + 
\vert u \vert^{2} u + i s \partial_{\tau} (\vert u 
\vert^{2} u) = - i \widetilde{\Gamma} u + i \breve{
\delta} \partial^{3}_{\tau} u + \frac{\widetilde{
\tau}_{n}}{\widetilde{\tau}_{0}} u \partial_{\tau} 
(\vert u \vert^{2}),
\end{eqnarray}
where $u$ is the slowly-varying amplitude of the complex field 
envelope, $z$ is the propagation distance along the fiber length, 
$\tau$ is the time measured in a frame of reference moving with 
the pulse at the group velocity, $s \! \in \! \Bbb R_{>0}$ governs 
the effects due to the intensity dependence of the group velocity 
(self-steepening), $\widetilde{\Gamma}$ is the intrinsic fiber 
loss, $\breve{\delta}$ governs the effects of the third-order 
linear dispersion, and $\widetilde{\tau}_{n} / \widetilde{\tau}_{
0}$ governs the soliton self-frequency shift effect.

Since, under typical operating conditions, $\widetilde{\Gamma}$, 
$\breve{\delta}$, and $\widetilde{\tau}_{n} / \widetilde{\tau}_{
0}$ are small parameters \cite{a1}, a strategy to study the 
solutions of Eq.~(1), for which the nonlinear effects dominate 
the higher-order linear dispersive one, is to set the right-hand 
side equal to zero, thus obtaining the following NLEE (integrable 
in the sense of the inverse scattering method (ISM) \cite{a3}),
\begin{eqnarray}
i \partial_{t} u + \frac{1}{2} \partial^{2}_{x} u + 
\vert u \vert^{2} u + i s \partial_{x} (\vert u 
\vert^{2} u)=0,
\end{eqnarray}
which, hereafter, is called the modified nonlinear 
Schr\"{o}dinger equation (MNLSE) (the physical variables, 
$z$ and $\tau$, have been mapped isomorphically onto the 
mathematical $t$ and $x$ variables, which are standard 
in the ISM context), and to treat Eq.~(1) as a non-integrable 
perturbation of the MNLSE. {}From the above discussion, 
it is clear that perturbations of multi-soliton solutions
of the MNLSE can be very important in the physical context 
related to optical fibers \cite{a1}. Since practical lasers 
excite not only the soliton(ic) mode(s), but also an entire 
continuum of linear-like dispersive (radiative) waves, to 
have physically meaningful and practically representative 
results, it is necessary to investigate solutions of the 
MNLSE under general initial (launching, in the optical 
fiber literature \cite{a1}) conditions, without any 
artificial restrictions and/or constraints, which have 
both soliton(ic) and non-soliton(ic) (continuum) 
components: it is towards such a solution that the 
initial pulse launched into an optical fiber is evolving 
asymptotically \cite{a4}. In physical terms, the pulse 
adjusts its width as it propagates along the optical 
fiber to evolve into a (multi-) soliton, and a part of 
the pulse energy is shed away in the form of dispersive 
waves in the process: normally, these dispersive waves 
form a low-level broadband background radiation that 
accompanies the (multi-) soliton \cite{a1,a4}. {}From 
the physical point of view, therefore, it is seminal to 
understand how the continuum and the (near-) soliton(s) 
interact, and to be able to derive an explicit functional 
form for this process. Since Eq.~(2) is integrable via 
the ISM, one can use one of the techniques developed in 
the framework of this approach to solve the aforementioned 
problem; in particular, in this paper, the Riemann-Hilbert 
(RH) factorization method is applied.

For several soliton-bearing equations, e.g., KdV, 
Landau-Lifshitz, NLS, sine-Gordon and MKdV, it is known 
that the dominant $({\cal O}(1))$ asymptotic $(t \! \to 
\! \pm \infty)$ effect of the continuous spectra on the 
multi-soliton solutions is a shift in phase and 
position of their constituent solitons \cite{a5}: 
as will be shown in this paper, an analogous, 
though analytically more complicated, situation takes 
place for the MNLSE (the additional complexification 
occurs due to the non-standard normalization of the 
associated RH problem). While the above-mentioned 
works deal only with the leading-order $({\cal 
O}(1))$ asymptotic term, in this paper, for the 
MNLSE, not only the leading-order, but the 
next-to-leading-order $({\cal O}(t^{-1/2}))$ term as 
well is derived; in particular, besides inducing an  
${\cal O}(1)$ position and phase shift on the 
multi-soliton solution, this ${\cal O}(t^{-1/2})$
term represents the evolution of the continuum
component (dispersive wavetrain \cite{a1}) as well as 
the non-trivial interaction (overlap) of the soliton 
and continuum components of the solution. It is worth 
mentioning that, even though there have been several 
papers \cite{a6} devoted to studying the soliton 
solutions of the MNLSE, to the best of our knowledge, 
very little, if anything, was known about its solution(s) 
for the class of non-reflectionless initial data until 
very recently \cite{a7}. In the framework of the ISM, an 
asymptotic analysis of the aforementioned solution for 
the MNLSE can be divided into two stages: (1) the 
investigation of the continuum (solitonless) component 
of the solution \cite{a8,a9,a10}; and (2) the inclusion 
of the (multi-) soliton component via the application 
of a ``dressing'' procedure \cite{a11,a12} to the 
continuum background. In this paper, the above-mentioned 
asymptotic paradigm is carried out systematically for the 
MNLSE: the results obtained in this paper are formulated 
as Theorems~2.1--2.3. 

This paper is organized as follows. In Sec.~2, a matrix 
RH problem for the solution of a NLEE gauge-equivalent 
to Eq.~(2) is stated, and the results of this paper are 
summarized as Theorems~2.1--2.3. In Sec.~3, an extended RH 
problem is formulated and shown to be equivalent to the 
original one stated in Sec.~2, and as $t \! \to \! +\infty$, 
it is shown that the solution of the extended RH problem 
converges, modulo exponentially decreasing terms, to the 
solution of a model RH problem. In Sec.~4, the Beals-Coifman 
\cite{a13} formulation for the solution of a RH problem on 
an oriented contour is succinctly recapitulated, and the model 
RH problem is solved asymptotically as $t \! \to \! +\infty$ 
for the Schwartz class of non-reflectionless generic potentials. 
In Sec.~5, a phase integral which is associated with the
non-standard normalization of the above-mentioned RH 
problem is evaluated asymptotically as $t \! \to \! +\infty$. 
Finally, in Sec.~6, the asymptotic analysis as $t \! \to \! 
-\infty$ is presented.
\section{The Riemann-Hilbert (RH) Problem and Summary of 
Results}
In this section, the matrix RH problem is stated, and the 
main results of the paper are formulated as Theorems~2.1--2.3. 
Before doing so, however, it is necessary to introduce some 
notation and definitions which are used throughout the paper:
\begin{flushleft}
{\bf Notational Conventions}
\end{flushleft}
\begin{enumerate}
{\em \item[(1)] $e_{\alpha \beta}$, $\alpha,\beta \! \in 
\! \{1,2\}$, denote $2 \! \times \! 2$ matrices with entry 
$1$ in $(\alpha \, \beta)$, $(e_{\alpha \beta})_{ij} \! 
:= \! \delta_{\alpha i} \delta_{\beta j}$, $i,j \! \in \! 
\{1,2\}$, where $\delta_{ij}$ is the Kronecker delta;
\item[(2)] ${\rm I} \! := \! e_{11} \! + \! e_{22}
\! = \! {\rm diag}(1,1)$ denotes the $2 \! \times \! 
2$ identity matrix;
\item[(3)] $\sigma_{3} \! := \! e_{11} 
\! - \! e_{22} \! = \! {\rm diag}(1,-1)$, $\sigma_{-} 
\! := \! e_{21}$, $\sigma_{+} \! := \! e_{12}$, and 
$\sigma_{1} \! := \! \sigma_{-} \! + \! \sigma_{+};$
\item[(4)] for a scalar $\varpi$ and a $2 \! \times \! 2$ 
matrix $\Upsilon$, $\varpi^{{\rm ad} (\sigma_{3})} \Upsilon 
\! := \! \varpi^{\sigma_{3}} \Upsilon \varpi^{-\sigma_{3}};$
\item[(5)] $\overline{(\bullet)}$ denotes complex conjugation 
of $(\bullet);$
\item[(6)] $M_{2}(\Bbb C)$ denotes the $2 \! \times 
\! 2$ complex matrix algebra with the following inner 
product $((\cdot,\! \cdot) \colon M_{2}(\Bbb C) \! \times 
\! M_{2}(\Bbb C) \! \rightarrow \! \Bbb C)$, $\forall \, 
a,b \! \in \! M_{2} (\Bbb C)$, $(a,b) \! := \! {\rm tr}
(\overline{b} a)$, and (for $a \! \in \! M_{2}(\Bbb C))$ 
the norm on $M_{2}(\Bbb C)$ is defined as $\vert a \vert 
\! := \! \sqrt{(a,\! a)};$
\item[(7)] ${\cal L}^{p}(D;\! M_{2}
(\Bbb C)) \! := \! \{\mathstrut f; \, f \colon D \! \to 
\! M_{2}(\Bbb C), \, \vert \vert f \vert \vert_{{\cal L}^{
p}(D;M_{2}(\Bbb C))} \! := \! (\int_{D} \vert f(\varrho) 
\vert^{p} \vert d \varrho \vert)^{1/p} \! < \! \infty, p 
\! \in \! \{1,2\}\};$
\item[(8)] ${\cal L}^{\infty}(D;\! M_{2}(\Bbb C)) \! := 
\! \{\mathstrut g; g \colon D \! \to \! 
M_{2}(\Bbb C), \vert \vert g \vert \vert_{{\cal L}^{\infty}
(D;M_{2}(\Bbb C))} \! := \! \max\limits_{1 \leq i,j \leq 2} 
\sup\limits_{\varrho \in D} \vert g_{ij}(\varrho) \vert \! 
< \! \infty\};$
\item[(9)] for $D$ an unbounded domain of $\Bbb R \cup i 
\Bbb R$, let ${\cal S}(D;\! \Bbb C)$ (resp.~${\cal S}(D;\! 
M_{2}(\Bbb C)))$ denote the Schwartz class on $D$, i.e., 
the class of smooth $\Bbb C$-valued (resp.~$M_{2}(\Bbb 
C)$-valued) functions $f(x) \colon D \! \rightarrow \! \Bbb 
C$ (resp.~$F(x) \colon D \! \rightarrow \! M_{2}(\Bbb C))$ 
which together with all derivatives tend to zero faster than 
any positive power of $\vert x \vert^{-1}$ as $\vert x \vert 
\! \rightarrow \! \infty$.\/}
\end{enumerate}

In this paper, as in \cite{a7}, along with the MNLSE, the 
following NLEEs are studied:
\begin{eqnarray}
&i {\partial}_{t} Q + {\partial}_{x}^{2} Q + i Q^{2} 
{\partial}_{x} \overline{Q} + \frac{1}{2} Q \vert Q 
\vert^{4} = 0,&
\end{eqnarray}
with initial condition $Q(x,\! 0) \! \in \! {\cal S}(\Bbb R;
\! \Bbb C)$, and the derivative nonlinear Schr\"{o}dinger 
equation (DNLSE),
\begin{eqnarray}
&i {\partial}_{t} q + {\partial}_{x}^{2} q - i {\partial}_{x} 
(\vert q \vert^{2} q) = 0,&
\end{eqnarray}
with initial condition $q(x,\! 0) \! \in \! {\cal S}(\Bbb R;\! 
\Bbb C)$. To recall the relations between the solutions of these 
NLEEs, the following propositions are formulated:
\begin{bbb}[{\rm \cite{a14}}]
The necessary and sufficient condition for the compatibility 
of the following system of linear ODEs (the Lax pair) for 
arbitrary $\lambda \! \in \! \Bbb C$,
\begin{eqnarray}
&\partial_{x} \Psi(x,t;\! \lambda) = U(x,t;\! \lambda) 
\Psi(x,t;\! \lambda), \, \, \, \, \, \, \, \, \, \partial_{
t} \Psi(x,t;\! \lambda) = V (x,t;\! \lambda) \Psi (x,t;\! 
\lambda),&
\end{eqnarray}
where
\begin{eqnarray*}
&U(x,t;\! \lambda) \! = \! - i \lambda^{2} \sigma_{3} \! + 
\! \lambda (\overline{Q} \sigma_{-} \! + \! Q \sigma_{+}) 
\! - \! \frac{i}{2} \vert Q \vert^{2} \sigma_{3},& \\ 
&V(x,t;\! \lambda) = 2 \lambda^{2} U(x,t;\! \lambda) \! - \! 
i \lambda ((\partial_{x} \overline{Q}) \sigma_{-} \! - \! 
(\partial_{x} Q) \sigma_{+}) \! + \! (\frac{i}{4} \vert 
Q \vert^{4} \! + \! \frac{1}{2}(\overline{Q} \partial_{x} 
Q \! - \! Q \partial_{x} \overline{Q})) \sigma_{3},& 
\end{eqnarray*}
is that $Q(x,t)$ satisfies Eq.~(3).
\end{bbb}

{\em Proof.\/} Eq.~(3) is the Frobenius compatibility 
condition for system~(5). \hfill \rule{6.5pt}{6.5pt}
\begin{bbb}
Let $Q(x,t)$ be a solution of Eq.~(3). Then there exists a 
corresponding solution of system~(5) such that $\Psi(x,t;0)$ 
is a diagonal matrix.
\end{bbb}

{\em Proof.\/}
For given $Q(x,t)$, let $\widehat{\Psi}(x,t;\! \lambda)$ be 
a solution of system~(5) which exists in accordance with 
Proposition~2.1. Setting $\lambda \! = \! 0$ in system~(5), 
one gets that $\widehat{\Psi}(x,t;\! \lambda) \! = \! \exp 
\{- \frac{i \sigma_{3}}{2} \! \int_{x_{0}}^{x} \! \vert Q
(\varrho,t) \vert^{2} d \varrho\} \widehat{{\cal K}}(\lambda)$, 
for some $x_{0} \! \in \! \Bbb R$ and non-degenerate matrix 
$\widehat{{\cal K}}(\lambda)$ which is independent of $x$ and 
$t$. The function $\Psi(x,t;\! \lambda) \! := \! \widehat{\Psi}
(x,t;\! \lambda)(\widehat{{\cal K}}(\lambda))^{-1}$ is the 
solution of system~(5) which is diagonal at $\lambda \! = \! 0$. 
\hfill \rule{6.5pt}{6.5pt}
\begin{bbb}[{\rm \cite{a15}}]
Let $Q(x,t)$ be a solution of Eq.~(3) and $\Psi(x,t;\! \lambda)$ 
the corresponding solution of system~(5) given in Proposition~2.2. 
Set $\Psi_{q}(x,t;\! \lambda) \! := \! \Psi^{-1}(x,t;0) \Psi(x,t;
\! \lambda)$. Then
\begin{eqnarray}
\partial_{x} \Psi_{q} (x,t;\! \lambda)={\cal U}_{q}(x,t;\! 
\lambda) \Psi_{q}(x,t;\! \lambda), \, \, \, \, \, \, \, \, \, 
\partial_{t} \Psi_{q}(x,t;\! \lambda)={\cal V}_{q}(x,t;\! 
\lambda) \Psi_{q}(x,t;\! \lambda),&
\end{eqnarray}
where 
\begin{eqnarray}
&{\cal U}_{q}(x,t;\! \lambda) = - i \lambda^{2} \sigma_{3} 
+ \lambda (\overline{q} \sigma_{-} + q \sigma_{+}),& 
\end{eqnarray}
\begin{eqnarray} 
&{\cal V}_{q}(x,t;\! \lambda) = \left(\begin{array}{cc}
- 2 i \lambda^{4} - i \lambda^{2} \vert q \vert^{2} & 2 
\lambda^{3} q + i \lambda \partial_{x} q + \lambda 
\vert q \vert^{2} q \\
2 \lambda^{3} \overline{q} - i \lambda \partial_{x} 
\overline{q} + \lambda \vert q \vert^{2} \overline{q} & 
2 i \lambda^{4} + i \lambda^{2} \vert q \vert^{2} 
\end{array} \right) \!,& 
\end{eqnarray}
with $q(x,t)$ defined by
\begin{eqnarray}
&q(x,t) := Q(x,t)((\Psi^{-1}(x,t;0))_{11})^{2},&
\end{eqnarray}
is the ``Kaup-Newell'' {\rm \cite{a16}} Lax pair for the DNLSE.
\end{bbb}

{\em Proof.\/} Differentiating $\Psi_{q}(x,t;\! \lambda) \! 
:= \! \Psi^{-1}(x,t;0) \Psi(x,t;\! \lambda)$ with respect 
to $x$ and $t$ and using the fact that $\Psi(x,t;0) \! = \! 
\exp \{- \frac{i \sigma_{3}}{2} \! \int_{x_{0}}^{x} \! \vert 
Q(\varrho,t) \vert^{2} d \varrho\}$, for some $x_{0} \! \in 
\! \Bbb R$, and $\Psi(x,t;\! \lambda)$ satisfy system~(5) for 
$\lambda \! = \! 0$ and $\lambda \! \in \! \Bbb C \! \setminus 
\! \{0\}$, respectively, defining $q(x,t)$ as in Eq.~(9), one 
gets that $\Psi_{q}(x,t;\! \lambda)$ satisfies system~(6), 
where ${\cal U}_{q}(x,t;\! \lambda)$ and ${\cal V}_{q}(x,t;\! 
\lambda)$ are given by Eqs.~(7) and (8): Eq.~(4) is the 
Frobenius compatibility condition for system~(6). 
\hfill \rule{6.5pt}{6.5pt}
\begin{bbb}
If $q(x,t)$ is a solution of the DNLSE such that $q(x,\! 0) \! 
\in \! {\cal S}(\Bbb R;\! \Bbb C)$, then
\begin{eqnarray}
&u(x,t) := \frac{1}{\sqrt{2 s}} \exp \{\frac{i}{s}(x-\frac{t}{
2s})\} q(\frac{t}{s}-x,\frac{t}{2})&
\end{eqnarray}
satisfies the MNLSE with initial condition $u(x,\! 0) \! \in 
\! {\cal S}(\Bbb R;\! \Bbb C)$.
\end{bbb}

{\em Proof.\/} Direct substitution. \hfill \rule{6.5pt}{6.5pt}
\begin{eee}
{\rm A convention is now adopted which is adhered to {\em 
sensus strictu\/} throughout the paper: for each segment of 
an oriented contour, according to the given orientation, the 
``$+$'' side is to the left and the ``$-$'' side is to the 
right as one traverses the contour in the direction of the 
orientation; hence, $(\bullet)_{+}$ and $(\bullet)_{-}$ denote, 
respectively, the non-tangential limits (boundary values) of 
$(\bullet)$ on an oriented contour {}from the ``$+$'' (left) 
and ``$-$'' (right) sides.\/}
\end{eee}

Before stating the matrix RH problem which is investigated 
asymptotically (as $t \! \rightarrow \! \pm \infty)$ in 
this paper (see Lemma~2.1), it will be convenient to 
introduce the following notation: {\em let ${\cal Z}_{d} 
\! := \! \cup_{i=1}^{N} (\{\pm \lambda_{i}\} \! \cup \! 
\{\pm \overline{\lambda_{i}}\})$ and $\widehat{\Gamma} 
\! := \! \{\mathstrut \lambda; \, \Im(\lambda^{2}) \! = 
\! 0\}$ (oriented as in Fig.~1) denote, respectively, the 
discrete and continuous spectra of the operator $\partial
_{x} \! - \! U(x,t;\! \lambda)$, and $\sigma_{\pounds} \! 
:= \! {\rm Spec}(\partial_{x} \! - \! U) \! = \! {\cal Z}
_{d} \! \cup \! \widehat{\Gamma}$ $({\cal Z}_{d} \! \cap 
\! \widehat{\Gamma} \! = \! \emptyset)$.\/}
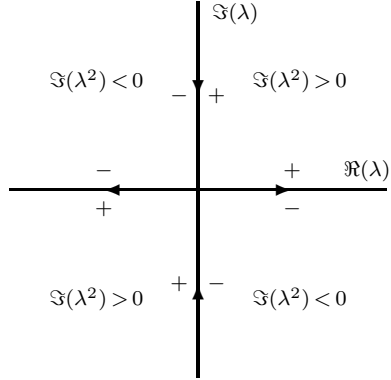
\begin{figure}[bht]
\begin{center}
\unitlength=1cm
\vspace{-0.825cm}
\begin{picture}(6,6)(0,0)
\thicklines
\put(2.5,0){\vector(0,1){1.25}}
\put(2.25,1.25){\makebox(0,0){$\scriptstyle{}+$}}
\put(2.75,1.25){\makebox(0,0){$\scriptstyle{}-$}}
\put(2.75,3.75){\makebox(0,0){$\scriptstyle{}+$}}
\put(2.25,3.75){\makebox(0,0){$\scriptstyle{}-$}}
\put(1.25,2.25){\makebox(0,0){$\scriptstyle{}+$}}
\put(1.25,2.75){\makebox(0,0){$\scriptstyle{}-$}}
\put(3.75,2.25){\makebox(0,0){$\scriptstyle{}-$}}
\put(3.75,2.75){\makebox(0,0){$\scriptstyle{}+$}}
\put(2.5,1.25){\line(0,1){1.25}}
\put(2.5,5){\vector(0,-1){1.25}}
\put(2.5,2.5){\line(0,1){1.25}}
\put(2.5,2.5){\vector(-1,0){1.25}}
\put(1.25,2.5){\line(-1,0){1.25}}
\put(2.5,2.5){\vector(1,0){1.25}}
\put(3.75,2.5){\line(1,0){1.25}}
\put(3.85,3.95){\makebox(0,0){$\scriptstyle{}\Im 
(\lambda^{2}) \, > \, 0$}}
\put(3.85,1.05){\makebox(0,0){$\scriptstyle{}\Im 
(\lambda^{2}) \, < \, 0$}}
\put(1.15,1.05){\makebox(0,0){$\scriptstyle{}\Im 
(\lambda^{2}) \, > \, 0$}}
\put(1.15,3.95){\makebox(0,0){$\scriptstyle{}\Im 
(\lambda^{2}) \, < \, 0$}}
\put(4.75,2.75){\makebox(0,0){$\scriptstyle{}\Re 
(\lambda)$}}
\put(3.0,4.85){\makebox(0,0){$\scriptstyle{}\Im 
(\lambda)$}}
\end{picture}
\vspace{-0.50cm}
\end{center}
\caption{Continuous spectrum ${\widehat \Gamma}$.}
\end{figure}
\begin{ccc}
Let $Q(x,t)$, as a function of $x$, $\! \in \! {\cal S}(\Bbb 
R;\! \Bbb C)$. Set $m(x,t;\! \lambda) \! := \! \Psi(x,t;\! 
\lambda) \exp \{i(\lambda^{2} x \! + \! 2 \lambda^{4} t) 
\sigma_{3}\}$. Then: (1) the bounded discrete set ${\cal Z}_{
d}$ is finite $({\rm card} ({\cal Z}_{d}) \! < \! \infty)$; 
(2) the poles of $m(x,t;\! \lambda)$ are simple; (3) the first 
(resp.~second) column of $m(x,t;\! \lambda)$ has poles at $\{
\pm \lambda_{i}\}_{i=1}^{N}$ (resp.~$\{\pm \overline{\lambda_{
i}}\}_{i=1}^{N})$; and (4) $\forall \, t \! \in \! \Bbb R$ the 
function $m(x,t;\! \lambda) \colon \Bbb C \! \setminus \! ({\cal 
Z}_{d} \! \cup \! \widehat{\Gamma}) \! \rightarrow \! {\rm SL\/}
(2,\! \Bbb C)$ solves the following RH problem:
\begin{enumerate}
\item[a.] $m(x,t;\! \lambda)$ is meromorphic $\forall \, \lambda 
\! \in \! \Bbb C \! \setminus \! \widehat{\Gamma};$
\item[b.] $m(x,t;\! \lambda)$ satisfies the following jump 
conditions,
\begin{eqnarray*}
&m_{+}(x,t;\! \lambda)=m_{-}(x,t;\! \lambda) v(x,t;\! \lambda), 
\, \, \, \, \, \, \, \lambda \! \in \! \widehat{\Gamma},& 
\end{eqnarray*}
where
\begin{eqnarray*}
&v(x,t;\! \lambda) = \exp \{-i(\lambda^{2} x \! + \! 2 \lambda^{
4} t) {\rm ad} (\sigma_{3})\} 
\! \left(\! \begin{array}{cc}
1 - r(\lambda) \overline{r(\overline{\lambda})} & 
r(\lambda) \\
- \overline{r (\overline{\lambda})} & 1
\end{array} \! \right) \!,& 
\end{eqnarray*}
$r(\lambda)$, the reflection coefficient associated with 
the direct scattering problem for the operator $\partial_{x} 
\! - \! U(x,t;\! \lambda)$, satisfies $r(-\lambda) \! = \! 
-r(\lambda)$, and $r(\lambda) \! \in \! {\cal S}(\widehat{
\Gamma};\! \Bbb C);$
\item[c.] for the simple poles of $m(x,t;\! \lambda)$ at $\{
\pm \lambda_{j}\}_{j=1}^{N}$ and $\{\pm \overline{\lambda_{
j}}\}_{j=1}^{N}$, there exist nilpotent matrices $\{v_{j}(x,t) 
\sigma_{-}\}_{j=1}^{N}$ and $\{\overline{v_{j}(x,t)} \sigma_{
+}\}_{j=1}^{N}$, respectively, such that the residues, for $1 
\! \leq \! j \! \leq \! N$, satisfy the (Beals-Coifman 
{\rm \cite{a13})} polar conditions,  
\begin{eqnarray*}
&{\rm res}(m(x,t;\! \lambda);\! \lambda_{j}) = \lim\limits_{
\lambda \rightarrow \lambda_{j}} \! m(x,t;\! \lambda) v_{j}
(x,t) \sigma_{-},& \\
&{\rm res} (m(x,t;\! \lambda);\! -\lambda_{j}) = - 
\sigma_{3} {\rm res} (m(x,t;\! \lambda);\! \lambda_{j}) 
\sigma_{3},& \\
&{\rm res} (m(x,t;\! \lambda);\! \overline{\lambda_{j}} \,) 
= \lim\limits_{\lambda \rightarrow \overline{\lambda_{
j}}} \! m(x,t;\! \lambda) \overline{v_{j}(x,t)} 
\sigma_{+},& \\
&{\rm res} (m(x,t;\! \lambda);\! -\overline{\lambda_{j}} \,) 
= - \sigma_{3} {\rm res} (m(x,t;\! \lambda);\! \overline{
\lambda_{j}} \,) \sigma_{3},& 
\end{eqnarray*}
where $v_{j}(x,t) \! := \! C_{j} \exp \{2 i(\lambda_{j}^{2} 
x \! + \! 2 \lambda_{j}^{4} t)\}$, and $C_{j}$ are 
complex constants associated with the direct scattering 
problem for the operator $\partial_{x} \! - \! U(x,t;\! 
\lambda);$
\item[d.] as $\lambda \! \rightarrow \! \infty$, $\lambda 
\! \in \! \Bbb C \! \setminus \! ({\cal Z}_{d} \! \cup \! 
\widehat{\Gamma})$,
\begin{eqnarray*}
&m(x,t;\! \lambda) = {\rm I} + {\cal O}(\lambda^{-1}).& 
\end{eqnarray*}
\end{enumerate}
\end{ccc}

{\em Proof.\/} Conditions {\em (1)--(4)\/} follow {}from the 
results given in \cite{a7,a13,a17}. \hfill \rule{6.5pt}{6.5pt}
\begin{ccc}
Let $\vert \vert r \vert \vert_{{\cal L}^{\infty}(\Bbb R;\Bbb 
C)} \! := \! \sup_{\lambda \in \Bbb R} \vert r(\lambda) 
\vert \! < \! 1$. Then: (1) the RH problem formulated in 
Lemma~2.1 is uniquely solvable; (2) $\Psi(x,t;\! \lambda) \! = 
\! m(x,t;\! \lambda) \exp \{-i(\lambda^{2} x \! + \! 2 
\lambda^{4} t) \sigma_{3}\}$ is the solution of system~(7) with
\begin{eqnarray}
&Q(x,t) := 2 i \! \lim\limits_{\lambda \rightarrow 
\infty} \! (\lambda m(x,t;\! \lambda))_{12};&
\end{eqnarray}
(3) the function $Q(x,t)$ defined by Eq.~(11) satisfies 
Eq.~(3), and 
\begin{eqnarray}
&q(x,t) := Q(x,t)((m^{-1}(x,t;0))_{11})^{2}&
\end{eqnarray}
satisfies the DNLSE; and (4) $m(x,t;\! \lambda)$ possesses 
the following symmetry reductions, $m(x,t; \linebreak[4] 
\lambda) \! = \! \sigma_{3} m(x,t;\! -\lambda) \sigma_{3}$ 
and $m(x,t;\! \lambda) \! = \! \sigma_{1} \overline{m(x,t;
\! \overline{\lambda})} \sigma_{1}$.

If $r(\lambda) \! \in \! {\cal S}(\widehat{\Gamma};\! \Bbb 
C)$, then, for any $t \! \in \! \Bbb R$, $Q(x,t)$ (resp.~$
q(x,t))$, as a function of $x$, $\! \in \! {\cal S}(\Bbb R;
\! \Bbb C)$.
\end{ccc}

{\em Proof.\/} The solvability of the RH problem (formulated 
in Lemma~2.1) is a consequence of Theorem~9.3 in \cite{a18} 
and the vanishing winding number of $1 \! - \! r(\lambda) 
\overline{r(\overline{\lambda})}$, $\int_{\widehat{\Gamma}} 
d(\arg(1 \! - \! r(\lambda) \overline{r(\overline{\lambda})})) 
\! = \! \sum_{l \in \{{> \atop <}\}} \! {\rm s}(l) {\rm n}(l) 
\! = \! 0$, where ${\rm s}(>) \! = \! - {\rm s}(<) \! = \! 1$, 
and ${\rm n}({> \atop <}) \! := \! {\rm card}(\{\mathstrut 
\lambda_{j}; \, \Im(\lambda_{j}^{2}) {> \atop <} 0\})$, items 
{\it (2)} and {\it (4)} can be verified through straightforward 
calculations, and the fact that $q(x,t)$ (Eq.~(12)) satisfies 
the DNLSE follows {}from Proposition~2.3 and the definition of 
$m(x,t;\! \lambda)$. \hfill \rule{6.5pt}{6.5pt}
\begin{eee}
{\rm In fact, in this paper, the solvability of the RH problem 
for $\vert \vert r \vert \vert_{{\cal L}^{\infty}(\Bbb R;\Bbb 
C)} \! < \! 1$ is proved for all sufficiently large $\vert t 
\vert$: the solvability of the RH problem for $\vert \vert r 
\vert \vert_{{\cal L}^{\infty}(\widehat{\Gamma})} \! < \! 1$ in 
the solitonless sector, ${\cal Z}_{d} \! \equiv \! \emptyset$, 
for all sufficiently large $\vert t \vert$ was proved in \cite{a7}. 
\underline{{\bf Note}}: the condition $\vert \vert r \vert \vert
_{{\cal L}^{\infty}(\widehat{\Gamma})} \! < \! 1$ which appears 
in \cite{a7} is restrictive, and can be replaced by the weaker 
condition $\vert \vert r \vert \vert_{{\cal L}^{\infty}(\Bbb R;
\Bbb C)} \! < \! 1$.\/}
\end{eee}

Before summarizing the main results of this paper, namely, 
Theorems~2.1--2.3, some further preamble is required: {\it
(1) the Kaup-Newell parametrization {\rm \cite{a16}\/} is 
adopted for the discrete eigenvalues, $\lambda_{j} \! := \! 
\Delta_{j} \exp \{\frac{i}{2} (\pi \! - \! \gamma_{j})\}$, 
$\Delta_{j} \! > \! 0$, $\gamma_{j} \! \in \! (0,\pi)$, $1 
\! \leq \! j \! \leq \! N$, and $\lambda_{j}^{2} \! := \! 
\xi_{j} \! + \! i \eta_{j}$, where $\xi_{j} \! = \! - 
\Delta_{j}^{2} \cos \gamma_{j}$ and $\eta_{j} \! = \! \Delta_{
j}^{2} \sin \gamma_{j}$ (note that, with this parametrization, 
$\{\pm \lambda_{i}\}_{i=1}^{N}$ (resp.~$\{\pm \overline{
\lambda_{i}}\}_{i=1}^{N})$ lie in the 1st and 3rd quadrants 
(resp.~2nd and 4th quadrants) of the complex plane of the 
auxiliary spectral parameter, $\lambda)$; and (2) it is supposed 
throughout that $\xi_{i} \! \not= \! \xi_{j}$, $1 \! \leq \! i \! 
\not= \! j \! \leq \! N$, so that it is convenient to choose the 
following enumeration for the points of the discrete spectrum 
(ordering of the solitons), $\xi_{1} \! > \! \cdots \! > \! \xi_{
n} \! > \! \cdots \! > \! \xi_{N}$.\/}
\begin{eee}
{\rm Even though the ``symbol'' (``notation'') $C(z)$ appearing 
in the various {\bf final} error estimations is \underline{{\bf 
not}} the same and should properly be denoted as $C_{1}(z)$, 
$C_{2}(z)$, etc., the simplified ``notation'' $C(z)$ is retained 
throughout since the principal concern here is not its explicit 
functional $z$-dependence, but rather, the functional class(es) 
to which it belongs. Throughout the paper, $M \! \in \! \Bbb 
R_{>0}$ is a fixed constant.\/}
\end{eee}
\begin{eee}
{\rm In Theorems~2.1--2.3 (see below), one should keep the upper 
signs as $t \! \to \! +\infty$ and the lower signs as $t \! 
\to \! -\infty$ everywhere.\/}
\end{eee}
\begin{ddd}
Let $m(x,t;\! \lambda)$ be the solution of the RH problem 
formulated in Lemma~2.1 with the condition $\vert \vert r \vert 
\vert_{{\cal L}^{\infty}(\Bbb R;\Bbb C)} \! < \! 1$ and $Q(x,t)$ 
be defined by Eq.~(11). Then as $t \! \to \! \pm \infty$ and 
$x \! \to \! \mp \infty$ such that $\lambda_{0} \! := \! 
\frac{1}{2} \sqrt{-\frac{x}{t}} \! > \! M$ and $(x,t) \! \in \! 
\Omega_{n} \! := \! \{\mathstrut (x,t); \, x \! - \! 4 t 
\Delta_{n}^{2} \cos \gamma_{n} \! := \! l_{n}(t) \! = \! 
{\cal O}(1)\}$, for those $\gamma_{n} \! \in \! (\frac{\pi}{2},
\pi)$,
\begin{eqnarray}
&Q(x,t) = Q^{\pm}_{{\rm as}}(x,t) + {\cal O} \! \left(
\frac{C(\lambda_{0}) \ln \vert t \vert}{t} \right) \!,&
\end{eqnarray}
where
\begin{eqnarray}
&Q^{\pm}_{{\rm as}}(x,t) := Q_{\pm}^{{\cal S}}(x,t) + 
Q_{\pm}^{{\cal C}}(x,t) + Q_{\pm}^{{\cal S}{\cal C}}(x,t),&
\end{eqnarray}
with
\begin{eqnarray}
&Q_{\pm}^{{\cal S}}(x,t) = \frac{2 i \Delta_{n} 
\sin(\gamma_{n}) \exp \{ \frac{i \gamma_{n}}{2} \} \exp 
\{2 i (\Delta_{n}^{2} (2 t \Delta_{n}^{2} + l_{n}(t) 
\cos \gamma_{n}) + \widehat{\phi}_{n}^{\pm}) \}}{ \cosh 
(\frac{i \gamma_{n}}{2} + 2 \Delta_{n}^{2} \sin(\gamma_{n}) 
l_{n}(t) - \widehat{x}_{n}^{\pm})},& \\
&\widehat{\phi}_{n}^{\pm} = - \frac{1}{2} \arg C_{n} + \arg 
\delta^{\pm}(\lambda_{n};\! \lambda_{0}) + \sum\limits_{l 
\in L_{\pm}} \! \! \arg \! \left(\! \frac{(\lambda_{n} - 
\overline{\lambda_{l}}) (\lambda_{n} + 
\overline{\lambda_{l}})}{(\lambda_{n} - \lambda_{l}) 
(\lambda_{n} + \lambda_{l})} \! \right) \!,& \\ 
&\widehat{x}_{n}^{\pm} = - \ln (\Delta_{n} \sin \gamma_{n}) 
+ \ln \! \vert C_{n} \vert - 2 \ln \! \vert \delta^{\pm} 
(\lambda_{n};\! \lambda_{0}) \vert + 2 \sum\limits_{l \in 
L_{\pm}} \! \! \ln \! \left(\! \frac{\vert \lambda_{n} - 
\lambda_{l} \vert \vert \lambda_{n} + \lambda_{l} 
\vert}{\vert \lambda_{n} - \overline{\lambda_{l}} \vert
\vert \lambda_{n} + \overline{\lambda_{l}} \vert} \! 
\right) \!,& \\
&\delta^{+}(\lambda;\! z) = \exp \{\int_{0}^{z} \frac{
\varrho \ln (1 - \vert r(\varrho) \vert^{2})}{(
\varrho^{2} - \lambda^{2})} \frac{d \varrho}{\pi i} - 
\int_{0}^{\infty} \frac{\varrho \ln (1 + \vert r(i 
\varrho) \vert^{2})}{(\varrho^{2} + \lambda^{2})} 
\frac{d \varrho}{\pi i} \},& \\
&\delta^{-}(\lambda;\! z) = \exp \{\int_{z}^{\infty} 
\frac{\varrho \ln (1 - \vert r(\varrho) \vert^{2})}{
(\varrho^{2} - \lambda^{2})} \frac{d \varrho}{\pi i} 
\},& \\ 
&Q_{\pm}^{{\cal C}} (x,t) = \sqrt{\pm \frac{\nu 
(\lambda_{0})}{2 \lambda_{0}^{2} t}} \exp \{ i (\phi^{\pm} 
(\lambda_{0}) + \widehat{\Phi}^{\pm}(\lambda_{0};t)
+ \frac{\pi}{2})\},& \\
&\nu (z) = - \frac{1}{2 \pi} \ln (1 - \vert r(z) 
\vert^{2}),& \\
&\phi^{+}(z) = \frac{1}{\pi} \int_{0}^{z} \ln \! \vert 
\varrho^{2} - z^{2} \vert d \ln (1 - \vert r (\varrho) 
\vert^{2}) - \frac{1}{\pi} \int_{0}^{\infty} \ln \! \vert 
\varrho^{2} + z^{2} \vert d \ln (1 + \vert r (i \varrho) 
\vert^{2} ),& \\ 
&\phi^{-}(z) = \frac{1}{\pi} \int_{z}^{\infty} \ln \! 
\vert \varrho^{2} - z^{2} \vert d \ln (1 - \vert r 
(\varrho) \vert^{2}),& \\
&\widehat{\Phi}^{\pm} (\lambda_{0};t) = 4 \lambda_{0}^{4} t 
\mp \nu (\lambda_{0}) \ln \! \vert t \vert \pm \arg \Gamma 
(i \nu (\lambda_{0})) + \arg r(\lambda_{0}) \mp 3 
\nu(\lambda_{0}) \ln 2& \nonumber \\
&\! \! \! \! \! \! \! \! \! \! \! \! \! \! \! \! \! \! \! 
\! \! \! \! \! \! \! \! \! \! + \, \, (2 \pm 1) 
\frac{\pi}{4} + 2 \sum\limits_{l \in L_{\pm}} \! \arg \! 
\left( \! \frac{(\lambda_{0} - \overline{\lambda_{l}}) 
(\lambda_{0} + \overline{\lambda_{l}})}{(\lambda_{0} - 
\lambda_{l}) (\lambda_{0} + \lambda_{l})} \! \right) 
\!,& \\
&Q_{\pm}^{{\cal S}{\cal C}}(x,t) = - \frac{4 
(\Xi^{\pm})^{2} \overline{g_{n}^{\pm}} \vert g_{n}^{\pm} 
\vert}{\eta_{n}} \sqrt{\pm \frac{\nu(\lambda_{0})}{2 
\lambda_{0}^{2} t}} \{ \exp (i \varphi_{n}^{\pm} 
(\lambda_{0};t)) + 2 i \cot(\gamma_{n}) \cos 
(\varphi_{n}^{\pm}(\lambda_{0};t)) \},& \\
&g^{\pm}_{n} := \vert g^{\pm}_{n} \vert \exp \{i \arg 
g^{\pm}_{n}\},& \nonumber \\ 
&\vert g^{\pm}_{n} \vert = \vert C_{n} \vert \vert 
\delta^{\pm}(\lambda_{n};\! \lambda_{0}) \vert^{-2} \exp 
\{-2 \Delta_{n}^{2} \sin (\gamma_{n}) l_{n}(t)\} \exp \! 
\left\{\! 2 \! \! \sum\limits_{l \in L_{\pm}} \! \ln \! 
\left(\! \frac{\vert \lambda_{n} - \lambda_{l} \vert \vert 
\lambda_{n} + \lambda_{l} \vert}{\vert \lambda_{n} - 
\overline{\lambda_{l}} \vert \vert \lambda_{n} + 
\overline{\lambda_{l}} \vert} \! \right) \! \right\} \!,& \\ 
&\arg g^{\pm}_{n} = \arg C_{n} - 2 \arg \delta^{\pm} 
(\lambda_{n};\! \lambda_{0}) + 2 \sum\limits_{l \in L_{\pm}} 
\! \! \arg \! \left( \! \frac{(\lambda_{n} - \lambda_{l}) 
(\lambda_{n} + \lambda_{l})}{(\lambda_{n} - 
\overline{\lambda_{l}}) (\lambda_{n} + 
\overline{\lambda_{l}})} \! \right)&
\nonumber \\
&\! \! \! \! \! \! \! \! \! \! \! \! \! \! \! \! \! \! 
\! \! \! \! \! \! \! \! \! \! \! \! \! \! \! \! \! \! 
\! \! \! \! \! \! \! \! \! \! \! \! \! \! \! - \, \, 
2 \Delta_{n}^{2} (2 t \Delta_{n}^{2} + l_{n}(t) \cos 
\gamma_{n}),& \\
&\Xi^{\pm} = \frac{\exp \{ \frac{i \gamma_{n}}{2} \} \exp 
\{ 2 \Delta_{n}^{2} \sin(\gamma_{n}) l_{n} (t) - 
\widehat{x}_{n}^{\pm} \}}{2 \cosh (\frac{i \gamma_{n}}{2} + 
2 \Delta_{n}^{2} \sin (\gamma_{n}) l_{n} (t) - 
\widehat{x}_{n}^{\pm})},& \\
&\varphi_{n}^{\pm} (\lambda_{0};t) := \arg g_{n}^{\pm} + 
\phi^{\pm} (\lambda_{0}) + \widehat{\Phi}^{\pm}(\lambda_{0};
t),&
\end{eqnarray}
$\sum_{l \in L_{+}} \! := \! \sum_{l=n+1}^{N}$, $\sum_{l \in 
L_{-}} \! := \! \sum_{l=1}^{n-1}$, $\Gamma(\cdot)$ is the 
gamma function {\rm \cite{a19}}, and $C(\lambda_{0}) \! \in \! 
{\cal S}(\Bbb R_{> M};\! \Bbb C)$, and, as $t \! \to \! \pm 
\infty$ and $x \! \rightarrow \! \pm \infty$ such that $\mu_{0} 
\! := \! \frac{1}{2} \sqrt{\frac{x}{t}} \! > \! M$ and $(x,
t) \! \in \! \mho_{n} \! := \! \{\mathstrut (x,t); \, -x \! 
+ \! 4 t \Delta_{n}^{2} \cos \gamma_{n} \! := \! - l_{n}
(t) \! = \! {\cal O}(1)\}$, for those $\gamma_{n} \! \in \! (0,
\frac{\pi}{2})$,
\begin{eqnarray}
&Q(x,t) = Q^{\pm \prime}_{{\rm as}}(x,t) + {\cal O} \! 
\left(\frac{C(\mu_{0}) \ln \vert t \vert}{t} \right) \!,&
\end{eqnarray}
where
\begin{eqnarray}
&Q^{\pm \prime}_{{\rm as}}(x,t) := Q_{\pm}^{{\cal S} \prime}
(x,t) + Q_{\pm}^{{\cal C} \prime} (x,t) + Q_{\pm}^{{\cal S}
{\cal C} \prime}(x,t),&
\end{eqnarray}
with
\begin{eqnarray}
&Q_{\pm}^{{\cal S} \prime}(x,t) = \frac{2 \Delta_{n} 
\sin(\gamma_{n}) \exp \{- \frac{i \gamma_{n}}{2}\} \exp 
\{2 i (\Delta_{n}^{2} (2 t \Delta_{n}^{2} + l_{n}(t) \cos 
\gamma_{n}) + \widehat{\phi}_{n}^{\pm \prime})\}}{\sinh 
(\frac{i \gamma_{n}}{2} + 2 \Delta_{n}^{2} \sin(\gamma_{n}) 
l_{n}(t) + \widehat{x}_{n}^{\pm \prime})},& \\
&\widehat{\phi}_{n}^{\pm \prime} = - \frac{1}{2} \arg C_{n} 
+ \arg \delta^{\pm}_{\flat} (\overline{\lambda_{n}};\mu_{0}) 
- \sum\limits_{l \in L_{\pm}} \! \! \arg \! \left(\! 
\frac{(\lambda_{n} - \overline{\lambda_{l}})(\lambda_{n} 
+ \overline{\lambda_{l}})}{(\lambda_{n} - \lambda_{l}) 
(\lambda_{n} + \lambda_{l})} \! \right) \!,& \\ 
&\widehat{x}_{n}^{\pm \prime} = - \ln (\Delta_{n} \sin 
\gamma_{n}) + \ln \! \vert C_{n} \vert - 2 \ln \! \vert 
\delta^{\pm}_{\flat}(\overline{\lambda_{n}};\mu_{0}) 
\vert + 2 \sum\limits_{l \in L_{\pm}} \! \! \ln \! 
\left(\! \frac{\vert \lambda_{n} - \lambda_{l} \vert 
\vert \lambda_{n} + \lambda_{l} \vert}{\vert \lambda_{n} 
- \overline{\lambda_{l}} \vert \vert \lambda_{n} + 
\overline{\lambda_{l}} \vert} \! \right) \!,& \\ 
&\delta^{+}_{\flat}(\lambda;\! z) = \exp \{ \int_{0}^{z} 
\frac{\varrho \ln(1 + \vert r(i \varrho) \vert^{2})}{
(\varrho^{2} - \lambda^{2})} \frac{d \varrho}{\pi i} - 
\int_{0}^{\infty} \frac{\varrho \ln (1 - \vert r(\varrho) 
\vert^{2})}{(\varrho^{2} + \lambda^{2})} \frac{d 
\varrho}{\pi i}\},& \\
&\delta^{-}_{\flat}(\lambda;\! z) = \exp \{ \int_{z}^{
\infty} \frac{\varrho \ln(1 + \vert r(i \varrho) 
\vert^{2})}{(\varrho^{2} - \lambda^{2})} \frac{d 
\varrho}{\pi i}\},& \\
&Q_{\pm}^{{\cal C} \prime}(x,t) = \sqrt{\mp \frac{\nu 
(i \mu_{0})}{2 \mu_{0}^{2} t}} \exp \{i (\phi^{\pm 
\prime} (\mu_{0}) + \widehat{\Phi}^{\pm \prime} 
(\mu_{0};t) + \pi)\},& \\
&\nu(iz) = - \frac{1}{2 \pi} \ln(1 + \vert r(iz) 
\vert^{2}),& \\
&\phi^{+ \prime}(z) = \frac{1}{\pi} \int_{0}^{z} \ln \! 
\vert \varrho^{2} - z^{2} \vert d \ln(1 + \vert r(i 
\varrho) \vert^{2}) - \frac{1}{\pi} \int_{0}^{\infty} 
\ln \! \vert \varrho^{2} + z^{2} \vert d \ln(1 - \vert 
r(\varrho) \vert^{2}),& \\
&\phi^{- \prime} (z) = \frac{1}{\pi} \int_{z}^{\infty}
\ln \! \vert \varrho^{2} - z^{2} \vert d \ln (1 + \vert 
r(i \varrho) \vert^{2}),& \\
&\widehat{\Phi}^{\pm \prime}(\mu_{0};t) = 4 \mu_{0}^{4} t 
\mp \nu(i \mu_{0}) \ln \! \vert t \vert \pm \arg \Gamma 
(i \nu(i \mu_{0})) + \arg r(i \mu_{0}) \mp 3 \nu(i \mu_{0}) 
\ln 2& \nonumber \\
&\! \! \! \! \! \! \! \! \! \! \! \! \! \! \! \! \! \! \! 
\! \! \! \! \! \! \! \! \! \! \! \! \! \! \! \! \!
\! \! - \, \, (2 \mp 1) \frac{\pi}{4} 
- 2 \sum\limits_{l \in L_{\pm}} \! \! \arg \! \left(\! 
\frac{(\mu_{0} - \overline{\lambda_{l}}) (\mu_{0} + 
\overline{\lambda_{l}})}{(\mu_{0} - \lambda_{l}) 
(\mu_{0} + \lambda_{l})} \! \right) \!,& \\
&Q_{\pm}^{{\cal S}{\cal C} \prime}(x,t) \! = \! - \frac{4 
i (\Xi^{\pm \prime})^{2} \overline{g_{n}^{\pm \prime}} 
\vert g_{n}^{\pm \prime} \vert}{\eta_{n}} \sqrt{\mp \frac{
\nu(i \mu_{0})}{2 \mu_{0}^{2} t}} \{\exp (i \varphi_{n}^{
\pm \prime}(\mu_{0};t)) \! - \! 2 i \cot(\gamma_{n}) \cos 
(\varphi_{n}^{\pm \prime}(\mu_{0};t))\} \!,& \\
&g^{\pm \prime}_{n} := \vert g^{\pm \prime}_{n} \vert 
\exp \{i \arg g^{\pm \prime}_{n}\},& \nonumber \\
&\vert g^{\pm \prime}_{n} \vert = \vert C_{n} \vert \vert 
\delta^{\pm}_{\flat}(\overline{\lambda_{n}};\mu_{0}) 
\vert^{-2} \exp \{2 \Delta_{n}^{2} \sin (\gamma_{n}) l_{n}
(t)\} \exp \! \left\{\! 2 \! \sum\limits_{l \in L_{\pm}} 
\! \! \ln \! \left(\! \frac{\vert \lambda_{n} - 
\lambda_{l} \vert \vert \lambda_{n} + \lambda_{l} 
\vert}{\vert \lambda_{n} - \overline{\lambda_{l}} \vert 
\vert \lambda_{n} + \overline{\lambda_{l}} \vert} \! 
\right) \! \right\} \!,& \\ 
&\arg g^{\pm \prime}_{n} = \arg C_{n} - 2 \arg 
\delta^{\pm}_{\flat}(\overline{\lambda_{n}};\mu_{0}) 
- 2 \sum\limits_{l \in L_{\pm}} \! \! \arg \! 
\left(\! \frac{(\lambda_{n} - \lambda_{l}) (\lambda_{n} 
+ \lambda_{l})}{(\lambda_{n} - \overline{\lambda_{l}}) 
(\lambda_{n} + \overline{\lambda_{l}})} \! \right)& 
\nonumber \\ 
&\! \! \! \! \! \! \! \! \! \! \! \! \! \! \! \! \! 
\! \! \! \! \! \! \! \! \! \! \! \! \! \! \! \! \! 
\! \! \! \! \! \! \! \! \! \! \! \! \! \! \! \! - 
\, \, 2 \Delta_{n}^{2} (2 t \Delta_{n}^{2} + l_{n}(t) 
\cos \gamma_{n}),& \\
&\Xi^{\pm \prime} = - \frac{\exp \{- \frac{i \gamma_{n}}
{2}\} \exp \{- 2 \Delta_{n}^{2} \sin(\gamma_{n}) l_{n}(t) 
- \widehat{x}_{n}^{\pm \prime} \}}{2 \sinh (\frac{i 
\gamma_{n}}{2} + 2 \Delta_{n}^{2} \sin (\gamma_{n}) 
l_{n} (t) + \widehat{x}_{n}^{\pm \prime})},& \\
&\varphi_{n}^{\pm \prime}(\mu_{0};t) := \arg g_{n}^{\pm 
\prime} + \phi^{\pm \prime}(\mu_{0})+\widehat{\Phi}^{\pm 
\prime}(\mu_{0};t),&
\end{eqnarray}
and $C(\mu_{0}) \! \in \! {\cal S}(\Bbb R_{> M};\! \Bbb C)$.
\end{ddd}
\begin{ddd}
Let $m(x,t;\! \lambda)$ be the solution of the RH problem 
formulated in Lemma~2.1 with the condition $\vert \vert r \vert 
\vert_{{\cal L}^{\infty}(\Bbb R;\Bbb C)} \! < \! 1$ and $q(x,t)$, 
the solution of the DNLSE (Eq.~(4)), be defined by Eq.~(12) in 
terms of the function $Q(x,t)$ given in Theorem~2.1. Then as $t 
\! \rightarrow \! \pm \infty$ and $x \! \rightarrow \! \mp 
\infty$ such that $\lambda_{0} \! := \! \frac{1}{2} \sqrt{
-\frac{x}{t}} \! > \! M$ and $(x,t) \! \in \! \Omega_{n} \! 
:= \! \{\mathstrut (x,t); \, x \! - \! 4 t \Delta_{n}^{2} 
\cos \gamma_{n} \! := \! l_{n} (t) \! = \! {\cal O}(1)\}$, 
for those $\gamma_{n} \! \in \! (\frac{\pi}{2},\pi)$,
\begin{eqnarray}
&q(x,t) = Q^{\pm}_{{\rm as}}(x,t) \exp \{i \arg q_{{\rm as}}^{
\pm}(x,t)\} + {\cal O} \! \left(\frac{C(\lambda_{0}) (\ln \vert 
t \vert)^{2}}{t} \right) \!,&
\end{eqnarray}
where $Q^{\pm}_{{\rm as}}(x,t)$ are given in Theorem~2.1, 
Eqs.~(14)--(29),
\begin{eqnarray}
&\arg q_{{\rm as}}^{\pm} (x,t) = - 4 \sum\limits_{l \in L_{\pm}} 
\! \gamma_{l} + 4 \arctan (\eta_{n} \vert g_{n}^{\pm} \vert^{-2} 
+ \cot \gamma_{n}) + {\cal Y}_{\pm}(\lambda_{0})& \nonumber \\
&+ \, 4 \sqrt{\pm \frac{\nu(\lambda_{0})}{2 \lambda_{0}^{2} t}} 
\frac{\vert g_{n}^{\pm} \vert \sin(\gamma_{n})(\vert g_{n}^{\pm} 
\vert^{2} \cos (\varphi_{n}^{\pm}(\lambda_{0};t) - \gamma_{n}) - 
\eta_{n} \sin(\gamma_{n}) \cos(\varphi_{n}^{\pm}(\lambda_{0};t))
)}{((\eta_{n} \sin \gamma_{n} + \vert g_{n}^{\pm} \vert^{2} \cos 
\gamma_{n})^{2} + \vert g_{n}^{\pm} \vert^{4} \sin^{2} \! \gamma_{
n})}& \nonumber \\
&- \sqrt{\pm \frac{2}{t}} \int_{\raise-0.5ex\hbox{$
\scriptstyle{}\lambda_{0}$}}^{\infty} \frac{\sqrt{\nu(
\mu)}}{\mu^{2}} \! \left(\Re \{R^{\pm}(0)\} \cos(\widehat{
\Theta}^{\pm}(\mu;t)) + \Im \{R^{\pm}(0)\} \sin(\widehat{
\Theta}^{\pm}(\mu;t)) \right) \! \frac{d \mu}{\pi},& \\
&{\cal Y}_{+}(z) \! = \! \frac{2}{\pi} \int_{0}^{z} \frac{\ln 
(1 - \vert r(\varrho) \vert^{2})}{\varrho} d \varrho \! - \! 
\frac{2}{\pi} \int_{0}^{\infty} \frac{\ln (1 + \vert r(i 
\varrho) \vert^{2})}{\varrho} d \varrho, \, \, \, \, \, \, 
{\cal Y}_{-}(z) \! = \! \frac{2}{\pi} \int_{z}^{\infty} 
\frac{\ln (1 - \vert r(\varrho) \vert^{2})}{\varrho} d 
\varrho,& \\
&R^{\pm}(0) \! := \! \left(\! \left. \frac{d(r(z) \vert_{z \in 
\Bbb R})}{dz} \right\vert_{z=0} - \left. \frac{d(r(z) \vert_{z 
\in i \Bbb R})}{dz} \right\vert_{z=0} \right) \! \exp \! 
\left\{4 i \sum\limits_{l \in L_{\pm}} \gamma_{l} \right\} 
\!,& \\
&\widehat{\Theta}^{\pm}(\lambda_{0};t) := \widehat{\Phi}^{\pm}(
\lambda_{0};t)+\phi^{\pm}(\lambda_{0})+{\cal Y}_{\pm}(\lambda_{
0}),&
\end{eqnarray}
with $C(\lambda_{0}) \! \in \! {\cal S}(\Bbb R_{>M};\! \Bbb C)$, 
and, as $t \! \rightarrow \! \pm \infty$ and $x \! \rightarrow 
\! \pm \infty$ such that $\mu_{0} \! := \! \frac{1}{2} 
\sqrt{\frac{x}{t}} \! > \! M$ and $(x,t) \! \in \! \mho_{n} \! 
:= \! \{\mathstrut (x,t); \, - x \! + \! 4 t \Delta_{n}^{2} 
\cos \gamma_{n} \! := \! - l_{n}(t) \! = \! {\cal O}(1)\}$, 
for those $\gamma_{n} \! \in \! (0,\frac{\pi}{2})$, 
\begin{eqnarray}
&q(x,t) = Q^{\pm \prime}_{{\rm as}} (x,t) \exp \{i \arg q_{{\rm 
as}}^{\pm \prime}(x,t) \} + {\cal O} \! \left(\frac{C(\mu_{0}) 
(\ln \vert t \vert)^{2}}{t} \right) \!,&
\end{eqnarray}
where $Q^{\pm \prime}_{{\rm as}}(x,t)$ are given in Theorem~2.1, 
Eqs.~(31)--(46),
\begin{eqnarray}
&\arg q_{{\rm as}}^{\pm \prime}(x,t)=4 \sum\limits_{l \in L_{
\pm}} \! \gamma_{l}+4 \arctan(\eta_{n} \vert g_{n}^{\pm \prime} 
\vert^{-2}-\cot \gamma_{n})+{\cal Y}_{\pm}^{\prime}(\mu_{0})& 
\nonumber \\
&- \, 4 \sqrt{\mp \frac{\nu(i \mu_{0})}{2 \mu_{0}^{2} t}} \frac{
\vert g_{n}^{\pm \prime} \vert \sin(\gamma_{n}) (\vert g_{n}^{\pm 
\prime} \vert^{2} \cos (\varphi_{n}^{\pm \prime}(\mu_{0};t) + 
\gamma_{n}) + \eta_{n} \sin(\gamma_{n}) \cos(\varphi_{n}^{\pm 
\prime}(\mu_{0};t)))}{((\eta_{n} \sin \gamma_{n} - \vert g_{n}^{
\pm \prime} \vert^{2} \cos \gamma_{n})^{2} + \vert g_{n}^{\pm 
\prime} \vert^{4} \sin^{2} \! \gamma_{n})}& \nonumber \\
&- \sqrt{\pm \frac{2}{t}} \int_{\raise-0.5ex\hbox{$
\scriptstyle{}\mu_{0}$}}^{\infty} \frac{\sqrt{-\nu(i\mu)
}}{\mu^{2}} \! \left(\Re \{R^{\pm \prime}(0)\} \cos(\widehat{
\Theta}^{\pm \prime}(\mu;t)) + \Im \{R^{\pm \prime}(0)\} \sin
(\widehat{\Theta}^{\pm \prime}(\mu;t)) \right) \! \frac{d 
\mu}{\pi},& \\
&{\cal Y}_{+}^{\prime}(z) \! = \! \frac{2}{\pi} \int_{0}^{z} 
\frac{\ln (1 + \vert r(i \varrho) \vert^{2})}{\varrho} 
d \varrho \! - \! \frac{2}{\pi} \int_{0}^{\infty} \frac{\ln 
(1 - \vert r(\varrho) \vert^{2})}{\varrho} d \varrho, 
\, \, \, \, \, \, {\cal Y}_{-}^{\prime}(z) \! = \! \frac{2}{
\pi} \int_{z}^{\infty} \frac{\ln (1 + \vert r(i \varrho) 
\vert^{2})}{\varrho} d \varrho,& \\
&R^{\pm \prime}(0) \! := \! \left(\! \left. \frac{
d(r(z) \vert_{z \in \Bbb R})}{dz} \right\vert_{z=0} - 
\left. \frac{d(r(z) \vert_{z \in i \Bbb R})}{dz} 
\right\vert_{z=0} \right) \! \exp \! \left\{-4i \sum\limits_{
l \in L_{\pm}} \gamma_{l} \right\} \!,& \\
&\widehat{\Theta}^{\pm \prime}(\mu_{0};t) := \widehat{\Phi}^{
\pm \prime}(\mu_{0};t)+\phi^{\pm \prime}(\mu_{0})+{\cal Y}_{
\pm}^{\prime}(\mu_{0}),&
\end{eqnarray}
and $C(\mu_{0}) \! \in \! {\cal S}(\Bbb R_{>M};\! \Bbb C)$.
\end{ddd}
\begin{ddd}
Let $m(x,t;\! \lambda)$ be the solution of the RH problem 
formulated in Lemma~2.1 with the condition $\vert \vert r 
\vert \vert_{{\cal L}^{\infty}(\Bbb R;\Bbb C)} \! < \! 1$ 
and $u(x,t)$, the solution of the MNLSE (Eq.~(2)), be defined 
by Eq.~(10) in terms of the function $q(x,t)$ given in 
Theorem~2.2. Then as $t \! \to \! \pm \infty$ and $x \! \to 
\! \pm \infty$ such that $\widehat{\lambda}_{0} \! := \! \sqrt{
\frac{1}{2}(\frac{x}{t} \! - \! \frac{1}{s})} \! > \! M$, 
$\frac{x}{t} \! > \! \frac{1}{s}$, $s \! \in \! \Bbb R_{>0}$, 
and $(x,t) \! \in \! \widetilde{\Omega}_{n} \! := \! \{\mathstrut 
(x,t); \, - x + t (\frac{1}{s} \! - \! 2 \Delta_{n}^{2} \cos 
\gamma_{n}) \! := \! \widehat{l}_{n} (t) \! = \! {\cal O}(1)\}$, 
for those $\gamma_{n} \! \in \! (\frac{\pi}{2},\pi)$,
\begin{eqnarray}
&u(x,t) = v_{{\rm as}}^{\pm}(x,t) w_{{\rm as}}^{\pm}(x,t) 
+ {\cal O} \! \left(\frac{C(\widehat{\lambda}_{0}) (\ln 
\vert t \vert)^{2}}{t} \right) \!,&
\end{eqnarray}
where
\begin{eqnarray}
&v^{\pm}_{{\rm as}}(x,t) := v_{\pm}^{{\cal S}}(x,t) + 
v_{\pm}^{{\cal C}}(x,t) + v_{\pm}^{{\cal S}{\cal C}}(x,t),&
\end{eqnarray}
with
\begin{eqnarray}
&v_{\pm}^{{\cal S}}(x,t) = \frac{\sqrt{2} i \Delta_{n} 
\sin(\gamma_{n}) \exp \{ \frac{i \gamma_{n}}{2} \} \exp 
\{ 2 i (\Delta_{n}^{2} (t \Delta_{n}^{2} + \widehat{l}_{n} 
(t) \cos \gamma_{n}) + \widetilde{\phi}_{n}^{\pm})\}}{ 
\sqrt{s} \cosh (\frac{i \gamma_{n}}{2} + 2 \Delta_{n}^{2} 
\sin(\gamma_{n}) \widehat{l}_{n}(t) - \widetilde{x}_{n}^{
\pm})},& \\
&\widetilde{\phi}_{n}^{\pm} = - \frac{1}{2} \arg C_{n} + 
\arg \delta^{\pm} (\lambda_{n};\widehat{\lambda}_{0}) + 
\sum\limits_{l \in L_{\pm}} \! \! \arg \! \left(\! 
\frac{(\lambda_{n} - \overline{\lambda_{l}})(\lambda_{n} 
+ \overline{\lambda_{l}})}{(\lambda_{n} - \lambda_{l}) 
(\lambda_{n} + \lambda_{l})} \! \right) \!,& \\ 
&\widetilde{x}_{n}^{\pm} = - \ln (\Delta_{n} \sin 
\gamma_{n}) + \ln \! \vert C_{n} \vert - 2 \ln \! 
\vert \delta^{\pm} (\lambda_{n};\widehat{\lambda}_{0}) 
\vert + 2 \sum\limits_{l \in L_{\pm}} \! \! \ln \! 
\left(\! \frac{\vert \lambda_{n} - \lambda_{l} \vert 
\vert \lambda_{n} + \lambda_{l} \vert}{\vert \lambda_{
n} - \overline{\lambda_{l}} \vert \vert \lambda_{n} + 
\overline{\lambda_{l}} \vert} \! \right) \!,& \\ 
&v_{\pm}^{{\cal C}} (x,t) = \sqrt{\pm \frac{\nu 
(\widehat{\lambda}_{0})}{2 \widehat{\lambda}_{0}^{2} s t}} 
\exp \{ i (\phi^{\pm} (\widehat{\lambda}_{0}) + 
\widetilde{\Phi}^{\pm} (\widehat{\lambda}_{0};t) + 
\frac{\pi}{2})\},& \\
&\widetilde{\Phi}^{\pm} (\widehat{\lambda}_{0};t) = 2 
\widehat{\lambda}_{0}^{4} t \mp \nu (\widehat{\lambda}_{0}) 
\ln \! \vert t \vert \pm \arg \Gamma (i \nu 
(\widehat{\lambda}_{0})) + \arg r (\widehat{\lambda}_{0}) 
\mp 2 \nu(\widehat{\lambda}_{0}) \ln 2& \nonumber \\
&\! \! \! \! \! \! \! \! \! \! \! \! \! \! \! \! \! \! \! 
\! \! \! \! \! \! \! \! \! \! \! + \, \, (2 \pm 1) 
\frac{\pi}{4} + 2 \sum\limits_{l \in L_{\pm}} \! \! \arg 
\! \left(\! \frac{(\widehat{\lambda}_{0} - 
\overline{\lambda_{l}}) (\widehat{\lambda}_{0} + 
\overline{\lambda_{l}})}{(\widehat{\lambda}_{0} - 
\lambda_{l}) (\widehat{\lambda}_{0} + \lambda_{l})} \! 
\right) \!,& \\
&v_{\pm}^{{\cal S} {\cal C}}(x,t) = - \frac{4 (\widetilde{
\Xi}^{\pm})^{2} \overline{\widetilde{g}_{n}^{\pm}} \vert 
\widetilde{g}_{n}^{\pm} \vert}{\eta_{n}} \sqrt{\pm \frac{ 
\nu (\widehat{\lambda}_{0})}{2 \widehat{\lambda}_{0}^{2} s 
t}} \{ \exp (i \widetilde{\varphi}_{n}^{\pm} (\widehat{
\lambda}_{0};t)) + 2 i \cot (\gamma_{n}) \cos (\widetilde{
\varphi}_{n}^{\pm}(\widehat{\lambda}_{0};t)) \},& \\
&\widetilde{g}^{\pm}_{n} := \vert \widetilde{g}^{\pm}_{n} 
\vert \exp \{i \arg \widetilde{g}^{\pm}_{n}\},& \nonumber \\
&\vert \widetilde{g}^{\pm}_{n} \vert = \vert C_{n} \vert 
\vert \delta^{\pm} (\lambda_{n};\widehat{\lambda}_{0}) 
\vert^{-2} \exp \{- 2 \Delta_{n}^{2} \sin (\gamma_{n}) 
\widehat{l}_{n}(t) \} \exp \! \left\{\! 2 \! \sum\limits_{
l \in L_{\pm}} \! \! \ln \! \left(\! \frac{\vert 
\lambda_{n} - \lambda_{l} \vert \vert \lambda_{n} + 
\lambda_{l} \vert}{\vert \lambda_{n} - \overline{\lambda_{
l}} \vert \vert \lambda_{n} + \overline{\lambda_{l}} \vert} 
\! \right) \! \right\} \!,& \\
&\arg \widetilde{g}^{\pm}_{n} = \arg C_{n} - 2 \arg 
\delta^{\pm} (\lambda_{n};\widehat{\lambda}_{0}) + 2 
\sum\limits_{l \in L_{\pm}} \! \! \arg \! \left(\! 
\frac{(\lambda_{n} - \lambda_{l}) (\lambda_{n} + 
\lambda_{l})}{(\lambda_{n} - \overline{\lambda_{l}}) 
(\lambda_{n} + \overline{\lambda_{l}})} \! \right)& 
\nonumber \\ 
&\! \! \! \! \! \! \! \! \! \! \! \! \! \! \! \! \! \! \! 
\! \! \! \! \! \! \! \! \! \! \! \! \! \! \! \! \! \! \! 
\! \! \! \! \! \! \! \! \! \! \! \! \! \! \! \! - \, \, 
2 \Delta_{n}^{2} (t \Delta_{n}^{2} + \widehat{l}_{n}(t) 
\cos \gamma_{n}),& \\
&\widetilde{\Xi}^{\pm} = \frac{\exp \{ \frac{i \gamma_{n}}{2} 
\} \exp \{ 2 \Delta_{n}^{2} \sin(\gamma_{n}) \widehat{l}_{n} 
(t) - \widetilde{x}_{n}^{\pm} \}}{2 \cosh (\frac{i 
\gamma_{n}}{2} + 2 \Delta_{n}^{2} \sin (\gamma_{n}) 
\widehat{l}_{n}(t) - \widetilde{x}_{n}^{\pm})},& \\
&\widetilde{\varphi}_{n}^{\pm}(\widehat{\lambda}_{
0};t) := \arg \widetilde{g}_{n}^{\pm} + \phi^{
\pm} (\widehat{\lambda}_{0}) + \widetilde{\Phi}^{
\pm} (\widehat{\lambda}_{0};t),& \\
&w_{{\rm as}}^{\pm} (x,t) = \exp \! \left\{\! i \!
\left(\! - 4 \sum\limits_{l \in L_{\pm}} \! \gamma_{
l} + 4 \arctan (\eta_{n} \vert \widetilde{g}_{n}^{
\pm} \vert^{-2} + \cot \gamma_{n}) + {\cal Y}_{\pm} 
(\widehat{\lambda}_{0}) + \frac{t}{2 s^{2}} (4 
\widehat{\lambda}_{0}^{2} s \right. \right.& \nonumber \\
&\left. \left. + 1) + 4 \sqrt{\pm 
\frac{\nu(\widehat{\lambda}_{0})}{\widehat{
\lambda}_{0}^{2} t}} \frac{ \vert \widetilde{g}_{n}^{\pm} 
\vert \sin(\gamma_{n}) (\vert \widetilde{g}_{n}^{\pm} 
\vert^{2} \cos (\widetilde{\varphi}_{n}^{\pm} (\widehat{
\lambda}_{0};t) - \gamma_{n}) - \eta_{n} \sin(\gamma_{n}) 
\cos(\widetilde{\varphi}_{n}^{\pm}(\widehat{\lambda}_{0};
t)))}{((\eta_{n} \sin \gamma_{n} + \vert 
\widetilde{g}_{n}^{\pm} \vert^{2} \cos \gamma_{n})^{2} + 
\vert \widetilde{g}_{n}^{\pm} \vert^{4} \sin^{2} \! 
\gamma_{n})} \right. \right.& \nonumber \\
&\left. \left. - \frac{2}{\sqrt{\pm t}} \int_{
\raise-0.75ex\hbox{$\scriptstyle{}\widehat{\lambda}_{0}$}
}^{\infty} \frac{\sqrt{\nu(\mu)}}{\mu^{2}} \! \left(\Re \{
R^{\pm}(0)\} \cos(\widetilde{\Theta}^{\pm}(\mu;t)) + \Im 
\{R^{\pm}(0)\} \sin(\widetilde{\Theta}^{\pm}(\mu;t)) \right) 
\! \frac{d \mu}{\pi} \! \right) \! \right\} \!,& \\
&\widetilde{\Theta}^{\pm}(\widehat{\lambda}_{0};t) := 
\widetilde{\Phi}^{\pm}(\widehat{\lambda}_{0};t) + \phi^{
\pm}(\widehat{\lambda}_{0}) + {\cal Y}_{\pm}(\widehat{
\lambda}_{0}),&
\end{eqnarray}
and $C(\widehat{\lambda}_{0}) \! \in \! {\cal S}(\Bbb R_{>M};\! 
\Bbb C)$, and, as $t \! \rightarrow \! \pm \infty$ and $x \! \to 
\! \mp \infty$ or $\pm \infty$ such that $\widehat{\mu}_{0} \! 
:= \! \sqrt{\frac{1}{2}(\frac{1}{s} \! - \! \frac{x}{t})} 
\! > \! M$, $\frac{x}{t} \! < \! \frac{1}{s}$, $s \! \in \! \Bbb 
R_{> 0}$, and $(x,t) \! \in \! \widetilde{\mho}_{n} \! := \! \{
\mathstrut (x,t); \, x - t (\frac{1}{s} \! - \! 2 \Delta_{n}^{
2} \cos \gamma_{n}) \! := \! - \widehat{l}_{n}(t) \! = \! 
{\cal O}(1)\}$, for those $\gamma_{n} \! \in \! (0,\frac{\pi}{
2})$,
\begin{eqnarray}
&u(x,t) = v_{{\rm as}}^{\pm \prime}(x,t) w_{{\rm as}}^{\pm 
\prime}(x,t) + {\cal O} \! \left(\frac{C(\widehat{\mu}_{0})(\ln 
\vert t \vert)^{2}}{t} \right) \!,&
\end{eqnarray}
where
\begin{eqnarray}
&v^{\pm \prime}_{{\rm as}}(x,t) := v_{\pm}^{{\cal S} \prime}
(x,t) + v_{\pm}^{{\cal C} \prime}(x,t) + v_{\pm}^{{\cal S}{
\cal C} \prime}(x,t),&
\end{eqnarray}
with
\begin{eqnarray}
&v_{\pm}^{{\cal S} \prime}(x,t) = \frac{\sqrt{2} \Delta_{n} 
\sin (\gamma_{n}) \exp \{- \frac{i \gamma_{n}}{2} \} \exp \{ 
2 i (\Delta_{n}^{2} (t \Delta_{n}^{2} + \widehat{l}_{n}(t) 
\cos \gamma_{n}) + \widetilde{\phi}_{n}^{\pm \prime})\}}{
\sqrt{s} \sinh (\frac{i \gamma_{n}}{2} + 2 \Delta_{n}^{2} 
\sin(\gamma_{n}) \widehat{l}_{n}(t) + \widetilde{x}_{n}^{\pm 
\prime})},& \\
&\widetilde{\phi}_{n}^{\pm \prime} = - \frac{1}{2} \arg 
C_{n} + \arg \delta^{\pm}_{\flat} (\overline{\lambda_{n}};
\widehat{\mu}_{0}) - \sum\limits_{l \in L_{\pm}} \! \! \arg 
\! \left(\! \frac{(\lambda_{n} - \overline{\lambda_{l}}) 
(\lambda_{n} + \overline{\lambda_{l}})}{(\lambda_{n} - 
\lambda_{l})(\lambda_{n} + \lambda_{l})} \! \right) 
\!,& \\ 
&\widetilde{x}_{n}^{\pm \, \prime} = - \ln (\Delta_{n} 
\sin \gamma_{n}) + \ln \! \vert C_{n} \vert - 2 \ln \! 
\vert \delta^{\pm}_{\flat} (\overline{\lambda_{n}};
\widehat{\mu}_{0}) \vert + 2 \sum\limits_{l \in L_{\pm}} 
\! \! \ln \! \left(\! \frac{\vert \lambda_{n} - 
\lambda_{l} \vert \vert \lambda_{n} + \lambda_{l} 
\vert}{\vert \lambda_{n} - \overline{\lambda_{l}} \vert 
\vert \lambda_{n} + \overline{\lambda_{l}} \vert} \! 
\right) \!,& \\
&v_{\pm}^{{\cal C} \prime}(x,t) = \sqrt{\mp \frac{\nu 
(i \widehat{\mu}_{0})}{2 \widehat{\mu}_{0}^{2} s t}} 
\exp \{i (\phi^{\pm \prime}(\widehat{\mu}_{0}) + 
\widetilde{\Phi}^{\pm \prime} (\widehat{\mu}_{0};t) +
\pi) \},& \\
&\widetilde{\Phi}^{\pm \prime} (\widehat{\mu}_{0};t) = 2 
\widehat{\mu}_{0}^{4} t \mp \nu (i \widehat{\mu}_{0}) \ln 
\! \vert t \vert \pm \arg \Gamma (i \nu (i \widehat{\mu}_{
0})) + \arg r(i \widehat{\mu}_{0}) \mp 2 \nu(i \widehat{
\mu}_{0}) \ln 2& \nonumber \\
&\! \! \! \! \! \! \! \! \! \! \! \! \! \! \! \! \! \! 
\! \! \! \! \! \! \! \! \! \! \! \! \! \! \! \! \! \! 
\! - \, \, (2 \mp 1) \frac{\pi}{4} - 2 \sum\limits_{l 
\in L_{\pm}} \! \! \arg \! \left(\! \frac{(\widehat{\mu}_{0} 
- \overline{\lambda_{l}})(\widehat{\mu}_{0} + \overline{
\lambda_{l}})}{(\widehat{\mu}_{0} - \lambda_{l}) 
(\widehat{\mu}_{0} + \lambda_{l})} \! \right) \!,& \\
&v_{\pm}^{{\cal S} {\cal C} \prime}(x,t) \! = \! - 
\frac{4 i (\widetilde{\Xi}^{\pm \prime})^{2} 
\overline{\widetilde{g}_{n}^{\pm \prime}} \vert 
\widetilde{g}_{n}^{\pm \prime} \vert}{\eta_{n}} \sqrt{\mp 
\frac{\nu(i \widehat{\mu}_{0})}{2 \widehat{\mu}_{0}^{2} s 
t}} \{ \exp (i \widetilde{\varphi}_{n}^{\pm \prime} 
(\widehat{\mu}_{0};t)) \! - \! 2 i \cot (\gamma_{n}) \cos 
(\widetilde{\varphi}_{n}^{\pm \prime} (\widehat{\mu}_{0};
t))\} \!,& \\
&\widetilde{g}^{\pm \prime}_{n} := \vert \widetilde{
g}^{\pm \prime}_{n} \vert \exp \{i \arg \widetilde{g}^{\pm 
\prime}_{n}\},& \nonumber \\ 
&\vert \widetilde{g}^{\pm \prime}_{n} \vert = \vert C_{n} 
\vert \vert \delta^{\pm}_{\flat} (\overline{\lambda_{n}};
\widehat{\mu}_{0}) \vert^{-2} \exp \{2 \Delta_{n}^{2} \sin 
(\gamma_{n}) \widehat{l}_{n}(t)\} \exp \! \left\{\! 2 \! 
\sum\limits_{l \in L_{\pm}} \! \! \ln \! \left(\! 
\frac{\vert \lambda_{n} - \lambda_{l} \vert \vert \lambda_{
n} + \lambda_{l} \vert}{\vert \lambda_{n} - \overline{
\lambda_{l}} \vert \vert \lambda_{n} + \overline{\lambda_{
l}} \vert} \! \right) \! \right\} \!,& \\
&\arg \widetilde{g}^{\pm \prime}_{n} = \arg C_{n} - 2 \arg 
\delta^{\pm}_{\flat}(\overline{\lambda_{n}};\widehat{\mu
}_{0}) - 2 \sum\limits_{l \in L_{\pm}} \! \! \arg \! \left(
\! \frac{(\lambda_{n} - \lambda_{l}) (\lambda_{n} + 
\lambda_{l})}{(\lambda_{n} - \overline{\lambda_{l}}) 
(\lambda_{n} + \overline{\lambda_{l}})} \! \right)& 
\nonumber \\
&\! \! \! \! \! \! \! \! \! \! \! \! \! \! \! \! \! \! \! 
\! \! \! \! \! \! \! \! \! \! \! \! \! \! \! \! \! \! \! 
\! \! \! \! \! \! \! \! \! \! \! \! \! \! \! - \, \, 2 
\Delta_{n}^{2} (t \Delta_{n}^{2} + \widehat{l}_{n} (t) \cos 
\gamma_{n}),& \\
&\widetilde{\Xi}^{\pm \prime} = - \frac{\exp \{- \frac{i 
\gamma_{n}}{2}\} \exp \{-2 \Delta_{n}^{2} \sin(\gamma_{n}) 
\widehat{l}_{n}(t) - \widetilde{x}_{n}^{\pm \prime}\}}{2 
\sinh (\frac{i \gamma_{n}}{2} + 2 \Delta_{n}^{2} \sin 
(\gamma_{n}) \widehat{l}_{n}(t) + \widetilde{x}_{n}^{\pm 
\prime})},& \\
&\widetilde{\varphi}_{n}^{\pm \prime}(\widehat{\mu}_{0};t) 
:= \arg \widetilde{g}_{n}^{\pm \prime} + \phi^{\pm 
\prime} (\widehat{\mu}_{0}) + \widetilde{\Phi}^{\pm \prime} 
(\widehat{\mu}_{0};t),& \\
&w_{{\rm as}}^{\pm \prime} (x,t) = \exp \! \left\{\! i \!
\left(\! 4 \sum\limits_{l \in L_{\pm}} \! \gamma_{l} + 4 
\arctan (\eta_{n} \vert \widetilde{g}_{n}^{\pm \prime} 
\vert^{-2} - \cot \gamma_{n}) + {\cal Y}_{\pm}^{\prime} 
(\widehat{\mu}_{0}) + \frac{t}{2 s^{2}} (- 4 \widehat{
\mu}_{0}^{2} s \right. \right.& \nonumber \\
&\left. \left. + 1) - 4 \sqrt{\mp \frac{\nu (i \widehat{
\mu}_{0})}{\widehat{\mu}_{0}^{2} t}} \frac{\vert 
\widetilde{g}_{n}^{\pm \prime} \vert \sin(\gamma_{n}) 
(\vert \widetilde{g}_{n}^{\pm \prime} \vert^{2} \cos 
(\widetilde{\varphi}_{n}^{\pm \prime}(\widehat{\mu}_{0};
t) + \gamma_{n}) + \eta_{n} \sin(\gamma_{n}) \cos(
\widetilde{\varphi}_{n}^{\pm \prime} (\widehat{\mu}_{
0};t)))}{((\eta_{n} \sin \gamma_{n} - \vert \widetilde{
g}_{n}^{\pm \prime} \vert^{2} \cos \gamma_{n})^{2} + 
\vert \widetilde{g}_{n}^{\pm \prime} \vert^{4} \sin^{2}
\! \gamma_{n})} \right. \right.& \nonumber \\
&\left. \left. - \frac{2}{\sqrt{\pm t}} \int_{
\raise-0.75ex\hbox{$\scriptstyle{}\widehat{\mu}_{0}$}}^{
\infty} \frac{\sqrt{-\nu(i\mu)}}{\mu^{2}} \! \left(\Re \{
R^{\pm \prime}(0)\} \cos(\widetilde{\Theta}^{\pm \prime}
(\mu;t)) + \Im \{R^{\pm \prime}(0)\} \sin(\widetilde{
\Theta}^{\pm \prime}(\mu;t)) \right) \! \frac{d\mu}{\pi} 
\! \right) \! \right\} \!,& \\
&\widetilde{\Theta}^{\pm \prime}(\widehat{\mu}_{0};t) 
:= \widetilde{\Phi}^{\pm \prime}(\widehat{\mu}_{0};
t) + \phi^{\pm \prime}(\widehat{\mu}_{0}) + {\cal Y}_{
\pm}^{\prime}(\widehat{\mu}_{0}),&
\end{eqnarray}
and $C(\widehat{\mu}_{0}) \! \in \! {\cal S}(\Bbb R_{>M};\! 
\Bbb C)$.
\end{ddd}

One possible application of the asymptotic results obtained 
in Theorems~2.1--2.3 is associated with the so-called ``soliton
scattering'', namely, the calculation of the position and phase 
shifts of the $n$th soliton $(1 \! \leq \! n \! \leq \! N)$ 
for $Q(x,t)$, $q(x,t)$, and $u(x,t)$ in the presence of the 
continuous spectrum: other physical applications of these 
asymptotic results include, for example, the calculation of the 
temporal and spectral intensities for the solutions of the DNLSE 
and MNLSE.
\begin{fff} \hfill
\begin{enumerate}
\item[{\rm {\bf (A)}}] $Q(x,t) \! :$
\begin{eqnarray*}
\Delta x_{n}^{Q^{{\cal S}}} & := & (2 \eta_{n})^{-1}(
\widehat{x}^{+}_{n} - \widehat{x}^{-}_{n}) \\
 & = & \eta_{n}^{-1} \left\{\sum\limits_{l=1 \atop \not= n}^{
N}{\rm sgn\/}(l-n) \ln \! \left(\! \frac{\vert \lambda_{n} - 
\lambda_{l} \vert \vert \lambda_{n} + \lambda_{l} \vert}{\vert 
\lambda_{n} - \overline{\lambda_{l}} \vert \vert \lambda_{n} + 
\overline{\lambda_{l}} \vert} \! \right) \! - \! \ln \! \left(
\! \frac{\vert \delta^{+}(\lambda_{n};\! \lambda_{0}) \vert}{
\vert \delta^{-}(\lambda_{n};\! \lambda_{0}) \vert} \! 
\right) \! \right\} \!,
\end{eqnarray*}
\begin{eqnarray*}
\Delta \phi_{n}^{Q^{{\cal S}}} & := & 2 (\widehat{\phi}_{
n}^{+} - \widehat{\phi}_{n}^{-}) \\
 & = & 2 \left\{\sum\limits_{l=1 \atop \not= n}^{N}{\rm sgn\/} 
(l-n) \arg \! \left(\! \frac{(\lambda_{n} - \overline{\lambda_{
l}})(\lambda_{n} + \overline{\lambda_{l}})}{(\lambda_{n} - 
\lambda_{l})(\lambda_{n} + \lambda_{l})} \! \right) \! + \! \arg 
\! \left(\! \frac{\delta^{+}(\lambda_{n};\! \lambda_{0})}{
\delta^{-}(\lambda_{n};\! \lambda_{0})} \! \right) \! \right\} 
\!,
\end{eqnarray*}
\begin{eqnarray*}
\Delta x_{n}^{Q^{{\cal S} \prime}} & := & (2 \eta_{n})^{
-1}(\widehat{x}^{+ \prime}_{n} - \widehat{x}^{- \prime}_{n}) \\
 & = & \eta_{n}^{-1} \left\{\sum\limits_{l=1 \atop \not= n}^{N}
{\rm sgn\/}(l-n) \ln \! \left(\! \frac{\vert \lambda_{n} - 
\lambda_{l} \vert \vert \lambda_{n} + \lambda_{l} \vert}{\vert 
\lambda_{n} - \overline{\lambda_{l}} \vert \vert \lambda_{n} + 
\overline{\lambda_{l}} \vert} \! \right) \! - \! \ln \! \left(
\! \frac{\vert \delta^{+}_{\flat}(\overline{\lambda_{n}};\mu_{
0}) \vert}{\vert \delta^{-}_{\flat}(\overline{\lambda_{n}};
\mu_{0}) \vert} \! \right) \! \right\} \!,
\end{eqnarray*}
\begin{eqnarray*}
\Delta \phi_{n}^{Q^{{\cal S} \prime}} & := & 2(\widehat{
\phi}_{n}^{+ \prime} - \widehat{\phi}_{n}^{- \prime}) \\
 & = & - 2 \left\{\sum\limits_{l=1 \atop \not= n}^{N} {\rm 
sgn\/} (l-n) \arg \! \left(\! \frac{(\lambda_{n} - \overline{
\lambda_{l}})(\lambda_{n} + \overline{\lambda_{l}})}{
(\lambda_{n} - \lambda_{l})(\lambda_{n} + \lambda_{l})} \! 
\right) \! - \! \arg \! \left(\! \frac{\delta^{+}_{\flat}
(\overline{\lambda_{n}};\mu_{0})}{\delta^{-}_{\flat}
(\overline{\lambda_{n}};\mu_{0})} \! \right) \! \right\} \!;
\end{eqnarray*}
\item[{\rm {\bf (B)}}] $q(x,t)$ {\rm (DNLSE):}
\begin{eqnarray*}
&\Delta x_{n}^{q^{{\cal S}}} = \Delta x_{n}^{Q^{{\cal S}}},& \\
&\Delta \phi_{n}^{q^{{\cal S}}} = \Delta \phi_{n}^{Q^{{\cal S}}}
-4 \sum\limits_{l=1 \atop \not= n}^{N} {\rm sgn\/}(l-n) \gamma_{
l}+{\cal Y}_{+}(\lambda_{0})-{\cal Y}_{-}(\lambda_{0}),& \\
&\Delta x_{n}^{q^{{\cal S} \prime}}=\Delta x_{n}^{Q^{{\cal S} 
\prime}},& \\
&\Delta \phi_{n}^{q^{{\cal S} \prime}}=\Delta \phi_{n}^{Q^{{
\cal S} \prime}}+4 \sum\limits_{l=1 \atop \not= n}^{N} {\rm 
sgn\/}(l-n) \gamma_{l}+{\cal Y}_{+}^{\prime}(\mu_{0})-{\cal Y}_{
-}^{\prime}(\mu_{0});&
\end{eqnarray*}
\item[{\rm {\bf (C)}}] $u(x,t)$ {\rm (MNLSE):}
\begin{eqnarray*}
&\Delta x_{n}^{u^{{\cal S}}} = \Delta x_{n}^{Q^{{\cal S}}} 
\vert_{\lambda_{0} \to \widehat{\lambda}_{0}},& \\
&\Delta \phi_{n}^{u^{{\cal S}}} = \Delta \phi_{n}^{q^{{\cal S}}} 
\vert_{\lambda_{0} \to \widehat{\lambda}_{0}},& \\
&\Delta x_{n}^{u^{{\cal S} \prime}}=\Delta x_{n}^{Q^{{\cal S} 
\prime}} \vert_{\mu_{0} \to \widehat{\mu}_{0}},& \\
&\Delta \phi_{n}^{u^{{\cal S} \prime}}=\Delta \phi_{n}^{q^{{
\cal S} \prime}} \vert_{\mu_{0} \to \widehat{\mu}_{0}}.&
\end{eqnarray*}
\end{enumerate}
\end{fff}

{\em Proof.\/} Follows {}from the definition of soliton position
and phase shifts given in \cite{a2} and Theorems~2.1--2.3, 
Eqs.~(16), (17), (33), (34), (48), (53), (60), (61), (69), 
(74), (75) and (83). \hfill \rule{6.5pt}{6.5pt}
\begin{eee}
{\rm The expressions for the soliton phase shifts given in 
Corollary~2.1, namely, $\Delta \phi_{n}^{Q^{{\cal S}}}$, 
$\Delta \phi_{n}^{Q^{{\cal S} \prime}}$, $\Delta \phi_{n}^{q^{{
\cal S}}}$, $\Delta \phi_{n}^{q^{{\cal S} \prime}}$, $\Delta 
\phi_{n}^{u^{{\cal S}}}$, and $\Delta \phi_{n}^{u^{{\cal S} 
\prime}}$, $1 \! \leq \! n \! \leq \! N$, are to be understood 
$\bmod (2 \pi)$.\/}
\end{eee}
\begin{eee}
{\rm For the asymptotics of the $\Bbb C$-valued functions 
$Q(x,t)$, $q(x,t)$, and $u(x,t)$, one must actually consider 
four different cases, depending, respectively, on the 
quadrant of the $(x,t)$-plane. In this paper, the proof of 
the asymptotic expansions for $Q(x,t)$ and $q(x,t)$ (resp.~$
u(x,t))$ is presented for the cases $(x,t) \! \rightarrow \! 
(\mp \infty,\pm \infty)$ (resp.~$(x,t) \! \rightarrow \! 
(\pm \infty,\pm \infty))$ such that $\lambda_{0} \! > \! M$ 
and $(x,t) \! \in \! \Omega_{n}$ (resp.~$\widehat{\lambda}_{
0} \! > \! M$ and $(x,t) \! \in \! \widetilde{\Omega}_{n})$ 
for those $\gamma_{n} \! \in \! (\frac{\pi}{2},\! \pi)$: the 
results for the remaining domains of the $(x,t)$-plane are 
obtained analogously. If the conditions on $\gamma_{n}$ stated 
in Theorems~2.1--2.3 are violated, then $(x,t) \! \in \! \{
\Bbb R^{2} \setminus \Omega_{n},\Bbb R^{2} \setminus \widetilde{
\Omega}_{n},\Bbb R^{2} \setminus \mho_{n},\Bbb R^{2} \setminus 
\widetilde{\mho}_{n}\}$, but the asymptotic results stated still 
remain valid although the second terms on the right-hand sides 
of the asymptotic expansions become the leading-order terms of 
the corresponding asymptotic expansions, while the remaining 
terms are exponentially small and negligible with respect to the 
given error estimations.\/}
\end{eee}
\section{The Model RH Problem}
In order to simplify the asymptotic analysis of the original 
RH problem formulated in Lemma~2.1, a simpler, model RH 
problem (see Lemma~3.3) is derived in this section. As an 
intermediate step towards the formulation of the model RH 
problem, it will be convenient to derive an ``extended'' RH 
problem (see Lemma~3.2): the general idea pertaining to the 
transformations {}from the original RH problem to the model 
one is elucidated in the paragraph following Lemma~3.1 (see 
below).
\begin{bbb}[{\rm \cite{a7}}]
In the solitonless sector $({\cal Z}_{d} \! \equiv \! 
\emptyset)$, as $t \! \rightarrow \! + \infty$ and $x 
\! \rightarrow \! - \infty$ such that $\lambda_{0} \! 
:= \! \frac{1}{2} \sqrt{- \frac{x}{t}} \! > \! M$,
\begin{eqnarray*}
&m(x,t;\! \lambda)=\Delta(\lambda)+{\cal O} \! \left
(\frac{C(\lambda_{0})}{\sqrt{t}} \right) \!,&
\end{eqnarray*}
where $\Delta(\lambda) \! := \! (\delta^{+}(\lambda;\! 
\lambda_{0}))^{\sigma_{3}}$,
\begin{eqnarray*}
&\delta^{+}(\lambda;\! \lambda_{0}) = \left(\! \left(\frac{ 
\lambda - \lambda_{0}}{\lambda} \right) \! \! \left(\frac{ 
\lambda + \lambda_{0}}{\lambda} \right) \! \right)^{i \nu} 
\exp \{ \sum\limits_{l \in \{\pm\}} (\rho_{l}(\lambda) + 
\widehat{\rho}_{l}
(\lambda)) \},& \\
&\rho_{\pm}(\lambda) = \frac{1}{2 \pi i} \int_{0}^{\pm 
\lambda_{0}} \ln \! \left(\frac{1 - \vert r(\varsigma) \vert^{
2}}{1 - \vert r(\lambda_{0}) \vert^{2}} \right) \! \frac{d 
\varsigma}{(\varsigma-\lambda)}, \, \, \, \, \, \, \, \, \, 
\widehat{\rho}_{\pm}(\lambda) = \int_{\pm i \infty}^{i 0} \! 
\frac{\ln (1 - r(\varsigma) \overline{r(\overline{\varsigma})}
)}{(\varsigma - \lambda)} \frac{d \varsigma}{2 \pi i},&
\end{eqnarray*}
$\nu \! := \! \nu(\lambda_{0})$ is given by Eq.~(21), 
$\vert \vert (\delta^{+}(\cdot;\! \lambda_{0}))^{\pm 1} \vert 
\vert_{{\cal L}^{\infty}(\Bbb C;\Bbb C)} \! := \! \sup_{
\lambda \in \Bbb C} \vert (\delta^{+}(\lambda;\! \lambda_{0})
)^{\pm 1} \vert \! < \! \infty$, $\left(\overline{\delta^{+}(
\pm \overline{\lambda};\! \lambda_{0})} \right)^{-1} \! = \! 
\delta^{+}(\lambda;\! \lambda_{0})$, the principal branch of 
the logarithmic function is taken, $\ln(\mu \! - \! \lambda) 
\! := \! \ln \! \vert \mu \! - \! \lambda \vert \! + \! i 
\arg (\mu \! - \! \lambda)$, $\arg(\mu \! - \! \lambda) \! \in 
\! (-\pi,\pi)$, and $C(\lambda_{0}) \! \in \! {\cal S}(\Bbb 
R_{>M};\! M_{2}(\Bbb C))$.
\end{bbb}
\begin{eee}
{\rm For notational convenience, until the end of Sec.~5, 
all explicit $x,t$ dependences are suppressed, except where 
absolutely necessary, and $\delta^{+}(\lambda;\! \lambda_{
0}) \! := \! \delta(\lambda)$.\/}
\end{eee}
\begin{ccc}
There exists a unique solution $m^{\Delta}(\lambda) \colon 
\Bbb C \! \setminus \! ({\cal Z}_{d} \! \cup \! \widehat{
\Gamma}) \! \rightarrow \! {\rm SL\/}(2,\! \Bbb C)$ of the 
following RH problem,
\begin{enumerate}
\item $m^{\Delta}(\lambda)$ is meromorphic $\forall \, 
\lambda \! \in \! \Bbb C \! \setminus \! 
\widehat{\Gamma},$
\item 
\begin{eqnarray*}
&m^{\Delta}_{+}(\lambda) = m^{\Delta}_{-}(\lambda) 
v^{\Delta}(\lambda), \, \, \, \, \, \, \lambda \in 
\widehat{\Gamma},& 
\end{eqnarray*}
where 
\begin{eqnarray*}
&v^{\Delta}(\lambda) = e^{-i \theta(\lambda) {\rm ad} 
(\sigma_{3})} 
\! \left(\! \begin{array}{cc}
(1 - r(\lambda) \overline{r (\overline{\lambda})} ) 
\delta_{-} (\lambda) (\delta_{+}(\lambda))^{-1} & 
r(\lambda) \delta_{-}(\lambda) \delta_{+} 
(\lambda) \\
- \overline{r (\overline{\lambda})} (\delta_{-} 
(\lambda))^{-1} (\delta_{+} (\lambda))^{-1} & 
(\delta_{-}(\lambda))^{-1} \delta_{+}(\lambda)
\end{array} \! \right) \!,&
\end{eqnarray*}
and $\theta(\lambda) \! := \! \lambda^{2} x \! + \! 2 
\lambda^{4} t,$ 
\item $m^{\Delta}(\lambda)$ has simple poles at $\{\pm 
\lambda_{i},\pm \overline{\lambda_{i}}\}_{i=1}^{N}$ with 
$(1 \! \leq \! i \! \leq \! N)$
\begin{eqnarray*}
&{\rm res}(m^{\Delta}(\lambda);\! \lambda_{i}) = 
\lim\limits_{\lambda \rightarrow \lambda_{i}} \! 
m^{\Delta}(\lambda) v_{i}(\delta(\lambda_{i}))^{-2} 
\sigma_{-},& \\
&{\rm res}(m^{\Delta} (\lambda);\! -\lambda_{i}) = - 
\sigma_{3} {\rm res}(m^{\Delta}(\lambda);\! \lambda_{i}) 
\sigma_{3},& \\
&{\rm res}(m^{\Delta}(\lambda);\! \overline{\lambda_{i}} \,) 
= \lim\limits_{\lambda \rightarrow \overline{\lambda_{i}}} 
\! m^{\Delta}(\lambda) \overline{v_{i}} (\delta(\overline{
\lambda_{i}}))^{2} \sigma_{+},& \\
&{\rm res}(m^{\Delta}(\lambda);\! -\overline{\lambda_{i}} \,) 
= - \sigma_{3}{\rm res}(m^{\Delta}(\lambda);\! \overline{
\lambda_{i}} \,) \sigma_{3},&
\end{eqnarray*}
\item as $\lambda \! \rightarrow \! \infty$, $\lambda \! 
\in \! \Bbb C \! \setminus \! ({\cal Z}_{d} \! \cup \! 
\widehat{\Gamma}),$
\begin{eqnarray*}
&m^{\Delta}(\lambda) = {\rm I} + {\cal O}(\lambda^{-1});& 
\end{eqnarray*}
\end{enumerate}
moreover, $Q(x,t) \! = \! 2 i \! \lim\limits_{\lambda \to \infty} 
\! (\lambda m^{\Delta}(x,t;\! \lambda))_{12}$ is equal to $Q(x,t)$ 
in Lemma~2.2, Eq.~(11).
\end{ccc}

{\em Proof.\/} Let $m(\lambda)$ be the solution of the RH 
problem formulated in Lemma~2.1. Define $m^{\Delta}(\lambda) 
\! := \! m(\lambda)(\Delta(\lambda))^{-1}$. 
\hfill \rule{6.5pt}{6.5pt}

In order to motivate Proposition~3.2 and Lemma~3.2 (see below), 
consider the trajectory of the $n$th soliton with $\gamma_{n} 
\! \in \! (\frac{\pi}{2},\! \pi)$ in the $(x,t)$-plane which 
belongs to the set $\Omega_{n} \! := \! \{\mathstrut (x,t); 
\, x \! - \! 4 t \Delta_{n}^{2} \cos \gamma_{n} \! = \! {\cal 
O}(1)\}$, and note {}from Lemma~2.1 and the soliton ordering in 
Sec.~2 that, as $t \! \rightarrow \! + \infty$ and $x \! 
\rightarrow \! - \infty$ such that $\lambda_{0} \! > \! M$ and 
$(x,t) \! \in \! \Omega_{n}$: (1) $\Re(v_{i} \vert_{\Omega_{n}}) 
\! \sim \! {\cal O} (\exp \{-8t \eta_{i}(\xi_{i} \! - \! \xi_{n}
)\}) \! \rightarrow \! 0 \, \forall \, i \! < \! n$ $(i \! \in 
\! \{1,2,\ldots,n \! - \! 1\})$; (2) $\Re(v_{i} \vert_{\Omega_{
n}}) \! \rightarrow \! \infty \, \forall \, i \! > \! n$ $(i \! 
\in \! \{n \! + \! 1,n \! + \! 2,\ldots,N\})$; and (3) $\Re(v_{
i} \vert_{\Omega_{n}}) \! \sim \! {\cal O}(1)$ for $i \! = \! n$. 
Thus, along the trajectory of the arbitrarily fixed 
$n$th soliton, there are exponentially growing polar 
conditions for solitons $i$ with $n \! + \! 1 \! \leq \! i 
\! \leq \! N$. One must effectively deal with such growing 
polar conditions in a self-consistent manner. In a recent 
paper \cite{a20} devoted to the asymptotics of the Toda 
rarefaction problem, Deift {\em et al.\/} showed how this 
could be done: they noticed that it is possible to replace the 
poles with the exponentially growing polar conditions by jump 
matrices on small, mutually disjoint ({\bf and} disjoint with 
respect to $\widehat{\Gamma})$ circles such that these jump 
matrices behave like ${\rm I}$ $+$  exponentially decreasing 
terms as $t \! \rightarrow \! + \infty$. Thus, instead of the 
original RH problem, one gets a new, ``extended'' RH problem 
with $4(N \! - \! n)$ fewer poles, and $4(N \! - \! n)$ 
additional circles with jump conditions stated on them. Finally, 
by removing the added circles {}from the specification of the 
extended RH problem, one arrives at the model RH problem: the 
estimation of the ``difference'' between the extended and model 
RH problems shows that the solution of the model RH problem 
approximates the solution of the original one modulo terms 
which are decaying exponentially as $t \! \rightarrow \! + 
\infty$.
\begin{bbb}
Introduce arbitrarily small, clockwise- and 
counter-clockwise-o\-r\-i\-e\-n\-t\-e\-d, mutually disjoint 
(and disjoint with respect to $\widehat{\Gamma})$ circles 
$K_{j}^{\pm}$ and $L_{j}^{\pm}$, $n \! + \! 1 \! \leq \! j 
\! \leq \! N$, around the eigenvalues $\{\pm \lambda_{j}
\}_{j=n+1}^{N}$ and $\{\pm \overline{\lambda_{j}}\}_{j=n+1}^{
N}$, respectively, and define
\begin{eqnarray}
&m^{\flat}(\lambda) := 
\left\{ \begin{array}{l} 
m^{\Delta} (\lambda), \, \, \, \, \, \, \lambda \in 
\Bbb C \setminus (\widehat{\Gamma} \cup 
(\bigcup\limits_{i=n+1}^{N} (K_{i}^{\pm} 
\cup L_{i}^{\pm}))), \\
m^{\Delta}(\lambda) \! \left({\rm I} - 
\frac{v_{i}(\delta(\pm \lambda_{i}))^{-2}}{(\lambda 
\mp \lambda_{i})} \sigma_{-} \right) \!, \, \, \, \, \, 
\lambda \in {\rm int} K_{i}^{\pm}, \, \, \, \, \, n+1 
\leq i \leq N, \\
m^{\Delta}(\lambda) \! \left({\rm I} + \frac{ 
\overline{v_{i}} (\delta(\pm \overline{\lambda_{i}}
))^{2} }{(\lambda \mp \overline{\lambda_{i}})} 
\sigma_{+} \right) \!, \, \, \, \, \, \, \, \lambda \in 
{\rm int} L_{i}^{\pm}, \, \, \, \, \, \, \, n+1 \leq i 
\leq N.
\end{array} \right.&
\end{eqnarray}
Then $m^{\flat}(\lambda)$ solves a RH problem on $(\sigma_{
\pounds} \! \setminus \! \cup_{i=n+1}^{N}(\{\pm \lambda_{i}\} 
\! \cup \! \{\pm \overline{\lambda_{i}}\})) \! \cup \! 
(\cup_{i=n+1}^{N} (K_{i}^{\pm} \! \cup \! L_{i}^{\pm}))$ with 
the same jumps as $m^{\Delta}(\lambda)$ on $\widehat{\Gamma}$, 
$m_{+}^{\flat}(\lambda) \! = \! m_{-}^{\flat}(\lambda) 
v^{\Delta}(\lambda)$, and
\begin{eqnarray}
&m_{+}^{\flat} (\lambda) = 
\left\{ \begin{array}{l}
m_{-}^{\flat} (\lambda) \! \left({\rm I} + 
\frac{v_{i}(\delta(\pm \lambda_{i}))^{-2}}{(\lambda 
\mp \lambda_{i})} \sigma_{-} \right) \!, \, \, \, \, \, 
\, \lambda \in K_{i}^{\pm}, \, \, \, \, \, n+1 \leq i 
\leq N,
\\ m_{-}^{\flat} (\lambda) \! \left({\rm I} + 
\frac{\overline{v_{i}} (\delta(\pm 
\overline{\lambda_{i}}))^{2} }{(\lambda \mp 
\overline{\lambda_{i}})} \sigma_{+} \right) \!, \, 
\, \, \, \, \, \, \, \lambda \in L_{i}^{\pm}, \, \, 
\, \, \, \, \, n+1 \leq i \leq N.  
\end{array} \right.& \nonumber 
\end{eqnarray}
\end{bbb}

{\em Proof.\/} Follows {}from Lemma~3.1 and the definition 
of $m^{\flat}(\lambda)$. \hfill \rule{6.5pt}{6.5pt}
\begin{eee}
{\rm The superscripts $\pm$ on $\{K_{i}^{\pm}\}_{i=n+1}^{N}$ 
and $\{L_{i}^{\pm}\}_{i=n+1}^{N}$, which are related with $\{
\pm \lambda_{i}\}_{i=n+1}^{N}$ and $\{\pm \overline{\lambda_{
i}}\}_{i=n+1}^{N}$, respectively, should \underline{{\bf not}} 
be confused with the subscripts $\pm$ 
appearing in the various RH problems in Secs.~3--5, namely, 
$m_{\pm} (\lambda)$, $m^{\Delta}_{\pm}(\lambda)$, $m^{
\flat}_{\pm}(\lambda)$, $m_{\pm}^{\sharp}(\lambda)$, 
$\chi_{\pm}(\lambda)$, $E_{\pm}(\lambda)$, and $\chi^{c}_{
\pm}(\lambda)$.\/}
\end{eee}
\begin{eee}
{\rm Even though the exponentially growing polar (residue) 
conditions have been replaced by jump matrices, it should be 
noted that, along the trajectory of soliton $n$, these jump 
matrices are also exponentially growing as $t \! \rightarrow 
\! + \infty$. These lower/upper diagonal, exponentially growing 
jump matrices are now replaced, through a sequence of $N \! - 
\! n$ similar transformations, by upper/lower diagonal jump 
matrices which converge, along the trajectory of soliton $n$, 
to ${\rm I}$ as $t \! \to \! + \infty$.\/}
\end{eee}
\begin{ccc}
Set
\begin{eqnarray}
&m^{\sharp} (\lambda) \! := \! 
\left\{ \begin{array}{l} 
m^{\flat}(\lambda) \prod\limits_{l=n+1}^{N} 
(d_{l_{+}} (\lambda))^{- \sigma_{3}}, \, \, \,
\, \, \, \lambda \in \Bbb C \setminus 
(\widehat{\Gamma} \cup (\bigcup\limits_{i=n+1}^{N}
(K_{i}^{\pm} \cup L_{i}^{\pm}))), \\
m^{\flat}(\lambda)(J_{K_{i}^{\pm}}(\lambda))^{-1} 
\prod\limits_{l=n+1}^{N} (d_{l_{-}}(\lambda))^{- 
\sigma_{3}}, \, \, \, \, \, \, \lambda \in {\rm 
int} K_{i}^{\pm}, \, \, \, \, \, n+1 \leq i \leq N, \\
m^{\flat} (\lambda) (J_{L_{i}^{\pm}}(\lambda))^{-1} 
\prod\limits_{l=n+1}^{N}(d_{l_{-}}(\lambda))^{-\sigma_{
3}}, \, \, \, \, \, \, \, \lambda \in {\rm int} L_{i}^{
\pm}, \, \, \, \, \, \, n+1 \leq i \leq N, 
\end{array} \right.& 
\end{eqnarray}
where
\begin{eqnarray}
&\left. \begin{array}{l}
d_{l_{+}}(\lambda) \! := \! \frac{(\lambda-\overline{
\lambda_{l}}) (\lambda + \overline{\lambda_{l}})}{(
\lambda - \lambda_{l})(\lambda + \lambda_{l}) }, \, \, \,
\, \, \, \lambda \in \Bbb C \setminus (\bigcup\limits_{
i=n+1}^{N}(K_{i}^{\pm} \cup L_{i}^{\pm})), \, \, \, \, \,
n+1 \leq l \leq N, \\
d_{l_{-}}(\lambda) \! := \! \left\{\begin{array}{l}
\frac{(\lambda - \overline{\lambda_{l}})(\lambda + 
\overline{\lambda_{l}}) }{ (\lambda \pm \lambda_{l})}, \,
\, \, \, \, \, \lambda \in \bigcup\limits_{i=n+1}^{N} 
{\rm int} K_{i}^{\pm}, \, \, \, \, \, n+1 \leq l \leq N, 
\\
\frac{ (\lambda \pm \overline{\lambda_{l}})}{(\lambda - 
\lambda_{l}) (\lambda + \lambda_{l})}, \, \, \, \, \, \, 
\, \lambda \in \bigcup\limits_{i=n+1}^{N} {\rm int} L_{i}
^{\pm}, \, \, \, \, \, \, n+1 \leq l \leq N,
\end{array} \right. 
\end{array} \right.&
\end{eqnarray}
and the ${\rm SL\/}(2,\! \Bbb C)$-valued, holomorphic in ${\rm 
int} K_{i}^{\pm}$ and ${\rm int} L_{i}^{\pm}$, respectively, 
functions $J_{K_{i}^{\pm}}(\lambda)$ and $J_{L_{i}^{\pm}}(
\lambda)$, $n \! + \! 1 \! \leq \! i \! \leq \! N$, are given 
by
\begin{eqnarray*}
&J_{K_{i}^{\pm}}(\lambda) \! = \! 
\left( \begin{array}{cc}
\frac{ \prod\limits_{l=n+1 \atop \not= i}^{N} \frac{ 
d_{l_{-}}^{-1} (\lambda)}{d_{l_{+}}^{-1}(\lambda)} 
\, \, - \, \, \frac{v_{i}(\delta(\pm \lambda_{i}))^{-2} 
C_{i}^{\sharp}}{(d_{i_{-}}(\lambda))^{2}} \prod\limits_{
l=n+1 \atop \not= i}^{N} \frac{ d_{l_{-}}^{-1} (\lambda)}{
d_{l_{+}}(\lambda)}}{(\lambda \mp \lambda_{i})} & \, \, \, 
\, \frac{C_{i}^{\sharp}}{(d_{i_{-}}(\lambda))^{2} } 
\prod\limits_{l=n+1 \atop \not= i}^{N} \frac{d_{l_{-}}^{
-1}(\lambda)}{d_{l_{+}}(\lambda)} \\
- v_{i} (\delta(\pm \lambda_{i}))^{-2} \prod\limits_{ 
l=n+1 \atop \not=i}^{N} \frac{d_{l_{-}}(\lambda) 
}{d_{l_{+}}(\lambda)} & \, \, \, \, (\lambda \mp \lambda_{
i}) \prod\limits_{l=n+1 \atop \not=i}^{N} \frac{d_{l_{-}}
(\lambda)}{d_{l_{+}} (\lambda)} 
\end{array} \right) \!,& \\
&J_{L_{i}^{\pm}}(\lambda) \! = \! 
\left( \begin{array}{cc}
(\lambda \mp \overline{\lambda_{i}}) \prod\limits_{l=n+1 
\atop \not= i}^{N} \frac{d_{l_{-}}^{-1}(\lambda)}{d_{l_{
+}}^{-1}(\lambda)} & \, \, \, \, \overline{v_{i}}(\delta(\pm 
\overline{\lambda_{i}}))^{2} \prod\limits_{l=n+1 \atop 
\not=i}^{N} \frac{d_{l_{-}}^{-1}(\lambda)}{d_{l_{+}}^{-1}
(\lambda)} \\ 
- \frac{\overline{C_{i}^{\sharp}}}{(d_{i_{-}}(\lambda))^{
-2}} \prod\limits_{l=n+1 \atop \not= i}^{N} \frac{d_{l_{-}
}(\lambda)}{d_{l_{+}}^{-1}(\lambda)} & \, \, \, \, \frac{
\prod\limits_{l=n+1 \atop \not= i}^{N} \frac{d_{l_{-}}
(\lambda)}{d_{l_{+}}(\lambda) } \, \, - \, \, \frac{
\overline{v_{i}}(\delta(\pm \overline{\lambda_{i}}))^{2} 
\overline{C_{i}^{\sharp}}}{(d_{i_{-}}(\lambda))^{-2}} 
\prod\limits_{l=n+1 \atop \not=i}^{N} \frac{d_{l_{-}}(
\lambda)}{d_{l_{+}}^{-1}(\lambda)}}{(\lambda \mp \overline{
\lambda_{i}})} 
\end{array} \right) \!,& 
\end{eqnarray*}
with
\begin{eqnarray}
&C_{i}^{\sharp} \! = \! (v_{i})^{-1}(\delta(\pm \lambda_{i})
)^{2}(d_{i_{-}}(\pm \lambda_{i}))^{2} \prod\limits_{l=n+1 
\atop \not= i}^{N} (d_{l_{+}} (\pm \lambda_{i}))^{2}, \, \,
\, \, \, \, \, \, n+1 \leq i \leq N.&
\end{eqnarray}
Then $m^{\sharp}(\lambda) \colon \Bbb C \! \setminus \! 
(({\cal Z}_{d} \! \setminus \! \cup_{i=n+1}^{N} (\{\pm 
\lambda_{i}\} \! \cup \! \{\pm \overline{\lambda_{i}}\})) 
\cup (\widehat{\Gamma} \! \cup \! (\cup_{i=n+1}^{N} 
(K_{i}^{\pm} \! \cup \! L_{i}^{\pm})))) \! \rightarrow \! 
{\rm SL\/}(2,\! \Bbb C)$ solves the following, extended 
RH problem on $(\sigma_{\pounds} \! \setminus \! \cup_{
i=n+1}^{N} \! (\{\pm \lambda_{i}\} \! \cup \! \{\pm 
\overline{\lambda_{i}}\})) \! \cup \! (\cup_{i=n+1}^{N} 
(K_{i}^{\pm} \! \cup \! L_{i}^{\pm}))$,
\begin{eqnarray*}
&m_{+}^{\sharp}(\lambda) = m_{-}^{\sharp}(\lambda) 
e^{- i \theta(\lambda) {\rm ad} (\sigma_{3})} 
v^{\sharp}(\lambda),&
\end{eqnarray*}
where 
\begin{eqnarray}
&v^{\sharp}(\lambda) \vert_{\widehat{\Gamma}} \! = \! 
\left(\begin{array}{cc}
\! \! \! (1 - r(\lambda) \overline{r(\overline{\lambda}
)}) \frac{\delta_{-}(\lambda)}{\delta_{+}(\lambda)} & \, 
\, \frac{r(\lambda)}{(\delta_{-}(\lambda) \delta_{+} 
(\lambda))^{-1}} \prod\limits_{l=n+1}^{N} \! \! \left(\! 
\frac{(\lambda - \overline{\lambda_{l}})(\lambda + 
\overline{\lambda_{l}})}{(\lambda - \lambda_{l})(\lambda 
+ \lambda_{l})} \! \right)^{2} \\
\! \! \! - \frac{ \overline{r (\overline{\lambda})} 
}{ \delta_{-}(\lambda) \delta_{+}(\lambda)} 
\prod\limits_{l=n+1}^{N} \! \! \left(\! \frac{ (\lambda 
- \overline{\lambda_{l}}) (\lambda + \overline{\lambda_{
l}})}{(\lambda - \lambda_{l}) (\lambda + \lambda_{l})} 
\! \right)^{-2} & \, \, \frac{\delta_{+}(\lambda)}{\delta_{
-}(\lambda)} 
\end{array} \right) \!,& \nonumber \\
&v^{\sharp}(\lambda) \! = \! 
\left\{ \begin{array}{l}
{\rm I} + \frac{(v_{i})^{-1}(\delta(\pm \lambda_{i}))^{2}
}{(\lambda \mp \lambda_{i})} \! \left( \! \frac{\lambda_{
i}^{2}-\overline{\lambda_{i}}^{2}}{2 \lambda_{i}} \! 
\right)^{2} \! \prod\limits_{l=n+1 \atop \not=i}^{N} \! 
\left( \! \frac{\overline{\lambda_{l}}^{2}-\lambda_{i}^{
2}}{\lambda_{l}^{2}-\lambda_{i}^{2}} \! \right)^{2} \! \!
\sigma_{+}, \, \, \, \, \, \, \, \, \lambda \in 
\bigcup\limits_{i=n+1}^{N} K_{i}^{\pm}, \\
{\rm I} + \frac{(\overline{v_{i}})^{-1}(\delta(\pm \overline{
\lambda_{i}}))^{-2}}{(\lambda \mp \overline{\lambda_{i}})} \! 
\left(\! \frac{ \lambda_{i}^{2} - \overline{\lambda_{i}}^{2}
}{2 \overline{\lambda_{i}}} \! \right)^{2} \! \prod\limits_{l
=n+1 \atop \not=i}^{N} \! \left(\! \frac{\lambda_{l}^{2} - 
\overline{\lambda_{i}}^{2}}{\overline{\lambda_{l}}^{2} - 
\overline{\lambda_{i}}^{2}} \! \right)^{2} \! \! \sigma_{-}, 
\, \, \, \, \, \, \lambda \in \bigcup\limits_{i=n+1}^{N} L_{
i}^{\pm},
\end{array} \right.& 
\end{eqnarray}
with polar (residue) conditions, 
\begin{eqnarray*}
&{\rm res}(m^{\sharp}(\lambda);\! \lambda_{i}) \! = \! 
\lim\limits_{\lambda \rightarrow \lambda_{i}} \! 
m^{\sharp} (\lambda) v_{i} (\delta(\lambda_{i}))^{-2} 
\prod\limits_{l=n+1}^{N} \! \left( \! \frac{ 
(\lambda_{i} - \lambda_{l}) (\lambda_{i} + \lambda_{l}) 
}{ (\lambda_{i} - \overline{\lambda_{l}}) (\lambda_{i} 
+ \overline{\lambda_{l}})} \!  \right)^{2} \! \!
\sigma_{-}, \, \, \, \, \, \, 1 \leq i \leq n,& \\
&{\rm res}(m^{\sharp} (\lambda);\! -\lambda_{i}) \! = \! 
- \sigma_{3} \mathstrut {\rm res}(m^{\sharp}(\lambda);\! 
\lambda_{i} ) \mathstrut \sigma_{3}, \, \, \, \, \, \,
1 \leq i \leq n,& \\
&{\rm res}(m^{\sharp}(\lambda);\! \overline{\lambda_{i}} 
\,) \! = \! \lim\limits_{\lambda \rightarrow \overline{
\lambda_{i}}} \! m^{\sharp}(\lambda) \overline{v_{i}} 
(\delta(\overline{\lambda_{i}}))^{2} \prod\limits_{l=n+1}^{
N} \! \left(\! \frac{(\overline{\lambda_{i}} - \overline{
\lambda_{l}})(\overline{\lambda_{i}}+\overline{\lambda_{l}}
)}{(\overline{\lambda_{i}} - \lambda_{l}) (\overline{
\lambda_{i}} + \lambda_{l}) } \! \right)^{2} \! \! 
\sigma_{+}, \, \, \, \, \, \, 1 \leq i \leq n,& \\
&{\rm res}(m^{\sharp}(\lambda);\! -\overline{\lambda_{i}} 
\,) \! = \! - \sigma_{3} \mathstrut {\rm res}(m^{\sharp}
(\lambda);\! \overline{\lambda_{i}} \,) \mathstrut 
\sigma_{3}, \, \, \, \, \, \, 1 \leq i \leq n,&
\end{eqnarray*}
and, as $\lambda \! \rightarrow \! \infty$, $\lambda \! \in 
\! \Bbb C \! \setminus \! (({\cal Z}_{d} \! \setminus \! 
\cup_{i=n+1}^{N} (\{\pm \lambda_{i}\} \! \cup \! \{\pm 
\overline{\lambda_{i}}\})) \cup (\widehat{\Gamma} \! \cup 
\! (\cup_{i=n+1}^{N}(K_{i}^{\pm} \! \cup \! L_{i}^{\pm}))))$,
\begin{eqnarray*}
&m^{\sharp}(\lambda) = {\rm I} + {\cal O}(\lambda^{-1});&
\end{eqnarray*}
moreover, $Q(x,t) \! = \! 2 i \! \lim\limits_{\lambda \to \infty} 
\! (\lambda m^{\sharp}(x,t;\! \lambda))_{12}$ is equal to $Q(x,t)$ 
in Lemma~2.2, Eq.~(11).
\end{ccc}

{\em Proof.\/} The proof is presented for the eigenvalues $\{
\lambda_{i}\}_{i=n+1}^{N}$, around which are defined the small, 
clockwise-oriented, mutually disjoint circles $\{K_{i}^{+}\}_{
i=n+1}^{N}$: the proof for the eigenvalues $\{-\lambda_{i}\}_{
i=n+1}^{N}$ and $\{\pm \overline{\lambda_{i}}\}_{i=n+1}^{N}$ 
follows in an analogous manner. {}From the definition of $m^{
\sharp}(\lambda)$ and Proposition~3.2, one sees that, on $\{
K_{i}^{+}\}_{i=n+1}^{N}$, $m^{\sharp}(\lambda)$ solves the 
following RH problem $(\lambda \! \in \! \cup_{i=n+1}^{N} 
K_{i}^{+})$,
\begin{eqnarray*}
&m_{+}^{\sharp}(\lambda) \! = \! m_{-}^{\sharp}(\lambda) 
\underbrace{\prod\limits_{l=n+1}^{N} (d_{l_{-}} (\lambda))^{
\sigma_{3}} J_{K_{i}^{+}}(\lambda) \! \left( {\rm I} + \frac{
v_{i} (\delta(\lambda_{i}))^{-2}}{(\lambda-\lambda_{i})} 
\sigma_{-} \right) \! \prod\limits_{l=n+1}^{N} (d_{l_{+}} 
(\lambda))^{- \sigma_{3}}}_{{\rm jump\/} \, \, \, 
{\rm matrix}}.& 
\end{eqnarray*}
Demanding that the above ``jump matrix'' be equal to the 
following upper triangular form, ${\rm I} \! + \! \frac{C_{
i}^{\sharp}}{(\lambda - \lambda_{i})} \sigma_{+}$, $n \! + 
\! 1 \! \leq \! i \! \leq \! N$,
one shows that
\begin{eqnarray*}
&J_{K_{i}^{+}}(\lambda) \! = \! 
\left( \begin{array}{cc}
\frac{d_{i_{+}} (\lambda)}{d_{i_{-}} (\lambda)} 
\prod\limits_{l=n+1 \atop \not= i}^{N} \frac{ 
d_{l_{+}}(\lambda)}{d_{l_{-}}(\lambda)} \, - \, 
\frac{v_{i} (\delta(\lambda_{i}))^{-2} C_{i}^{\sharp} 
\prod\limits_{l=n+1 \atop \not= i}^{N} \frac{ d_{l_{
-}}^{-1} (\lambda)}{d_{l_{+}}(\lambda)}}{(\lambda 
- \lambda_{i})^{2} d_{i_{-}} (\lambda) d_{i_{+}} 
(\lambda)} & \, \, \, \, \frac{C_{i}^{\sharp}}{(\lambda - 
\lambda_{i})} \frac{\prod\limits_{l=n+1 \atop \not= 
i}^{N} \frac{d_{l_{-}}^{-1}(\lambda)}{d_{l_{+}} 
(\lambda)}}{d_{i_{-}}(\lambda) d_{i_{+}}(\lambda)} \\
- \frac{v_{i}(\delta(\lambda_{i}))^{-2}}{(\lambda 
- \lambda_{i})} \frac{d_{i_{-}}(\lambda)}{d_{i_{+}} 
(\lambda)} \prod\limits_{l=n+1 \atop \not= i}^{N} 
\frac{d_{l_{-}}(\lambda)}{d_{l_{+}}(\lambda)} & \, \, 
\, \, \frac{d_{i_{-}}(\lambda)}{d_{i_{+}}(\lambda)} 
\prod\limits_{l=n+1 \atop \not= i}^{N} \frac{d_{l_{
-}}(\lambda)}{d_{l_{+}}(\lambda)}
\end{array} \right) \!.& 
\end{eqnarray*}
Note that $\det(J_{K_{i}^{+}}(\lambda)) \! = \! 1$ $(n \! + 
\! 1 \! \leq \! i \! \leq \! N)$. Defining, for $n \! + \! 
1 \! \leq \! l \! \leq \! N$, $d_{l_{+}}(\lambda)$ and $d_{
l_{-}}(\lambda)$ as in Eqs.~(87), and choosing $C_{i}^{\sharp}$, 
$n \! + \! 1 \! \leq \! i \! \leq \! N$, as in Eq.~(88) (with 
$\! + \lambda_{i})$, one gets the expression for $J_{K_{i}^{+}}
(\lambda)$ (which is holomorphic $\forall \, \lambda \! \in \! 
\cup_{i=n+1}^{N} {\rm int} K_{i}^{+})$ given in the Lemma; also, 
because of the symmetry properties of $\delta(\lambda)$ 
(Proposition~3.1), $\overline{C_{i}^{\sharp}} \! = \! 
(\overline{v_{i}})^{-1} (\delta(\pm \overline{\lambda_{i}}))^{
-2} \linebreak[4] \cdot (d_{i_{-}}(\pm \overline{\lambda_{i}})
)^{-2} \prod\limits_{l=n+1 \atop \not= i}^{N} (d_{l_{+}}(\pm 
\overline{\lambda_{i}}))^{-2}$. The remainder of the proof is 
a consequence of Lemma~3.1, Proposition~3.2, and the definition 
of $m^{\sharp}(\lambda)$. \hfill \rule{6.5pt}{6.5pt}
\begin{eee}
{\rm Even though, along the trajectory of soliton $n$, all the 
initial, exponentially growing nilpotent residue matrices have 
been replaced by jump matrices which tend to ${\rm I}$ as $t \! 
\to \! +\infty$, i.e., $\exists \, \varepsilon \! \in \! \Bbb 
R_{>0}$ such that $\forall \, i \! \in \! \{n \! + \! 1,n \! + 
\! 2,\ldots,N\}$ $\vert (v_{i} \vert_{\Omega_{n}})^{-1} \vert \! 
\sim \! {\cal O}(\exp \{-\varepsilon t\})$, it does not necessarily 
follow that elements in the solution of the extended RH problem for 
$m^{\sharp}(\lambda)$ cannot grow exponentially; for example, note 
that the $(2 \, 1)$-elements of $J_{K_{i}^{\pm}}(\lambda)$ and the 
$(1 \, 2)$-elements of $J_{L_{i}^{\pm}}(\lambda)$, $n \! + \! 1 \! 
\leq \! i \! \leq \! N$, grow exponentially.\/}
\end{eee}

By estimating the error, along the trajectory of soliton $n$ $(
1 \! \leq \! n \! \leq \! N)$, when the jump matrices on $\{K_{
i}^{\pm},L_{i}^{\pm}\}_{i=n+1}^{N}$ are removed {}from the 
specification of the RH problem for $m^{\sharp}(\lambda)$, one 
gets the following---asymptotically solvable---model RH problem.
\begin{ccc}
Let $\chi(\lambda)$ solve the following RH problem on $\sigma_{
\pounds} \! \setminus \! \cup_{i=n+1}^{N}(\{\pm \lambda_{i}\} \! 
\cup \! \{\pm \overline{\lambda_{i}}\})$,
\begin{eqnarray*}
\chi_{+}(\lambda)=\chi_{-}(\lambda) e^{-i \theta(\lambda) {\rm 
ad}(\sigma_{3})} v^{\sharp}(\lambda) \vert_{\widehat{\Gamma}}, \, 
\, \, \, \, \, \, \lambda \in \widehat{\Gamma}, 
\end{eqnarray*}
with polar (residue) conditions,
\begin{eqnarray*}
&{\rm res} (\chi(\lambda);\! \lambda_{i}) \! = \! \lim\limits_{
\lambda \rightarrow \lambda_{i}} \! \chi (\lambda) v_{i} 
(\delta(\lambda_{i}))^{-2} \! \prod\limits_{l=n+1}^{N} \! 
\left( \! \frac{ (\lambda_{i} - \lambda_{l}) (\lambda_{i} 
+ \lambda_{l}) }{ (\lambda_{i} - \overline{\lambda_{l}}) 
(\lambda_{i} + \overline{\lambda_{l}}) } \! \right)^{2} 
\! \! \sigma_{-}, \, \, \, \, \, \, 1 \leq i \leq n,& \\
&{\rm res}(\chi (\lambda);\! -\lambda_{i}) \! = \! - \sigma_{3} 
\mathstrut {\rm res} (\chi(\lambda);\! \lambda_{i}) \mathstrut 
\sigma_{3}, \, \, \, \, \, \, 1 \leq i \leq n,& \\
&{\rm res}(\chi(\lambda);\! \overline{\lambda_{i}} \,) \! = \! 
\lim\limits_{\lambda \rightarrow \overline{\lambda_{i}}} \chi
(\lambda) \overline{v_{i}} (\delta(\overline{\lambda_{i}}))^{2} 
\! \prod\limits_{l=n+1}^{N} \! \left(\! \frac{(\overline{
\lambda_{i}} - \overline{\lambda_{l}}) (\overline{\lambda_{i}} 
+ \overline{\lambda_{l}})}{(\overline{\lambda_{i}} - \lambda_{
l})(\overline{\lambda_{i}} + \lambda_{l}) } \! \right)^{2} \! \! 
\sigma_{+}, \, \, \, \, \, \, 1 \leq i \leq n,& \\
&{\rm res}(\chi(\lambda);\! -\overline{\lambda_{i}} \,) \! = \! 
-\sigma_{3} \mathstrut {\rm res} (\chi (\lambda);\! \overline{
\lambda_{i}} \,) \mathstrut \sigma_{3}, \, \, \, \, \, \, 1 
\leq i \leq n,&
\end{eqnarray*}
and, as $\lambda \! \rightarrow \! \infty$, $\lambda \! \in \! 
\Bbb C \! \setminus \! (\cup_{i=1}^{n} (\{\pm \lambda_{i}\} \! 
\cup \! \{\pm \overline{\lambda_{i}}\}) \! \cup \! \widehat{
\Gamma})$,
\begin{eqnarray*}
&\chi(\lambda) = {\rm I} + {\cal O} (\lambda^{-1}).&
\end{eqnarray*}
Then as $t \! \rightarrow \! + \infty$ and $x \! \rightarrow \! 
-\infty$ such that $\lambda_{0} \! > \! M$ and $(x,t) \! \in \! 
\Omega_{n}$, the function $E(\lambda) \! := \! m^{\sharp}
(\lambda)(\chi(\lambda))^{-1}$ has the following asymptotics,
\begin{eqnarray}
E(\lambda) = {\rm I} + {\cal O} \! \left(F(\lambda;\! 
\lambda_{0}) \exp \{- a b t\} \right) \!,
\end{eqnarray}
where $\vert \vert F(\cdot;\! \lambda_{0}) \vert \vert_{{\cal 
L}^{\infty}(\Bbb C;M_{2}(\Bbb C))} \! < \! \infty$, $\vert \vert 
F(\lambda;\cdot) \vert \vert_{{\cal L}^{\infty}(\Bbb R_{> M};M_{
2}(\Bbb C))} \! < \! \infty$, $F(\lambda;\! \lambda_{0}) \! \sim 
\! {\cal O} \! \left(\frac{C(\lambda_{0})}{\lambda} \right)$ as 
$\lambda \! \rightarrow \! \infty$ with $C(\lambda_{0}) \! \in \! 
{\cal L}^{\infty}(\Bbb R_{> M};\! M_{2}(\Bbb C))$, $a \! := \! 8 
\min \{\eta_{i}\}_{i=n+1}^{N} \, (> \! 0)$, and $b \! := \! \min 
\{\vert \xi_{n} \! - \! \xi_{i} \vert\}_{i=n+1}^{N}$.
\end{ccc}

{\em Proof.\/} Writing, for $n \! + \! 1 \! \leq \! i \! \leq \! 
N$, Eqs.~(89) in the following form,
\begin{eqnarray*}
&v^{\sharp}(\lambda) \! := \! 
\left\{ \begin{array}{l}
{\rm I} + (v_{i})^{-1} \widetilde{{\cal W}}_{K_{i}^{\pm}} 
(\lambda) \sigma_{+}, \, \, \, \, \, \, \, \lambda \in 
\bigcup\limits_{i=n+1}^{N} K_{i}^{\pm}, \\
{\rm I} + (\overline{v_{i}})^{-1} \widetilde{{\cal 
W}}_{L_{i}^{\pm}} (\lambda) \sigma_{-}, \, \, \, \, \, \,
\, \, \lambda \in \bigcup\limits_{i=n+1}^{N} L_{i}^{\pm},
\end{array} \right.&
\end{eqnarray*}
consider the ``error function'' $E(\lambda)$ defined 
in the Lemma. One notes that: (1) $\det(E(\lambda)) 
\! = \! 1$; (2) $E(\lambda)$ has no poles; and (3) 
$E(\lambda)$ solves the following RH problem on the 
oriented contour $\Sigma_{E} \! := \! \cup_{
i=n+1}^{N}(K_{i}^{+} \! \cup \! K_{i}^{-} \! \cup \! 
L_{i}^{+} \! \cup \! L_{i}^{-})$,
\begin{eqnarray*}
&E_{+}(\lambda) \! = \! E_{-}(\lambda) \! \left({\rm I} + 
(v_{i})^{-1} \widetilde{{\cal W}}_{ K_{i}^{\pm}} 
(\lambda) \! \left(\begin{array}{cc}
\! \! - \chi_{11}(\lambda) \chi_{21}(\lambda) & \! 
(\chi_{11} (\lambda))^{2} \\
\! \! - (\chi_{21} (\lambda) )^{2} & \! \chi_{11}(\lambda) 
\chi_{21}(\lambda)
\end{array} \! \right) \! \right) \!, 
\, \, \, \, \, \, \, \lambda \in K_{i}^{\pm},& \\ 
&E_{+}(\lambda) \! = \! E_{-}(\lambda) \! \left( {\rm I} 
+ (\overline{v_{i}})^{-1} \widetilde{{\cal W}}_{ 
L_{i}^{\pm}} (\lambda) \! \left( \begin{array}{cc} 
\! \! \chi_{12} (\lambda) \chi_{22} (\lambda) & \! 
- (\chi_{12} (\lambda))^{2} \\
\! \! (\chi_{22} (\lambda))^{2} & \! - \chi_{12}(\lambda) 
\chi_{22} (\lambda) \end{array} \! \right) \! \right) \!, 
\, \, \, \, \, \, \, \lambda \in L_{i}^{\pm},&
\end{eqnarray*}
$n \! + \! 1 \! \leq \! i \! \leq \! N$, and, as $\lambda \! 
\rightarrow \! \infty$, $\lambda \! \in \! \Bbb C \! \setminus 
\! \Sigma_{E}$, $E(\lambda) \! = \! {\rm I} \! + \! {\cal O}
(\lambda^{-1})$. Now, writing the RH problem for $E(\lambda)$ 
on the oriented contour $\Sigma_{E}$ in terms of an equivalent 
system of linear singular integral equations, using the explicit 
asymptotic solution of the model RH problem for $\chi(\lambda)$ 
given in Sec.~4, recalling that, as $t \! \to \! +\infty$ and $x 
\! \to \! -\infty$ such that $\lambda_{0} \! > \! M$ and $(x,t) 
\! \in \! \Omega_{n}$, $(v_{i} \vert_{\Omega_{n}})^{-1} \! \sim 
\! {\cal O}(\exp \{-8t \eta_{i} \vert \xi_{n} \! - \! \xi_{i} 
\vert\})$, $n \! + \! 1 \! \leq \! i \! \leq \! N$, and 
proceeding as in the proof of Lemma~3.3 in \cite{a20}, one 
deduces the estimate in Eq.~(90). \hfill \rule{6.5pt}{6.5pt}
\section{Asymptotic Solution of the Model RH Problem}
In this section, the asymptotic (as $t \! \to \! +\infty$ and $x 
/ t \! \sim \! {\cal O}(1))$ solution of the model RH problem 
(Lemma~3.3) for the Schwartz class of non-reflectionless generic 
potentials $(r(\lambda) \! \in \! {\cal S}(\widehat{\Gamma};\! 
\Bbb C))$ is presented. Before doing so, however, recall the 
following well-known fact {}from matrix RH theory \cite{a11,a13}.
\begin{bbb}
The solution of the model RH problem (Lemma~3.3), $\chi(\lambda) 
\colon \Bbb C \setminus \! (\widehat{\Gamma} \! \cup \! (\cup_{
i=1}^{n}(\{\pm \lambda_{i}\} \! \cup \! \{\pm \overline{\lambda_{
i}}\}))) \! \rightarrow \! {\rm SL\/}(2,\! \Bbb C)$, has the 
following representation,
\begin{eqnarray}
\chi(\lambda) = \chi_{d}(\lambda) + \int_{
\raise-0.5ex\hbox{$\scriptstyle{}\widehat{\Gamma}$}} \frac{
\chi_{-}(\varrho)(v^{\sharp}(\varrho) \vert_{\widehat{
\Gamma}}-{\rm I})}{(\varrho-\lambda)} \frac{d \varrho}{2 
\pi i},
\end{eqnarray}
where
\begin{eqnarray}
&\chi_{d}(\lambda) = {\rm I} + \sum\limits_{i=1}^{n} \left(
\frac{{\rm res}(\chi(\lambda);\lambda_{i})}{(\lambda - \lambda_{
i})} - \frac{\sigma_{3} \mathstrut {\rm res} (\chi(\lambda);
\lambda_{i}) \mathstrut \sigma_{3}}{(\lambda + \lambda_{i})} + 
\frac{{\rm res} (\chi(\lambda);\overline{\lambda_{i}})}{(\lambda 
- \overline{\lambda_{i}})} - \frac{\sigma_{3} \mathstrut 
{\rm res}(\chi(\lambda);\overline{\lambda_{i}}) \mathstrut 
\sigma_{3}}{(\lambda + \overline{\lambda_{i}})} \right) \!.&
\end{eqnarray}
The solution of Eq.~(91) can be written as the 
following ordered product,
\begin{eqnarray}
\chi(\lambda) = \chi_{d}(\lambda) \chi^{c}(\lambda), 
\end{eqnarray}
where $\chi_{d}(\lambda)$ is given by Eq.~(92), and
$\chi^{c}(\lambda)$ solves the following RH problem: 
(1) $\chi^{c}(\lambda)$ is piecewise holomorphic 
$\forall \, \lambda \! \in \! \Bbb C \! \setminus \! 
\widehat{\Gamma}$; (2) $\chi^{c}_{+}(\lambda) = 
\chi^{c}_{-}(\lambda) \exp \{- i \theta(\lambda)
{\rm ad}(\sigma_{3})\} (v^{\sharp}(\lambda) \vert_{
\widehat{\Gamma}})$, $\lambda \! \in \! \widehat{
\Gamma}$; and (3) as $\lambda \! \rightarrow \! 
\infty$, $\lambda \! \in \! \Bbb C \! \setminus \! 
\widehat{\Gamma}$, $\chi^{c}(\lambda) \! = \! {\rm I} 
\! + \! {\cal O}(\lambda^{-1})$.
\end{bbb}
\begin{eee}
{\rm {}From Proposition~4.1, Eq.~(93), it is seen that, in 
order to solve the model RH problem, explicit knowledge of 
$\chi_{d}(\lambda)$ and $\chi^{c}(\lambda)$ is necessary. The 
determination of $\chi^{c}(\lambda)$ is technically the more 
complicated of the two: actually, the determination of $\chi
_{d}(\lambda)$ depends on the explicit knowledge of $\chi^{c}
(\lambda)$ (see Proposition~4.2); hence, the asymptotic 
solution of $\chi^{c}(\lambda)$ is presented first (see 
Lemma~4.1).\/}
\end{eee}

In order to more fully comprehend certain elements of the proof 
of Lemma~4.1 given below, the Beals-Coifman \cite{a13} formulation 
for the solution of a (matrix) RH problem on an oriented contour 
is requisite: a self-contained synopsis of this formulation as it 
applies to the solution of the RH problem for $\chi^{c}(\lambda)$ 
stated in Proposition~4.1 now follows. Writing the jump matrix in 
the following factorized form, $v^{\sharp}(\lambda) \vert_{
\widehat{\Gamma}} \! := \! ({\rm I} \! - \! w^{-}_{x,t}
(\lambda))^{-1} ({\rm I} \! + \! w^{+}_{x,t}(\lambda))$, $\lambda 
\! \in \! \widehat{\Gamma}$, where $w^{\pm}_{x,t}(\lambda) \! \in 
\! \cap_{k \in \{2,\infty\}} {\cal L}^{k}(\widehat{\Gamma};\! 
M_{2}(\Bbb C))$ (with $\vert \vert w^{\pm}_{x,t}(\cdot) \vert 
\vert_{\cap_{k \in \{2,\infty\}} {\cal L}^{k}(\widehat{\Gamma};
M_{2}(\Bbb C))} \! := \! \sum_{k \in \{2,\infty\}} \vert \vert 
w^{\pm}_{x,t}(\cdot) \vert \vert_{{\cal L}^{k} (\widehat{\Gamma};
M_{2}(\Bbb C))})$, respectively, are nilpotent off-diagonal 
upper/lower triangular matrices, define $w_{x,t}(\lambda) \! := 
\! w^{-}_{x,t}(\lambda) \! + \! w^{+}_{x,t}(\lambda)$, and 
introduce the operator $C_{w_{x,t}}$ on ${\cal L}^{2}(\widehat{
\Gamma};\! M_{2}(\Bbb C))$ as $C_{w_{x,t}} f \! := \! C_{+}
(f \, w^{-}_{x,t}) \! + \! C_{-}(f \, w^{+}_{x,t})$, where $f \! 
\in \! {\cal L}^{2}(\widehat{\Gamma};\! M_{2}(\Bbb C))$, and 
$C_{\pm} \colon {\cal L}^{2}(\widehat{\Gamma};\linebreak[4] 
M_{2}(\Bbb C)) \! \to \! {\cal L}^{2}(\widehat{\Gamma};\! M_{2}
(\Bbb C))$ denote the Cauchy operators, $(C_{\pm}f)(\lambda) \! 
:= \! \! \lim\limits_{\lambda^{\prime} \to \lambda \atop 
\lambda^{\prime} \in {\pm} \, {\rm side} \, {\rm of} \, \widehat{
\Gamma}} \! \int_{\raise-0.45ex\hbox{$\scriptstyle{}\widehat{
\Gamma}$}} \frac{f(\varrho)}{(\varrho-\lambda^{\prime})} \frac{d 
\varrho}{2 \pi i}$.
\begin{ddd}[{\rm \cite{a13}}]
If $\mu^{c}(\lambda) \! \in \! {\rm I} \oplus {\cal L}^{2}
(\widehat{\Gamma};\! M_{2}(\Bbb C))$ solves the following linear 
singular integral equation,
\begin{eqnarray*}
(\underline{{\bf Id}} - C_{w_{x,t}}) \mu^{c} = {\rm I},
\end{eqnarray*}
where $\underline{{\bf Id}}$ is the identity operator on ${\rm I} 
\! \oplus \! {\cal L}^{2}(\widehat{\Gamma};\! M_{2}(\Bbb C))$, then 
the solution of the RH problem for $\chi^{c}(\lambda)$ is 
\begin{eqnarray*}
\chi^{c}(\lambda) = {\rm I} + \int_{
\raise-0.5ex\hbox{$\scriptstyle{}\widehat{\Gamma}$}} 
\frac{\mu^{c}(\varrho) w_{x,t}(\varrho)}{(\varrho - \lambda)} 
\frac{d \varrho}{2 \pi i}, \, \, \, \, \, \, \, \, \lambda 
\in \Bbb C \! \setminus \! \widehat{\Gamma},
\end{eqnarray*}
where $\mu^{c}(\lambda) \! = \! \chi^{c}_{+}(\lambda)({\rm 
I} \! + \! w^{+}_{x,t}(\lambda))^{-1} \! = \! \chi^{c}_{-}
(\lambda)({\rm I} \! - \! w^{-}_{x,t}(\lambda))^{-1}$.
\end{ddd}
\begin{ccc}
Let $\epsilon_{0}$ denote an arbitrarily fixed, sufficiently 
small positive real number. For $\aleph \! \in \! \{0,\pm 
\lambda_{0}\}$, set ${\cal N}(\aleph;\! \epsilon_{0}) \! 
:= \! \{\mathstrut \lambda; \, \vert \lambda \! - \! \aleph 
\vert \! \leq \! \epsilon_{0}\}$. Then as $t \! \to \! +\infty$ 
and $x \! \to \! -\infty$ such that $\lambda_{0} \! > \! M$ and 
$\lambda \! \in \! \Bbb C \! \setminus \! \cup_{\aleph \in \{0,
\pm \lambda_{0}\}} {\cal N}(\aleph;\! \epsilon_{0})$, $\chi^{c}
(\lambda)$ has the following asymptotic expansion,
\begin{eqnarray*}
\chi^{c}(\lambda) & = & {\rm I} + \frac{
1}{4} \sqrt{\frac{\nu (\lambda_{0})}{2 \lambda_{0}^{2} 
t}} \left(\frac{1}{\lambda - \lambda_{0}} + \frac{1}{
\lambda + \lambda_{0}} \right) \! \left( \exp \{- i 
(\phi^{+}(\lambda_{0}) + \widehat{\Phi}^{+}(\lambda_{
0};t))\} \sigma_{-} \right. \\
 & + & \left. \exp \{i (\phi^{+} (\lambda_{0}) + \widehat{
\Phi}^{+}(\lambda_{0};t))\} \sigma_{+} \right) + {\cal O} 
\! \left(\frac{G(\lambda;\! \lambda_{0}) \ln t}{t} \right) \!,
\end{eqnarray*}
where $\nu(\lambda_{0})$, $\phi^{+}(\lambda_{0})$, and 
$\widehat{\Phi}^{+}(\lambda_{0};t)$ are given in Theorem~2.1, 
Eqs.~(21), (22), and (24), $\vert \vert G(\cdot;\! \lambda_{
0}) \vert \vert_{{\cal L}^{\infty}(\Bbb C \setminus \cup_{
\aleph \in \{0,\pm \lambda_{0}\}} {\cal N}(\aleph;\epsilon_{
0});M_{2}(\Bbb C))} \! < \! \infty$, $G(\lambda;\cdot) \! \in 
\! {\cal S}(\Bbb R_{> M};\! M_{2}(\Bbb C))$, $G(\lambda;\! 
\lambda_{0}) \! \sim \! {\cal O} \! \left(\frac{C(\lambda_{0}
)}{\lambda} \right)$ as $\lambda \! \to \! \infty$ with $C(
\lambda_{0}) \! \in \! {\cal S}(\Bbb R_{>M};\! M_{2}(\Bbb C))$, 
and satisfies the following involutions, $\chi^{c}(-\lambda) 
\! = \! \sigma_{3} \chi^{c}(\lambda) \sigma_{3}$ and $\chi^{c}
(\lambda) \! = \! \sigma_{1} \overline{\chi^{c}(\overline{
\lambda})} \sigma_{1}$.
\end{ccc}

{\em Proof.\/} In Secs.~5 and 6 of \cite{a7}, it was shown that, 
for $\lambda \! \in \! \Bbb C \setminus \cup_{\aleph \in \{0,
\pm \lambda_{0}\}} {\cal N}(\aleph;\! \epsilon_{0})$, as $t \! 
\to \! +\infty$ and $x \! \to \! -\infty$ such that $\lambda_{0} 
\! > \! M$, for arbitrary $l^{\prime} \! \in \! \Bbb Z_{\geq 1}$, 
\begin{eqnarray*}
&\chi^{c}(\lambda) = {\rm I} + \sum\limits_{\vec{m}_{3} \in \Bbb 
M_{3}} \int\limits_{a_{l}(\vec{m}_{3})}^{a_{u}(\vec{m}_{3})} 
\frac{m_{3} \mu^{c} (\varsigma) \vert_{{\rm sgn}(m_{3})} (\delta
(\varsigma))^{2 m_{3}} e^{-2im_{3} t \theta (\varsigma)} {\cal 
R}^{\vec{m}_{3}}(\varsigma) \sigma_{{\rm sgn}(m_{3})}}{(\varsigma 
- \lambda)} \frac{d \varsigma}{2 \pi i}& \\
&\, \, \, \, \, \, \, \, \, \, \, \, \, \, \, \, \, \, \, \, 
+ \, \sum\limits_{\vec{m}_{2} \in \Bbb M_{2}} \int\limits_{0}^{
b(\vec{m}_{2})} \frac{m_{2} \mu^{c}(\varsigma) \vert_{{\rm sgn}
(m_{2})} (\delta(\varsigma))^{2 m_{2}} e^{-2im_{2}t \theta 
(\varsigma)} {\cal R}^{\vec{m}_{2}}(\varsigma) \sigma_{{\rm sgn}
(m_{2})}}{(\varsigma - \lambda)} \frac{d \varsigma}{2 \pi i} + 
{\cal O} \! \left(\frac{\underline{c}(\lambda_{0})}{(\lambda_{
0}^{2} t)^{l^{\prime}}} \right) \!,&
\end{eqnarray*}
where: (1) $\vec{m}_{k} \! \in \! \Bbb M_{k}$ denotes the set of 
vectors with $k$ components each of which take the values $\pm 1$ 
$({\rm card\/}(\Bbb M_{k}) \! = \! 2^{k})$; (2) $a_{l}(\vec{m}_{3}
) \! = \! m_{1} \lambda_{0} \! + \! m_{2} \varepsilon \exp\{-
\frac{i \pi m_{3}}{4}\}$, $a_{u}(\vec{m}_{3}) \! = \! m_{1} 
\lambda_{0}$, and $b(\vec{m}_{2}) \! = \! m_{1} \varepsilon 
\exp \{\frac{i \pi m_{2}}{4}\}$, where $\varepsilon$ is an 
arbitrarily fixed, sufficiently small positive real number; 
(3) $\mu^{c}(\varsigma) \vert_{+} \! := \! \mu^{c}(\varsigma) 
\vert_{L^{\prime}}$, $\mu^{c}(\varsigma) \vert_{-} \! := \! 
\mu^{c}(\varsigma) \vert_{\overline{L^{\prime}}}$, with $L^{
\prime} \! = \! \{\mathstrut \lambda; \, \lambda \! 
= \! \widehat{u} \exp \{\frac{i\pi}{4}\}, \, 
\widehat{u} \! \in \! (-\varepsilon,\varepsilon)\} 
\! \cup \! (\cup_{l \in \{\pm 1\}} \{\mathstrut \lambda; \, 
\lambda \! = \! l \lambda_{0} \! + \! \widehat{u} \exp \{- 
\frac{i \pi}{4}\}, \, \widehat{u} \! \in \! (-\varepsilon,
\varepsilon)\})$, and where $\mu^{c}(\cdot)$ is the solution 
of the Beals-Coifman \cite{a13} linear singular integral 
equation (Theorem~4.1); (4) $\theta(\varsigma) \! = \! 
2 \varsigma^{2} (\varsigma^{2} \! - \! 2 \lambda_{0}^{2})$; 
(5) $\delta(\varsigma) \! = \! ((\frac{\varsigma - \lambda_{
0}}{\varsigma})(\frac{\varsigma + \lambda_{0}}{\varsigma})
)^{i \nu} \exp \{\sum_{l \in \{\pm\}} (\rho_{l}(\varsigma) 
\! + \! \widehat{\rho}_{l} (\varsigma))\}$, $\nu \! := 
\! \nu(\lambda_{0})$, $\rho_{\pm}(\varsigma) \! = \! \frac{
1}{2 \pi i} \int_{0}^{\pm \lambda_{0}} \! \ln \! \left(\! 
\frac{1 - \vert r(\varrho) \vert^{2}}{1 - \vert r(
\lambda_{0}) \vert^{2}} \! \right) \! \frac{d \varrho}{(
\varrho - \varsigma)}$, and $\widehat{\rho}_{\pm}(\varsigma) 
\! = \! \int_{\pm i \infty}^{i 0} \! \frac{\ln(1-r(\varrho) 
\overline{r(\overline{\varrho})})}{(\varrho - \varsigma)} 
\frac{d \varrho}{2 \pi i}$; (6) ${\cal R}^{-1,-1,1}(
\varsigma) \! = \! {\cal R}^{1,1,1}(\varsigma) \! = \! - 
r(\varsigma) {\cal P}(\varsigma)$, ${\cal R}^{-1,1,1}(
\varsigma) \! = \! {\cal R}^{1,-1,1}(\varsigma) \! = \! 
\frac{r(\varsigma) {\cal P}(\varsigma)}{(1-r(\varsigma) 
\overline{r(\overline{\varsigma})})}$, ${\cal R}^{-1,-1,-1}
(\varsigma) \! = \! {\cal R}^{1,1,-1}(\varsigma) \! = \! 
({\cal R}^{1,1,1} (\varsigma))^{\ast}$, and ${\cal R}^{-1,
1,-1}(\varsigma) \! = \! {\cal R}^{1,-1,-1}(\varsigma) \! = 
\! ({\cal R}^{1,-1,1}(\varsigma))^{\ast}$, where ${\cal P}
(z) \! := \! \prod_{l=n+1}^{N} \! \left(\! \frac{(z - 
\overline{\lambda_{l}})(z + \overline{\lambda_{l}})}{(z-
\lambda_{l})(z+\lambda_{l})} \! \right)^{2}$, and $\alpha
(\cdot) \! = \! (\beta(\cdot))^{\ast}$ means that $\alpha
(\cdot)$ is the same piecewise-rational function as $\beta
(\cdot)$ except with the complex conjugated coefficients; 
(7) ${\cal R}^{1,1}(\varsigma) \! = \! {\cal R}^{-1,1}
(\varsigma) \! = \! \frac{r(\varsigma) {\cal P}(\varsigma)
}{(1-\vert r(\varsigma) \vert^{2})}-\frac{r(i\varsigma) 
{\cal P}(i \varsigma)}{(1 + \vert r(i \varsigma) \vert^{2})}$ 
and ${\cal R}^{1,-1}(\varsigma) \! = \! {\cal R}^{-1,-1}
(\varsigma) \! = \! -({\cal R}^{1,1}(\varsigma))^{\ast}$; and 
(8) $\underline{c}(\lambda_{0}) \! \in \! {\cal L}^{\infty}
(\Bbb R_{> M};\! M_{2}(\Bbb C))$. Since, in the above 
expression for $\chi^{c}(\lambda)$, the estimation of all 
the integrals is analogous, without loss of generality, the 
following integral is considered,
\begin{eqnarray*}
&I_{0} := \int\limits_{ \lambda_{0} + 
\varepsilon e^{- \frac{i \pi}{4}}}^{\lambda_{0}} 
{\cal A}_{0} (\varsigma) {\cal B}_{0}(\varsigma) 
\frac{d \varsigma}{2 \pi i},&
\end{eqnarray*}
where ${\cal A}_{0}(\varsigma) \! := \! \mu^{c}(\varsigma) 
\vert_{L^{\prime}}$, and ${\cal B}_{0}(\varsigma) \! := \! 
- \frac{\delta^{2}(\varsigma) \exp \{- 2 i t \theta(\varsigma)\} 
r(\varsigma) {\cal P}(\varsigma)}{(\varsigma-\lambda)} \sigma_{
+}$. Begin by estimating ${\cal B}_{0}(\varsigma) \linebreak[4] 
\cdot \! \frac{d \varsigma}{2 \pi i}$: (1) expand ${\cal B}_{0}
(\varsigma)$ in a Taylor series about $\lambda_{0}$; (2) make 
the following change of variable \cite{a7}, $\varsigma \! := \! 
\varsigma(\widetilde{w}) \! = \! \lambda_{0} \! + \! 
\widetilde{w}(16 \lambda_{0}^{2} t)^{-1/2}$, and express the 
expansion obtained in (1) above in terms of $\widetilde{w}$; and 
(3) use the following identity, $ab \! = \! (a \! - \! 1)(b \! - 
\! 1) \! + \! (a \! - \! 1) \! + \! (b \! - \! 1) \! +1$. 
Carrying out steps~(1)--(3), one gets that,
\begin{eqnarray*}
&{\cal B}_{0}(\varsigma) \frac{d \varsigma}{2 \pi 
i} \vert_{\varsigma(\widetilde{w})} = - \frac{(
\widetilde{w})^{2 i \nu} \lambda_{0}^{-4 i \nu} 
t^{- i \nu} 2^{- 2 i \nu} e^{2 s(\lambda_{0})} 
e^{4 i \lambda_{0}^{4} t} e^{- i \widetilde{w}^{2}} 
d \widetilde{w} \sigma_{+}}{(\lambda - \lambda_{0}) 
(2 \pi i) \sqrt{16 \lambda_{0}^{2} t}} \{{\cal R}
(\lambda_{0}) + ({\cal R}^{\prime} (\lambda_{0})& \\
&- \frac{3 i \nu {\cal R} (\lambda_{0})}{\lambda_{0}}) 
\frac{\widetilde{w}}{\sqrt{16 \lambda_{0}^{2} t}}\} 
\sum\limits_{\vec{l}_{4} \in \Bbb L_{4}} 
\prod\limits_{k=1}^{4} (p_{k}(\widetilde{w}))^{l_{k}} 
- \frac{(\widetilde{w})^{2 i \nu + 1} \lambda_{0}^{-4 
i \nu} t^{- i \nu} 2^{- 2 i \nu} e^{2 s(\lambda_{0})} 
e^{4 i \lambda_{0}^{4} t} e^{- i \widetilde{w}^{2}} d 
\widetilde{w} \sigma_{+}}{(\lambda - \lambda_{0})^{2} 
(2 \pi i) (16 \lambda_{0}^{2} t)}& \\
&\times \, {\cal R}(\lambda_{0}) \sum\limits_{\vec{l}_{4} 
\in \Bbb L_{4}} \prod\limits_{k=1}^{4} (p_{k}
(\widetilde{w}))^{l_{k}} + {\cal O} \! \left(\{\frac{C_{
1}^{\flat}(\lambda_{0})}{(\lambda-\lambda_{0})}+\frac{C_{
2}^{\flat}(\lambda_{0})}{(\lambda - \lambda_{0})^{2}}\} 
\frac{e^{4 i \lambda_{0}^{4} t} \widetilde{w}^{2 i \nu + 
2} e^{-i \widetilde{w}^{2}} d \widetilde{w} \sigma_{+}}{
t^{3/2+i \nu}} \right) \!,&
\end{eqnarray*}
where $\vec{l}_{k} \! \in \! \Bbb L_{k}$ denotes the 
set of vectors with $k$ components each of which take
the values $0$ and $1$ $({\rm card\/}(\Bbb L_{k}) \! 
= \! 2^{k})$, $s(\lambda_{0}) \! := \! \sum_{l 
\in \{\pm\}} (\rho_{l}(\lambda_{0}) \! + \! \widehat{
\rho}_{l} (\lambda_{0}))$, ${\cal R}(\lambda_{0}) \!
:= \! - r(\lambda_{0}) {\cal P} (\lambda_{0})$, 
${\cal R}^{\prime}(\lambda_{0}) \! = \! - r^{\prime}
(\lambda_{0}) {\cal P}(\lambda_{0}) \! - \! r(\lambda_{
0}) {\cal P}(\lambda_{0}) \sum_{k=n+1}^{N} \{\frac{4i 
\sin (\arg (\lambda_{0} - \lambda_{k}))}{\vert \lambda_{
0} - \lambda_{k} \vert} + \frac{4 i \sin (\arg(\lambda_{
0} + \lambda_{k}))}{\vert \lambda_{0} + \lambda_{k} 
\vert}\}$, $(\bullet)^{\prime}(\lambda_{0}) \! 
:= \! \frac{d(\bullet)(z)}{dz} \vert_{z=\lambda_{0}}$,
\begin{eqnarray*}
&p_{k}(\widetilde{w}) := \exp \{2 \widetilde{
\Delta}_{k}(\lambda_{0} + \frac{\widetilde{w}}{
\sqrt{16 \lambda_{0}^{2} t}}) \} - 1, \, \, \, 
\, \, \, \, \, k \in \{1,2\},& \\
&p_{3}(\widetilde{w}) := \exp \{- \frac{i 
\widetilde{w}^{3}}{\lambda_{0} \sqrt{16 \lambda_{
0}^{2} t}}\} - 1, \, \, \, \, \, \, \, \, p_{4} 
(\widetilde{w}) := \exp \{- \frac{i \widetilde{
w}^{4}}{4^{3} \lambda_{0}^{4} t}\} - 1,& \\
&\widetilde{\Delta}_{1}(\lambda_{0} + \frac{
\widetilde{w}}{\sqrt{16 \lambda_{0}^{2} t}}) =
\sum\limits_{l \in \{\pm\}} (\rho_{l}(\lambda_{0} 
+ \frac{\widetilde{w}}{\sqrt{16 \lambda_{0}^{2} t}}) 
- \rho_{l}(\lambda_{0})),& \\
&\widetilde{\Delta}_{2}(\lambda_{0} + \frac{
\widetilde{w}}{\sqrt{16 \lambda_{0}^{2} t}}) 
= \sum\limits_{l \in \{\pm\}} (\widehat{
\rho}_{l}(\lambda_{0} + \frac{ \widetilde{w}}{
\sqrt{16 \lambda_{0}^{2} t}}) - \widehat{\rho}_{l} 
(\lambda_{0})),& 
\end{eqnarray*}
and $C_{i}^{\flat}(\lambda_{0}) \! \in \! {\cal S}(\Bbb R_{
>M};\! \Bbb C)$, $i \! \in \! \{1,2\}$. Now, proceed to 
estimate $\mu^{c}(\varsigma) \vert_{L^{\prime}}$ for 
$\varsigma \! \in \! (\lambda_{0},\lambda_{0} \! + \! 
\varepsilon \exp \{-\frac{i \pi}{4}\})$: for this, the 
Beals-Coifman \cite{a13} formulation for the solution 
of a RH problem on an oriented contour is necessary 
(Theorem~4.1 and the paragraph preceding it for discussion 
and notation); in particular, one has to estimate the 
functions $w^{\pm}_{x,t}(\varsigma)$ on $L^{\prime} 
\cup \overline{L^{\prime}}$. In Sec.~5 of \cite{a7}, 
it was shown that, on $(\lambda_{0},\lambda_{0} \! + \! 
\varepsilon \exp \{-\frac{i \pi}{4}\})$, $w^{+}_{x,t}
(\varsigma) \! = \! -(\delta(\varsigma))^{{\rm ad}(\sigma
_{3})} \exp \{-it\theta (\varsigma){\rm ad}(\sigma_{3})\} 
\linebreak[4] \cdot 
r(\varsigma) {\cal P}(\varsigma) \sigma_{+}$ and $w^{-}_{x,
t}(\varsigma) \! = \! 0$; hence, {}from the Beals-Coifman 
\cite{a13} formulation, for any $f \! \in \! {\cal L}^{2}(
L^{\prime};\! M_{2}(\Bbb C))$, $C_{w_{x,t}} f \! = \! C_{-}
(f \, w^{+}_{x,t})$. To estimate $w^{+}_{x,t}(\varsigma)$, 
one proceeds as follows: (1) recalling that $r(\varsigma) 
\! \in \! {\cal S}(\widehat{\Gamma};\! \Bbb C)$ and $\vert 
\vert r \vert \vert_{{\cal L}^{\infty}(\Bbb R;\Bbb C)} \! 
< \! 1$, expand $(\delta(\varsigma))^{{\rm ad}(\sigma_{3})}$, 
for $\varsigma \! \in \! (\lambda_{0},\lambda_{0} \! + \! 
\varepsilon \exp \{- \frac{i \pi}{4}\})$, via an integration 
by parts argument; (2) expand $\exp \{-it\theta(\varsigma)
{\rm ad}(\sigma_{3})\} r(\varsigma) {\cal P}(\varsigma)$ in 
a Taylor series about $\lambda_{0}$; and (3) change 
variables \cite{a7}, $\varsigma \! := \! \varsigma(
\widetilde{w})$. Carrying out steps~(1)--(3), one shows 
that, 
\begin{eqnarray*}
&w^{+}_{x,t}(\varsigma) \vert_{\varsigma(\widetilde{
w})} = e^{i \{4 \lambda_{0}^{4} t - \widetilde{w}^{
2} + 2 \nu \ln(\widetilde{w}/2) - \nu \ln t + 2 \phi
(\lambda_{0})\}} (v_{00}(\widetilde{w};\! \lambda_{0})
+ \frac{v_{10}(\widetilde{w};\lambda_{0}) + v_{11}
(\widetilde{w};\lambda_{0}) \ln t}{\sqrt{t}})& \\ 
&\! \! \! \! \! \! \! \! \! \! \! \! \! \! \! \! 
\! \! \! \! \! \! \! \! \! \! \! \! \! \! \! \! \! \! \! \! 
\! \! \! \! \! \! \! \! \! \! \! \! \! \! \! \! \! \! \! \! 
\! \! \! \! \! \! \! \! \! \! \! \! \! \! \! \! \! \! \! \! 
\! \! \! \! \! \! \! \! \! \! \! \! \! \! \! \! \! \! \! + 
\, \, {\cal O} \! \left(\frac{v_{22}(\widetilde{
w};\lambda_{0})(\ln t)^{2}}{t} \right) \!,&
\end{eqnarray*}
where $\phi(\lambda_{0}) \! = \! 2 \nu \ln \!
\lambda_{0} \! + \! \frac{1}{2 \pi} \! \int_{0}^{
\lambda_{0}} \ln \! \vert z^{2} \! - \! \lambda_{0}^{
2} \vert d \ln (1 \! - \! \vert r(z) \vert^{2}) \! 
- \! \frac{1}{2 \pi} \! \int_{0}^{\infty} \ln \! \vert 
z^{2} + \lambda_{0}^{2} \vert d \ln (1 + \vert r(iz) 
\vert^{2})$, $v_{00}(\widetilde{w};\! \lambda_{0}) \! 
:= \! {\cal R}(\lambda_{0})$, $v_{10}(\widetilde{w};\! 
\lambda_{0}), v_{11}(\widetilde{w};\! \lambda_{0})$, 
and $v_{22}(\widetilde{w};\! \lambda_{0})$ are nilpotent 
matrix polynomials whose elements are sums of products 
of terms of the type $\widetilde{w}^{j}$ and $(\ln \! 
\widetilde{w})^{k}$, $j \! \in \! \Bbb Z_{\geq 1}$, $k 
\! \in \! \Bbb Z_{\geq 0}$, with $\lambda_{0}$-dependent 
coefficients, and, for $0 \! \leq \! j \! \leq \! 2$, 
$0 \! \leq \! k \! \leq \! j$,
\begin{eqnarray*}
&\vert \vert \exp \{- i (\cdot)^{2}\} \exp \{2 i \nu 
\ln(\cdot)\} v_{jk}(\cdot;\! \lambda_{0}) \vert \vert_{
\cap_{l \in \{1,2,\infty\}} {\cal L}^{l}(L_{s}^{\prime} 
\setminus \{0\};\Bbb C)} < \infty,&
\end{eqnarray*}
with $\vert \vert (\cdot) \vert \vert_{\cap_{l \in 
\{1,2,\infty\}} {\cal L}^{l} (L_{s}^{\prime} \setminus 
\{0\};\Bbb C)} \! := \! \sum_{l \in \{1,2,\infty\}} 
\vert \vert (\cdot) \vert \vert_{{\cal L}^{l}(L_{s}^{
\prime} \setminus \{0\};\Bbb C)}$, where $L_{s}^{\prime} 
\! \setminus \! \{0\}$ denotes the scaled and shifted 
version of $L^{\prime} \! \setminus \! \{\lambda_{0}\}$. 
Hence, for any $f \! \in \! {\cal L}^{l}(L_{s}^{\prime};
\! M_{2}(\Bbb C))$, $l \! \in \! \{2,\infty\}$,
\begin{eqnarray*}
&(C_{w_{x,t}}f)(\widetilde{w};\! \lambda_{0}) = \sum\limits_{
j=0}^{1} \sum\limits_{k=0}^{j} \frac{(\ln t)^{k}}{t^{j/2}} 
(C_{jk}^{\natural}f)(\widetilde{w};\! \lambda_{0}) + {\cal O} 
\! \left(\frac{(\ln t)^{2}}{t}(C_{22}^{\natural}f)(\widetilde{
w};\! \lambda_{0}) \right) \!,&
\end{eqnarray*}
where $\frac{\exp \{2 i \nu \ln 2\}}{\exp\{2 i \phi(\lambda_{
0})\}}(C_{jk}^{\natural}f)(\widetilde{w};\! \lambda_{0}) \! 
:= \! \lim\limits_{\lambda^{\prime} \rightarrow \widetilde{
w} \atop \lambda^{\prime} \in - \, {\rm side} \, {\rm of} \, 
L_{s}^{\prime}} \! \int_{\raise-0.5ex\hbox{$\scriptstyle{}L_{
s}^{\prime}$}} \frac{f(z) \exp \{-i z^{2}\} z^{2 i \nu} v_{jk}
(z;\lambda_{0})}{(z - \lambda^{\prime})} \frac{d z}{2 \pi i}$, 
$0 \! \leq \! j \! \leq \! 2$, $0 \! \leq \! k \! \leq \! j$, 
and $\vert \vert C_{jk}^{\natural}(\cdot;\! \lambda_{0}) \vert 
\vert_{{\cal M}(L_{s}^{\prime} \setminus \{0\};M_{2}(\Bbb C))} 
\! \leq \! {\cal K}_{1}^{\natural}(\lambda_{0}) \! < \! \infty$, 
with ${\cal M}(\bullet;\! M_{2}(\Bbb C))$ denoting the space of 
bounded linear operators acting {}from ${\cal L}^{l}(\bullet;\! 
M_{2}(\Bbb C))$ into ${\cal L}^{2}(\bullet;\! M_{2}(\Bbb C))$, 
$l \! \in \! \{2,\infty\}$. According to Theorem~4.1, $\mu^{c}
(\cdot)$ satisfies the following linear singular integral 
equation on $L^{\prime} \! \cup \! \overline{L^{\prime}}$, 
$(\underline{{\bf Id}} - C_{w_{x,t}}) \mu^{c} \! = \! {\rm I}$; 
hence, $\mu^{c} \! = \! (\underline{{\bf Id}} - C_{w_{x,t}})^{
-1} {\rm I}$. It was shown in \cite{a7} that, as $t \! \to \! 
+\infty$ and $x \! \to \! -\infty$ such that $\lambda_{0} \! > 
\! M$, $\ker(\underline{{\bf Id}} - C_{w_{x,t}}) \! = \! 
\emptyset$ and $\vert \vert (\underline{{\bf Id}} - C_{00}^{
\natural}(\cdot;\! \lambda_{0}))^{-1} \vert \vert_{{\cal M}
(L_{s}^{\prime} \setminus \{0\};M_{2}(\Bbb C))} \! \leq \! 
{\cal K}_{2}^{\natural}(\lambda_{0}) \! < \! \infty$. Using the 
method of successive approximations, one shows that, as $t \! 
\to \! +\infty$ and $x \! \to \! -\infty$ such that $\lambda_{0} 
\! > \! M$, $\mu^{c}(\cdot)$ can be expanded in the following 
Neumann-type series (see, also, Part~II of \cite{a12}, and 
\cite{a21}),
\begin{eqnarray*}
&{\cal A}_{0}(\varsigma) \vert_{\varsigma(\widetilde{w})} = 
\mu^{c}(\widetilde{w}) \vert_{L_{s}^{\prime} \setminus \{0\}} 
= \mu^{c}_{00}(\widetilde{w};\! \lambda_{0}) + \frac{\mu^{c}_{
10} (\widetilde{w};\lambda_{0}) + \mu^{c}_{11} (\widetilde{w};
\lambda_{0}) \ln t}{\sqrt{t}} + {\cal O} \! \left(\frac{\mu^{
c}_{22}(\widetilde{w};\lambda_{0})(\ln t)^{2}}{t} \right) \!,&
\end{eqnarray*}
where $\mu^{c}_{00}(\widetilde{w};\! \lambda_{0}) \! := \! 
(\underline{{\bf Id}} - C_{00}^{\natural}(\widetilde{w};\! 
\lambda_{0}))^{-1} {\rm I}$, $\vert \vert (\underline{{\bf Id}} 
- C_{00}^{\natural}(\cdot;\! \lambda_{0}))^{-1} {\rm I} \vert 
\vert_{{\cal L}^{2}(L_{s}^{\prime} \setminus \{0\};M_{2}(\Bbb 
C))} \! < \! \infty$, and $\vert \vert \mu^{c}_{jk} (\cdot;\! 
\lambda_{0}) \vert \vert_{{\cal L}^{2}(L_{s}^{\prime} \setminus 
\{0\};M_{2}(\Bbb C))} \! < \! \infty$, $1 \! \leq \! j \! \leq 
\! 2$, $0 \! \leq \! k \! \leq \! j$: an explicit expression 
for $(\underline{{\bf Id}}-C_{00}^{\natural}(\widetilde{w};\! 
\lambda_{0}))^{-1} {\rm I}$ in terms of parabolic-cylinder 
functions was given in Sec.~7 of \cite{a7} (see below). Making 
one more change of variable, $\varrho \! = \! \sqrt{2} 
\widetilde{w} \exp \{\frac{i \pi}{4}\}$, and recalling the 
definition of $I_{0}$, one shows that
\begin{eqnarray*}
&I_{0} - I_{1/2} = I_{0,a} + I_{0,b} + I_{0,c} + I_{0,d} + 
{\cal E}_{r},&
\end{eqnarray*}
where
\begin{eqnarray*}
&I_{1/2} := Y_{a}(\lambda,\! \lambda_{0};t) \int_{0}^{
\widehat{\alpha}} \mu^{c}_{00}(\frac{\varrho e^{- \frac{i 
\pi}{4}}}{\sqrt{2}};\! \lambda_{0}) \varrho^{2 i \nu} e^{- 
\varrho^{2}/2} \sigma_{+} d \varrho,& \\
&I_{0,a} := Y_{a}(\lambda,\! \lambda_{0};t) \sum\limits_{
\vec{l}_{4} \in \Bbb L_{4}}^{\prime} \int_{0}^{\widehat{
\alpha}} \mu^{c}_{00}(\frac{\varrho e^{- \frac{i \pi}{4}}}{
\sqrt{2}};\! \lambda_{0}) \prod\limits_{k=1}^{4} (p_{k}
(\frac{\varrho e^{-\frac{i \pi}{4}}}{\sqrt{2}}))^{l_{k}} 
\varrho^{2 i \nu} e^{-\varrho^{2}/2} \sigma_{+} d \varrho,& 
\\
&I_{0,b} := Y_{b}(\lambda,\! \lambda_{0};t) 
\sum\limits_{\vec{l}_{4} \in \Bbb L_{4}} \int_{0}^{
\widehat{\alpha}} \mu^{c}_{00}(\frac{\varrho e^{-\frac{
i \pi}{4}}}{\sqrt{2}};\! \lambda_{0}) \prod\limits_{k
=1}^{4}(p_{k}(\frac{\varrho e^{- \frac{i \pi}{4}}
}{\sqrt{2}}))^{l_{k}} \varrho^{2 i \nu + 1} e^{- 
\varrho^{2}/2} \sigma_{+} d \varrho,& \\
&I_{0,c} := Y_{c}(\lambda,\! \lambda_{0};t) 
\sum\limits_{\vec{l}_{4} \in \Bbb L_{4}} \int_{0}^{
\widehat{\alpha}} \{\mu^{c}_{10}(\frac{\varrho e^{- 
\frac{i \pi}{4}}}{\sqrt{2}};\! \lambda_{0}) + \mu^{
c}_{11}(\frac{\varrho e^{- \frac{i \pi}{4}}}{\sqrt{
2}};\! \lambda_{0}) \ln t\}& \\ 
&\! \! \! \! \! \! \! \! \! \! \! \! \! \! \! \! \! 
\! \! \! \! \! \! \! \! \! \! \! \! \! \! \! \! \! 
\! \! \! \! \! \! \! \! \! \! \! \! \! \! \times \, 
\, \prod\limits_{k=1}^{4}(p_{k}(\frac{\varrho e^{- 
\frac{i \pi}{4}}}{\sqrt{2}}))^{l_{k}} \varrho^{2 
i \nu} e^{- \varrho^{2}/2} \sigma_{+} d \varrho,& \\
&I_{0,d} := Y_{d}(\lambda,\! \lambda_{0};t) 
\sum\limits_{\vec{l}_{4} \in \Bbb L_{4}} \int_{0}^{
\widehat{\alpha}} \mu^{c}_{00}(\frac{\varrho e^{-\frac{
i \pi}{4}}}{\sqrt{2}};\! \lambda_{0}) \prod\limits_{k
=1}^{4}(p_{k}(\frac{\varrho e^{- \frac{i \pi}{4}}}{
\sqrt{2}}))^{l_{k}} \varrho^{2 i \nu + 1} e^{- 
\varrho^{2}/2} \sigma_{+} d \varrho,& \\
&{\cal E}_{r} := {\cal O} \! \left(\frac{y(\lambda_{0};t)}
{\lambda_{0}^{3} t^{3/2}} \{\sum\limits_{k=1}^{2} \frac{
C_{k}^{\sharp}(\lambda_{0})}{(\lambda-\lambda_{0})^{k}}\} 
\int_{0}^{\widehat{\alpha}} \mu^{c}_{00}(\frac{\varrho e^{
-\frac{i \pi}{4}}}{\sqrt{2}};\! \lambda_{0}) \varrho^{2 i 
\nu + 2} e^{- \varrho^{2}/2} \sigma_{+} d \varrho 
\right)& \\
&\! \! \! \! \! + \, \, {\cal O} \! \left(\frac{y(\lambda_{
0};t)}{\lambda_{0}^{2}t^{3/2}} \{\sum\limits_{k=1}^{2} \frac{
C_{k+2}^{\sharp}(\lambda_{0})}{(\lambda-\lambda_{0})^{k}}\} 
\int_{0}^{\widehat{\alpha}} \{\mu^{c}_{10}(\frac{\varrho e^{
- \frac{i \pi}{4}}}{\sqrt{2}};\! \lambda_{0}) + \mu^{c}_{11}
(\frac{\varrho e^{-\frac{i \pi}{4}}}{\sqrt{2}};\! \lambda_{0}
) \ln t\} \varrho^{2i\nu+1} e^{-\varrho^{2}/2} \sigma_{+} d 
\varrho \right) \!,&
\end{eqnarray*}
$\widehat{\alpha} \! := \! (32 \varepsilon^{2} \lambda_{0}^{2} 
t)^{1/2}$, the prime on the summation in the expression for 
$I_{0,a}$ means that the term corresponding to $(l_{1},l_{
2},l_{3},l_{4}) \! = \! (0,0,0,0)$ is omitted {}from the sum,
\begin{eqnarray*}
&Y_{a}(\lambda,\! \lambda_{0};t) = \frac{y(\lambda_{
0};t) e^{- \frac{i \pi}{4}} {\cal R} (\lambda_{0})}{(
\lambda - \lambda_{0}) \sqrt{32 \lambda_{0}^{2} t}}, \, 
\, \, \, \, \, \, \, \, Y_{b}(\lambda,\! \lambda_{0};t)=
\frac{i y(\lambda_{0};t) \{3i\nu {\cal R}(\lambda_{0})
-\lambda_{0} {\cal R}^{\prime}(\lambda_{0})\}}{(\lambda 
- \lambda_{0})(32 \lambda_{0}^{3} t)},& \\
&Y_{c}(\lambda,\! \lambda_{0};t)=\frac{y(\lambda_{0};t) 
e^{- \frac{i \pi}{4}} {\cal R}(\lambda_{0})}{(\lambda 
- \lambda_{0}) \sqrt{32 \lambda_{0}^{2}} t}, \, \, \, \, 
\, \, \, \, \, Y_{d}(\lambda,\! \lambda_{0};t)=-\frac{
i y(\lambda_{0};t) {\cal R}(\lambda_{0})}{(\lambda - 
\lambda_{0})^{2}(32 \lambda_{0}^{2} t)},&
\end{eqnarray*}
$y(\lambda_{0};t) \! := \! \frac{e^{\frac{\pi \nu}{2}} 
e^{2 s(\lambda_{0})} e^{4 i \lambda_{0}^{4}t}}{(2\pi i) 
\lambda_{0}^{4 i \nu} 2^{3i\nu} t^{i\nu}}$, and $C_{i}^{
\sharp}(\lambda_{0}) \! \in \! {\cal S}(\Bbb R_{> M};\! 
\Bbb C)$, $1 \! \leq \! i \! \leq \! 4$. As will be shown 
below, $I_{1/2}$ gives rise to the leading-order $({\cal O}
(t^{-1/2}))$ term: towards the proof of this statement, one 
proceeds by estimating the difference, $I_{0} \! - \! I_{1/2}$. 
Recall, first, the following inequality, $\vert \! \exp \{
(\cdot)\}-1 \vert \! \leq \! \vert (\cdot) \vert \sup_{s \in 
[0,1]} \vert \! \exp \{s(\cdot)\} \vert$; hence, $\vert \! 
\exp \{\widetilde{\Delta}_{i}^{\flat}\} \! - \! 1 \vert \! \leq 
\! \vert \widetilde{\Delta}_{i}^{\flat} \vert \sup_{s \in [0,1]} 
\vert \! \exp \{s \widetilde{\Delta}_{i}^{\flat}\} \vert$, 
where $\widetilde{\Delta}_{i}^{\flat} \! \equiv \! 2 
\widetilde{\Delta}_{i}(\lambda_{0} + \frac{\varrho \exp \{- 
\frac{i \pi}{4}\}}{\sqrt{32 \lambda_{0}^{2} t}})$, $i \! \in 
\! \{1,2\}$. Since, as shown in \cite{a7}, $\vert \vert (\delta
(\cdot))^{\pm 1} \vert \vert_{{\cal L}^{\infty}(\Bbb C;\Bbb C)} 
\! < \! \infty$, {}from the definitions of $\rho_{\pm}(\lambda
_{0})$, $\widehat{\rho}_{\pm} (\lambda_{0})$, and $\widetilde{
\Delta}_{i}^{\flat}$, $i \! \in \! \{1,2\}$, it follows that 
$\sup_{s \in [0,1]} \vert \! \exp \{s \widetilde{\Delta}
_{i}^{\flat}\} \vert \! < \! \infty$; furthermore, using the 
Lipschitz property of $\ln \! \left(\! \frac{1-\vert r(\lambda) 
\vert^{2}}{1- \vert r(\lambda_{0}) \vert^{2}} \! \right)$, 
$\vert \lambda \vert \! < \! \lambda_{0}$, and the fact that 
$r(\lambda) \! \in \! {\cal S}(\widehat{\Gamma};\! \Bbb C)$ 
and $\vert \vert r \vert \vert_{{\cal L}^{\infty}(\Bbb R;\Bbb 
C)} \! < \! 1$, via an integration by parts argument, one 
deduces that
\begin{eqnarray*}
&\vert \widetilde{\Delta}_{1}^{\flat} \vert \leq \frac{
K_{1}^{\flat}(\lambda_{0}) \varrho + K_{2}^{\flat}(
\lambda_{0}) \varrho \ln \varrho + K_{3}^{\flat}(
\lambda_{0}) \varrho \ln t}{\sqrt{\lambda_{0}^{2} t}}, 
\, \, \, \, \, \, \, \, \, \, \, \vert \widetilde{\Delta}_{
2}^{\flat} \vert \leq \frac{K_{4}^{\flat}(\lambda_{0}) 
\varrho}{\sqrt{\lambda_{0}^{2} t}},&
\end{eqnarray*}
with $K_{i}^{\flat}(\lambda_{0}) \! \in \! {\cal L}^{\infty}
(\Bbb R_{> M};\! \Bbb R_{> 0})$, $i \! \in \! \{1,4\}$. 
Similarly, one gets that,
\begin{eqnarray*}
&\vert \! \exp \{- \frac{i \varrho^{3} \exp (-\frac{3 \pi i}{
4})}{8 \sqrt{2} \lambda_{0}^{2} \sqrt{t}}\} - 1 \vert \leq 
\frac{\varrho^{3}}{8 \sqrt{2} \lambda_{0}^{2} \sqrt{t}} 
\underbrace{\sup\limits_{s \in [0,1]} \vert \! \exp \{\frac{
-s(1-i) \varrho^{3}}{16 \lambda_{0}^{2} \sqrt{t}}\} \vert}_{
< \, \infty} := \frac{\widetilde{K}_{1} (\lambda_{0}) 
\varrho^{3}}{\sqrt{\lambda_{0}^{2} t}},& \\
&\vert \! \exp \{ \frac{i \varrho^{4}}{4^{4} \lambda_{
0}^{4} t} \} - 1 \vert \leq \frac{\varrho^{4}}{4^{4} 
\lambda_{0}^{4} t} \underbrace{\sup\limits_{s \in 
[0,1]} \vert \! \exp \{ \frac{i s \varrho^{4}}{4^{4} 
\lambda_{0}^{4} t}\} \vert}_{< \, \infty} := 
\frac{\widetilde{K}_{2} (\lambda_{0}) \varrho^{
4}}{\lambda_{0}^{2} t},& 
\end{eqnarray*}
with $\widetilde{K}_{i}(\lambda_{0}) \! \in \! {\cal L}^{\infty}
(\Bbb R_{> M};\! \Bbb R_{> 0})$, $i \! \in \! \{1,2\}$. Although 
the expression for the difference, $I_{0} \! - \! I_{1/2}$, 
contains many terms, estimations for the respective terms are 
analogous. Consider, say, the bound for the term corresponding 
to $(l_{1},l_{2},l_{3},l_{4}) \! = \! (1,0,0,0)$ in $I_{0,a}$, 
which is denoted by $I_{0,a}^{1}$:
\begin{eqnarray*}
&I_{0,a}^{1} := Y_{a}(\lambda,\! \lambda_{0};t) \int_{
0}^{\widehat{\alpha}} \mu^{c}_{00}(\frac{\varrho e^{- 
\frac{i \pi}{4}}}{\sqrt{2}};\! \lambda_{0}) p_{1} 
(\frac{\varrho e^{- \frac{i \pi}{4}}}{\sqrt{2}}) 
\varrho^{2 i \nu} e^{- \varrho^{2}/2} \sigma_{+} 
d \varrho.& 
\end{eqnarray*}
Using the fact that $0 \! < \! \nu \! \leq \! \nu_{\max} \! 
:= \! - \frac{1}{2 \pi} \ln(1 \! - \! \sup_{\lambda \in \Bbb 
R} \vert r(\lambda) \vert^{2}) \! < \! \infty$, and recalling 
the definitions of $s(\lambda_{0})$ and ${\cal R}(\lambda_{
0})$, one gets that, for $\lambda \! \in \! \Bbb C \! 
\setminus \! {\cal N}(\lambda_{0};\! \epsilon_{0})$, $\vert 
Y_{a}(\lambda,\! \lambda_{0};t) \vert \! \leq \! \frac{\exp 
\{\frac{\pi \nu_{\max}}{2}\} \vert r(\lambda_{0}) \vert}{2 
\pi \vert \lambda-\lambda_{0} \vert \sqrt{32 \lambda_{0}^{2} 
t}}$; hence, letting the upper limit of integration tend to 
$+\infty$ (for brevity, the following notation is used: for 
{\bf matrices} $A$ and $B$, the inequality $\vert A \vert \! 
\leq \! \vert B \vert$ means that $\vert A_{ij} \vert \! 
\leq \! \vert B_{ij} \vert \, \, \forall \, i,j)$,
\begin{eqnarray*}
&\vert I_{0,a}^{1} \vert \leq \frac{\exp \{\frac{\pi \nu_{\max}
}{2}\} \vert r(\lambda_{0}) \vert}{2 \pi \vert \lambda-\lambda_{
0} \vert \sqrt{32 \lambda_{0}^{2} t}} \int_{0}^{\infty} \vert 
\mu^{c}_{00}(\frac{\varrho e^{-\frac{i \pi}{4}}}{\sqrt{2}};\! 
\lambda_{0}) \vert \vert p_{1}(\frac{\varrho e^{-\frac{i \pi}{4}
}}{\sqrt{2}}) \vert e^{-\varrho^{2}/2} \sigma_{+} d \varrho.&
\end{eqnarray*}
In \cite{a7}, it was shown that ${\cal U}^{c}_{00} \! := \! 
\vert \vert \mu^{c}_{00}(\frac{(\cdot) e^{- \frac{i \pi}{4}}}{
\sqrt{2}};\! \lambda_{0}) \vert \vert_{{\cal L}^{2}(\Bbb R_{\geq 
0};M_{2}(\Bbb C))} \! < \! \infty$; hence, {}from this estimate 
and the Cauchy-Schwarz inequality for integrals,
\begin{eqnarray*}
&\vert I_{0,a}^{1} \vert \leq \frac{\exp \{\frac{\pi \nu_{\max}
}{2}\} \vert r(\lambda_{0}) \vert \, {\cal U}^{c}_{00} \sigma_{
+}}{2 \pi \vert \lambda-\lambda_{0} \vert \sqrt{32 \lambda_{0}^{
2} t}} \vert \vert p_{1}(\frac{(\cdot) e^{-\frac{i \pi}{4}}}{
\sqrt{2}}) \exp \{-(\cdot)^{2}/2\} \vert \vert_{{\cal L}^{2}(\Bbb 
R_{\geq 0};\Bbb C)}.&
\end{eqnarray*}
Recalling the estimate for $\vert \! \exp \{\widetilde{\Delta}_{
1}^{\flat}\} \! - \! 1 \vert$, one shows that,
\begin{eqnarray*}
&\vert \vert p_{1}(\frac{(\cdot) e^{-\frac{i \pi}{4}}}{\sqrt{2}
}) \exp \{-(\cdot)^{2}/2\} \vert \vert_{{\cal L}^{2} (\Bbb R_{
\geq 0};\Bbb C)} \leq \frac{\widetilde{K}^{\flat}(\lambda_{0}) 
\ln t}{\sqrt{\lambda_{0}^{2} t}},&
\end{eqnarray*}
where $\widetilde{K}^{\flat}(\lambda_{0}) \! \in \! {\cal L}^{
\infty}(\Bbb R_{>M};\! \Bbb R_{>0})$; hence, uniformly for 
$\lambda \! \in \! \Bbb C \! \setminus \! {\cal N}(\lambda_{0};
\! \epsilon_{0})$,
\begin{eqnarray*}
&\vert I_{0,a}^{1} \vert \leq \frac{\widetilde{K}^{\sharp}_{1}
(\lambda_{0}) \ln t}{\vert \lambda - \lambda_{0} \vert \lambda_{
0}^{2} t},&
\end{eqnarray*}
with $\widetilde{K}^{\sharp}_{1}(\lambda_{0}) \! \in \! {\cal S}
(\Bbb R_{>M};\! M_{2}(\Bbb R_{>0}))$. Similarly, recalling the 
estimates for $\vert \! \exp \{-\frac{i \varrho^{3} \exp(-\frac{
3 \pi i}{4})}{8 \sqrt{2} \lambda_{0}^{2} \sqrt{t}}\} \linebreak[4] 
- 1 \vert$, $\vert \! \exp \{\frac{i \varrho^{4}}{4^{4} \lambda_{
0}^{4} t}\} - 1 \vert$, and $\vert \! \exp \{\widetilde{\Delta}_{
2}^{\flat}\} \! - \! 1 \vert$, and using the triangle inequality 
for ${\cal L}^{2}$-norms, one shows that the remaining terms for 
$I_{0,a}$ are of the type ${\cal O}(t^{-\frac{m}{2}})$ and ${\cal 
O}(t^{-\frac{n}{2}} \ln \! t)$, $2 \! \leq \! m \! \leq \! 5$, $3 
\! \leq \! n \! \leq \! 6$. Estimating the remaining terms of $I_{
0} \! - \! I_{1/2}$ analogously, one shows that, as $t \! \to \! 
+\infty$ and $x \! \to \! -\infty$ such that $\lambda_{0} \! > \! 
M$, uniformly for $\lambda \! \in \! \Bbb C \! \setminus \! {\cal 
N}(\lambda_{0};\! \epsilon_{0})$,
\begin{eqnarray*}
&\vert I_{0} - I_{1/2} \vert \leq \frac{\widetilde{
K}^{\sharp}_{2}(\lambda;\lambda_{0}) \ln t}{\lambda_{
0}^{2} t},&
\end{eqnarray*}
where $\vert \vert \widetilde{K}^{\sharp}_{2}(\cdot;\! \lambda_{
0}) \vert \vert_{{\cal L}^{\infty}(\Bbb C \setminus {\cal N}
(\lambda_{0};\epsilon_{0});M_{2}(\Bbb R_{>0}))} \! < \! \infty$, 
$\widetilde{K}^{\sharp}_{2}(\lambda;\cdot) \! \in \! {\cal S}
(\Bbb R_{> M};\! M_{2}(\Bbb R_{>0}))$, and, as $\lambda \! \to \! 
\infty$, $\widetilde{K}^{\sharp}_{2}(\lambda;\! \lambda_{0}) \sim 
\! {\cal O}(\widetilde{k}^{\sharp}_{2}(\lambda_{0}) \vert \lambda 
\vert^{-1})$, with $\widetilde{k}^{\sharp}_{2}(\lambda_{0}) \! 
\in \! {\cal S}(\Bbb R_{>M};\! M_{2}(\Bbb R_{>0}))$. 

Repeating the whole of the above analysis {\em mutatis mutandis\/} 
for each term on the right-hand side of the original integral 
expression for $\chi^{c}(\lambda)$ which appears at the very 
beginning of the proof, one shows that, as $t \! \to \! 
+\infty$ and $x \! \to \! -\infty$ such that $\lambda_{0} \! > \! 
M$, uniformly for $\lambda \! \in \! \Bbb C \! \setminus \! \cup_{
\aleph \in \{0,\pm \lambda_{0}\}}{\cal N}(\aleph;\! \epsilon_{0})$,
\begin{eqnarray*}
&\vert \chi^{c}(\lambda) - \chi^{c}_{1/2}(\lambda) \vert 
\leq \frac{(h_{1}^{+}(\lambda;\lambda_{0}) + h_{2}^{+}(\lambda;
\lambda_{0})) \ln t}{\lambda_{0}^{2} t},&
\end{eqnarray*}
where $h_{1}^{+}(\lambda;\! \lambda_{0}) \! := \! \sum_{l^{
\prime} \in \{0,i0\}} \! e_{l^{\prime}}^{+}(\lambda;\! \lambda_{
0}) \vert r(l^{\prime}) \vert$, $h_{2}^{+}(\lambda;\! \lambda_{0
}) \! := \! \sum_{l \in \{\pm \lambda_{0}\}} \! e_{l}^{+}(
\lambda;\! \lambda_{0})$, and the functions $e_{l^{\prime}}^{+}
(\lambda;\! \lambda_{0})$ and $e_{l}^{+}(\lambda;\! \lambda_{0})$ 
have the following property as $\lambda \! \to \! \infty$, $h_{1}
^{+}(\lambda;\! \lambda_{0}) \! + \! h_{2}^{+}(\lambda;\! \lambda
_{0}) \! \sim \! {\cal O}(\{\sum_{l^{\prime} \in \{0,i0\}} \! e^{
\sharp}_{l^{\prime}}(\lambda_{0}) \vert r(l^{\prime}) \vert \! + 
\! \sum_{l \in \{\pm \lambda_{0}\}} \! e^{\sharp}_{l}(\lambda_{0}
)\} \vert \lambda \vert^{-1})$, $e^{\sharp}_{l^{\prime}}(\lambda_
{0}) \! \in \! {\cal L}^{\infty}(\Bbb R_{>M};\! M_{2}(\Bbb R_{>0}
))$, $l^{\prime} \! \in \! \{0,i0\}$, and $e^{\sharp}_{l}(\lambda
_{0}) \! \in \! {\cal S}(\Bbb R_{>M};\! M_{2}(\Bbb R_{>0}))$, $l 
\! \in \! \{\pm \lambda_{0}\}$; moreover, $\vert \vert h_{1}^{+}
(\cdot;\! \lambda_{0}) \vert \vert_{{\cal L}^{\infty}(\Bbb C 
\setminus {\cal N}(0;\epsilon_{0});M_{2}(\Bbb R_{>0}))} \! < \! 
\infty$, $\vert \vert h_{2}^{+}(\cdot;\! \lambda_{0}) \vert \vert
_{{\cal L}^{\infty}(\Bbb C \setminus \cup_{\aleph \in \{\pm 
\lambda_{0}\}} {\cal N}(\aleph;\epsilon_{0});M_{2}(\Bbb R_{>0}))} 
\! < \! \infty$, $h_{1}^{+}(\lambda;\cdot) \! \in \! {\cal L}^{
\infty}(\Bbb R_{>M};\! M_{2}(\Bbb R_{>0}))$, $h_{2}^{+}(\lambda;
\cdot) \! \in \! {\cal S}(\Bbb R_{>M};\! M_{2}(\Bbb R_{>0}))$, 
and $\chi^{c}_{1/2}(\lambda)$ represents the sum over all $I_{
1/2}$-like terms in which the upper limits of integration tend 
to $+\infty$. One can write $\chi^{c}_{1/2}(\lambda)$ in the 
following form,
\begin{eqnarray*}
&\chi^{c}_{1/2}(\lambda)={\rm I}+\frac{(\widetilde{\Lambda}^{0})
^{{\rm ad}(\sigma_{3})}}{\sqrt{16 \lambda_{0}^{2} t}} {\cal X}^{
\Sigma_{r}}(\lambda)+\frac{(\widehat{\Lambda}_{{\cal C}}^{0})^{{
\rm ad}(\sigma_{3})}}{\sqrt{8 \lambda_{0}^{2} t}} {\cal X}^{
\Sigma_{{\cal C}}}(\lambda),&
\end{eqnarray*}
where
\begin{eqnarray*}
&\widetilde{\Lambda}^{0} = \frac{\exp \{2 i \lambda_{0}^{4} t\}}
{(16 \lambda_{0}^{4} t)^{\frac{i \nu}{2}}} \exp \{\sum\limits_{l 
\in \{\pm\}} \! (\rho_{l}(\lambda_{0}) + \widehat{\rho}_{l} 
(\lambda_{0}))\},& \\
&\rho_{\pm}(\lambda_{0}) = \frac{1}{2 \pi i} \int_{
0}^{\pm \lambda_{0}} \! \ln \! \! \left(\! \frac{1 - 
\vert r(\varsigma) \vert^{2}}{1 - \vert r(\lambda_{0}) 
\vert^{2}} \! \right) \! \frac{d \varsigma}{(\varsigma 
-\lambda_{0})}, \, \, \, \, \, \, \, \, \, \widehat{\rho}_{
\pm}(\lambda_{0})=\int_{\pm i \infty}^{i0} \frac{\ln 
(1 - r(\varsigma) \overline{r(\overline{\varsigma})}
)}{(\varsigma - \lambda_{0})} \frac{d \varsigma}{2 
\pi i},& \\
&{\cal X}^{\Sigma_{r}}_{11}(\lambda) = \frac{
\overline{r_{{\cal B}} (\lambda_{0})}}{2^{2 i \nu}} 
\int_{0}^{\varepsilon_{1}} \frac{{\cal X}^{\Sigma_{
B,r}}_{+,12}(\varsigma) \varsigma^{- 2 i \nu} e^{i 
\varsigma^{2}}}{(\lambda - \lambda_{0})} \frac{
d \varsigma}{2 \pi i} - \frac{\overline{r_{{\cal B}} 
(\lambda_{0})} e^{2 \pi \nu}}{2^{2 i \nu}} \int_{0}^{
\varepsilon_{2}} \frac{{\cal X}^{\Sigma_{B,r}}_{-,12} 
(\varsigma) \varsigma^{- 2 i \nu} e^{i \varsigma^{2}}
}{(\lambda - \lambda_{0})} \frac{d \varsigma}{2 \pi 
i}& \\
&\, \, \, \, \, \, \, \, \, \, \, \, \, \, \, \, \, \, 
\, \, \, \, \, + \, \, \frac{\overline{r_{{\cal B}}
(\lambda_{0})}}{2^{2 i \nu}} \int_{0}^{\varepsilon_{1}} 
\frac{{\cal X}^{\Sigma_{A,r}}_{+,12}(\varsigma) 
\varsigma^{- 2 i \nu} e^{i \varsigma^{2}}}{(\lambda + 
\lambda_{0})} \frac{d \varsigma}{2 \pi i} - \frac{
\overline{r_{{\cal B}} (\lambda_{0})} e^{2 \pi \nu}}{
2^{2 i \nu}} \int_{0}^{\varepsilon_{2}} \frac{{\cal 
X}^{\Sigma_{A,r}}_{-,12} (\varsigma) \varsigma^{- 2 i 
\nu} e^{i \varsigma^{2}}}{(\lambda + \lambda_{0})} 
\frac{d \varsigma}{2 \pi i},& \\
&{\cal X}^{\Sigma_{r}}_{12} (\lambda) = - \frac{r_{{\cal 
B}} (\lambda_{0})}{2^{-2 i \nu}} \int_{0}^{\overline{
\varepsilon_{1}}} \frac{{\cal X}^{\Sigma_{B,r}}_{-,11} 
(\varsigma) \varsigma^{2 i \nu} e^{- i \varsigma^{2}}
}{(\lambda - \lambda_{0})} \frac{d \varsigma}{2 \pi i}
+ \frac{r_{{\cal B}}(\lambda_{0}) e^{2 \pi \nu}}{2^{
-2 i \nu}} \int_{0}^{\overline{\varepsilon_{2}}} 
\frac{{\cal X}^{\Sigma_{B,r}}_{+,11}(\varsigma) 
\varsigma^{2 i \nu} e^{- i \varsigma^{2}}}{(\lambda - 
\lambda_{0})} \frac{d \varsigma}{2 \pi i}& \\
&\, \, \, \, \, \, \, \, \, \, \, \, \, \, \, \, \, \, 
- \, \, \frac{r_{{\cal B}}(\lambda_{0})}{
2^{-2 i \nu}} \int_{0}^{\overline{\varepsilon_{1}}} 
\frac{{\cal X}^{\Sigma_{A,r}}_{-,11} (\varsigma) 
\varsigma^{2 i \nu} e^{- i \varsigma^{2}}}{(\lambda + 
\lambda_{0})} \frac{d \varsigma}{2 \pi i} + \frac{r_{{
\cal B}}(\lambda_{0}) e^{2 \pi \nu}}{2^{-2 i \nu}} 
\int_{0}^{\overline{\varepsilon_{2}}} \frac{{\cal X}^{
\Sigma_{A,r}}_{+,11}(\varsigma) \varsigma^{2 i \nu} e^{- 
i \varsigma^{2}}}{(\lambda + \lambda_{0})} \frac{d 
\varsigma}{2 \pi i},& \\
&{\cal X}^{\Sigma_{r}}_{21} (\lambda) = \frac{
\overline{r_{{\cal B}}(\lambda_{0})}}{2^{2 i \nu}} 
\int_{0}^{\varepsilon_{1}} \frac{{\cal X}^{\Sigma_{
B,r}}_{+,22} (\varsigma) \varsigma^{- 2 i \nu} e^{i 
\varsigma^{2}}}{(\lambda - \lambda_{0})} \frac{d 
\varsigma}{2 \pi i} - \frac{\overline{r_{{\cal B}} 
(\lambda_{0})} e^{2 \pi \nu}}{2^{2 i \nu}} \int_{
0}^{\varepsilon_{2}} \frac{{\cal X}^{\Sigma_{B,
r}}_{-,22} (\varsigma) \varsigma^{- 2 i \nu} e^{i 
\varsigma^{2}}}{(\lambda - \lambda_{0})} \frac{d 
\varsigma}{2 \pi i}& \\
&\, \, \, \, \, \, \, \, \, \, \, \, \, \, \, \, \, \, 
\, \, \, \, \, + \, \, \frac{\overline{r_{{\cal B}} 
(\lambda_{0})}}{2^{2 i \nu}} \int_{0}^{\varepsilon_{1}} 
\frac{{\cal X}^{\Sigma_{A,r}}_{+,22} (\varsigma) 
\varsigma^{- 2 i \nu} e^{i \varsigma^{2}}}{(\lambda + 
\lambda_{0})} \frac{d \varsigma}{2 \pi i} - \frac{
\overline{r_{{\cal B}} (\lambda_{0})} e^{2 \pi \nu}}{
2^{2 i \nu}} \int_{0}^{\varepsilon_{2}} \frac{{\cal 
X}^{\Sigma_{A,r}}_{-,22}(\varsigma) \varsigma^{- 2 i 
\nu} e^{i \varsigma^{2}}}{(\lambda + \lambda_{0})} 
\frac{d \varsigma}{2 \pi i},& \\
&{\cal X}^{\Sigma_{r}}_{22} (\lambda) = - \frac{r_{{
\cal B}}(\lambda_{0})}{2^{-2 i \nu}} \int_{0}^{
\overline{\varepsilon_{1}}} \frac{{\cal X}^{\Sigma_{
B,r}}_{-,21}(\varsigma) \varsigma^{2 i \nu} e^{- i 
\varsigma^{2}}}{(\lambda - \lambda_{0})} \frac{d 
\varsigma}{2 \pi i} + \frac{r_{{\cal B}}(\lambda_{0}
) e^{2 \pi \nu}}{2^{-2 i \nu}} \int_{0}^{\overline{
\varepsilon_{2}}} \frac{{\cal X}^{\Sigma_{B,r}}_{
+,21} (\varsigma) \varsigma^{2 i \nu} e^{- i 
\varsigma^{2}}}{(\lambda - \lambda_{0})} \frac{d 
\varsigma}{2 \pi i}& \\
&\, \, \, \, \, \, \, \, \, \, \, \, \, \, \, \, \, \, 
- \, \, \frac{r_{{\cal B}}(\lambda_{0}
)}{2^{-2 i \nu}} \int_{0}^{\overline{\varepsilon_{1}}} 
\frac{{\cal X}^{\Sigma_{A,r}}_{-,21}(\varsigma) 
\varsigma^{2 i \nu} e^{- i \varsigma^{2}}}{(\lambda + 
\lambda_{0})} \frac{d \varsigma}{2 \pi i} + \frac{r_{
{\cal B}}(\lambda_{0}) e^{2 \pi \nu}}{2^{- 2 i \nu}} 
\int_{0}^{\overline{\varepsilon_{2}}} \frac{{\cal X}^{
\Sigma_{A,r}}_{+,21}(\varsigma) \varsigma^{2 i \nu} 
e^{-i \varsigma^{2}}}{(\lambda + \lambda_{0})} 
\frac{d \varsigma}{2 \pi i},& \\
&r_{{\cal B}}(\lambda_{0}) := r(\lambda_{0}) \!
\prod\limits_{l=n+1}^{N} \! \left(\! \frac{(\lambda_{
0} - \overline{\lambda_{l}})(\lambda_{0}+\overline{
\lambda_{l}})}{(\lambda_{0} - \lambda_{l})(\lambda_{0}
+ \lambda_{l})} \! \right)^{2} \!,& \\
&{\cal X}^{\Sigma_{B,r}}_{-,11} (\varsigma) = \frac{
\varsigma^{- i \nu} e^{\frac{i \varsigma^{2}}{2}}}{2^{
\frac{i \nu}{2}} e^{- \frac{\pi \nu}{4}}} D_{i \nu}(\sqrt{
2} \varsigma e^{\frac{i \pi}{4}}), \, \, \, \, \, \,
\, \, {\cal X}^{\Sigma_{B,r}}_{+,11} (\varsigma) = \frac{
\varsigma^{- i \nu} e^{\frac{i \varsigma^{2}}{2}}}{2^{
\frac{i \nu}{2}} e^{\frac{3 \pi \nu}{4}}} D_{i \nu} 
(\sqrt{2} \varsigma e^{- \frac{3 \pi i}{4}}),& \\
&{\cal X}^{\Sigma_{B,r}}_{+,12} (\varsigma) = \frac{
2^{\frac{i \nu}{2}} \varsigma^{i \nu} e^{- \frac{i 
\varsigma^{2}}{2} } e^{\frac{\pi \nu}{4}}}{\beta_{21}} 
\{ \partial_{\varsigma} D_{-i \nu} (\sqrt{2} \varsigma 
e^{- \frac{i \pi}{4}}) - i \varsigma D_{-i \nu} 
(\sqrt{2} \varsigma e^{- \frac{i \pi}{4}}) \},& \\
&{\cal X}^{\Sigma_{B,r}}_{-,12} (\varsigma) = \frac{
2^{\frac{i \nu}{2}} \varsigma^{i \nu} e^{- \frac{i 
\varsigma^{2}}{2}} e^{- \frac{3 \pi \nu}{4}}}{\beta_{
21}} \{\partial_{\varsigma} D_{-i \nu} (\sqrt{2} 
\varsigma e^{\frac{3 \pi i}{4}}) - i \varsigma D_{-i 
\nu} (\sqrt{2} \varsigma e^{\frac{3 \pi i}{4}})\},& 
\\
&{\cal X}^{\Sigma_{B,r}}_{-,21} (\varsigma) = \frac{
2^{- \frac{i \nu}{2}} \varsigma^{- i \nu} e^{\frac{i 
\varsigma^{2}}{2}} e^{\frac{\pi \nu}{4}}}{\beta_{12}} 
\{\partial_{\varsigma} D_{i \nu} (\sqrt{2} \varsigma 
e^{\frac{i \pi}{4}}) + i \varsigma D_{i \nu} (\sqrt{2} 
\varsigma e^{\frac{i \pi}{4}})\},& \\
&{\cal X}^{\Sigma_{B,r}}_{+,21} (\varsigma) = \frac{
2^{- \frac{i \nu}{2}} \varsigma^{-i \nu} e^{\frac{i 
\varsigma^{2}}{2}} e^{- \frac{3 \pi \nu}{4}}}{\beta_{
12}} \{\partial_{\varsigma} D_{i \nu} (\sqrt{2} 
\varsigma e^{- \frac{3 \pi i}{4}}) + i \varsigma D_{i 
\nu} (\sqrt{2} \varsigma e^{- \frac{3 \pi i}{4}})\},& 
\\
&{\cal X}^{\Sigma_{B,r}}_{+,22} (\varsigma) = \frac{
\varsigma^{i \nu} e^{- \frac{i \varsigma^{2}}{2}}}{
2^{- \frac{i \nu}{2}} e^{- \frac{\pi \nu}{4}}} D_{-i\nu} 
(\sqrt{2} \varsigma e^{- \frac{i \pi}{4}}), \, \, \, \,
\, \, \, \, {\cal X}^{\Sigma_{B,r}}_{-,22} (\varsigma) 
= \frac{\varsigma^{i \nu} e^{- \frac{i \varsigma^{2}}{
2}}}{2^{- \frac{i \nu}{2}} e^{\frac{3 \pi \nu}{4}}} 
D_{- i \nu} (\sqrt{2} \varsigma e^{\frac{3 \pi i}{4}
}),& \\
&{\cal X}^{\Sigma_{A,r}}_{\pm,11} (\varsigma) = {\cal 
X}^{\Sigma_{B,r}}_{\pm,11} (\varsigma), \, \, \, \, \,
\, \, \, {\cal X}^{\Sigma_{A,r}}_{\pm,12} (\varsigma) 
= - {\cal X}^{\Sigma_{B,r}}_{\pm,12}(\varsigma),& \\
&{\cal X}^{\Sigma_{A,r}}_{\pm,21} (\varsigma) = - {
\cal X}^{\Sigma_{B,r}}_{\pm,21} (\varsigma), \, \, \, 
\, \, \, \, \, \, {\cal X}^{\Sigma_{A,r}}_{\pm,22} 
(\varsigma) = {\cal X}^{\Sigma_{B,r}}_{\pm,22} 
(\varsigma),& \\
&\beta_{12} = - \frac{ 2^{1 + i \nu} \sqrt{\pi} e^{-
\frac{\pi \nu}{2}} e^{\frac{i \pi}{4}} }{ \overline{
r_{{\cal B}}(\lambda_{0})} \, \Gamma(-i\nu)}, \, \, 
\, \, \, \, \, \, \, \, \beta_{21} = \overline{\beta_{
12}},& 
\end{eqnarray*}
the integrals are evaluated along the rays $(0,\varepsilon_{
k})$ (and their complex conjugates), $\varepsilon_{1} \! := 
\! \infty \exp\{\frac{i\pi}{4}\}$, $\varepsilon_{2} \! := \! 
\infty \exp\{-\frac{3 \pi i}{4}\}$, $D_{\pm i \nu}(\cdot)$ is 
the parabolic-cylinder function \cite{a19}, and
\begin{eqnarray*}
&\widehat{\Lambda}_{{\cal C}}^{0} = \exp \{\sum\limits_{l \in 
\{\pm\}} \! (\rho^{{\cal C}}_{l}(0) + \widehat{\rho}^{{\cal C}
}_{l} (0))\},& \\
&\rho^{{\cal C}}_{\pm}(0) \! = \! -\frac{1}{2 \pi i} \int_{0}^{
\pm \lambda_{0}} \ln \! \vert \varsigma \vert \, d \ln (1-\vert 
r(\varsigma) \vert^{2}), \, \, \, \, \, \, \, \, \widehat{\rho}
^{{\cal C}}_{\pm}(0) \! = \! \int_{\pm \infty}^{0} \frac{\ln (1
+\vert r(i \varsigma) \vert^{2})}{\varsigma} \frac{d \varsigma}
{2 \pi i},& \\
&{\cal X}_{11}^{\Sigma_{{\cal C}}}(\lambda) = 
\widehat{{\cal R}}_{{\cal C}}^{(-)} (0) \int_{
0}^{\overline{\varepsilon_{1}}} \frac{{\cal 
X}_{-,12}^{\Sigma_{{\cal C},r}^{2}} (\varsigma) 
e^{-i \varsigma^{2}}}{\lambda} \frac{d \varsigma}{
2 \pi i} - \widehat{{\cal R}}_{{\cal C}}^{(-)} (0) 
\int_{0}^{\overline{\varepsilon_{2}}} \frac{{\cal 
X}_{-,12}^{\Sigma_{{\cal C},r}^{4}} (\varsigma) 
e^{-i \varsigma^{2}}}{\lambda} \frac{d \varsigma}{
2 \pi i},& \\
&{\cal X}_{12}^{\Sigma_{{\cal C}}}(\lambda) = - 
\widehat{{\cal R}}_{{\cal C}}^{(+)} (0) \int_{0}^{
\varepsilon_{1}} \frac{{\cal X}_{-,11}^{\Sigma_{{
\cal C},r}^{1}} (\varsigma) e^{i \varsigma^{2}}}{
\lambda} \frac{d \varsigma}{2 \pi i} + \widehat{{
\cal R}}_{{\cal C}}^{(+)} (0) \int_{0}^{\varepsilon_{
2}} \frac{{\cal X}_{-,11}^{\Sigma_{{\cal C},r}^{3}} 
(\varsigma) e^{i \varsigma^{2}}}{\lambda} \frac{d 
\varsigma}{2 \pi i },& \\
&{\cal X}_{21}^{\Sigma_{{\cal C}}}(\lambda) = 
\widehat{{\cal R}}_{{\cal C}}^{(-)}(0) \int_{0}^{
\overline{\varepsilon_{1}}} \frac{{\cal X}_{-,22}^{
\Sigma_{{\cal C},r}^{2}} (\varsigma) e^{-i \varsigma^{
2}}}{\lambda} \frac{d \varsigma}{2 \pi i} - \widehat{{
\cal R}}_{{\cal C}}^{(-)} (0) \int_{0}^{\overline{
\varepsilon_{2}}} \frac{{\cal X}_{-,22}^{\Sigma_{{\cal 
C},r}^{4}} (\varsigma) e^{-i \varsigma^{2}}}{\lambda} 
\frac{d \varsigma}{2 \pi i},& \\
&{\cal X}_{22}^{\Sigma_{{\cal C}}}(\lambda) = - 
\widehat{{\cal R}}_{{\cal C}}^{(+)}(0) \int_{0}^{
\varepsilon_{1}} \frac{{\cal X}_{-,21}^{\Sigma_{{\cal 
C},r}^{1}} (\varsigma) e^{i \varsigma^{2}}}{\lambda} 
\frac{d \varsigma}{2 \pi i} + \widehat{{\cal R}}_{{
\cal C}}^{(+)} (0) \int_{0}^{\varepsilon_{2}} 
\frac{{\cal X}_{-,21}^{\Sigma_{{\cal C},r}^{3}} 
(\varsigma) e^{i \varsigma^{2}}}{\lambda} \frac{d 
\varsigma}{2 \pi i },& \\
&\widehat{{\cal R}}_{{\cal C}}^{(+)}(0) = (-1)^{
2 i \nu} \lambda_{0}^{4 i \nu} \{\frac{r_{{\cal C}}(
0)}{(1 - \vert r(0) \vert^{2})} - \frac{r_{{\cal C}}
(i0)}{(1 + \vert r(i0) \vert^{2})}\}, \, \, \, \, \, 
\, \, \, \, \widehat{{\cal R}}_{{\cal C}}^{(-)}(0) = 
\overline{\widehat{{\cal R}}_{{\cal C}}^{(+)}(0)},& \\
&r_{{\cal C}}(0) := r(0) \exp \{4 i \sum\limits_{l=n+1}^{N} 
\gamma_{l}\}, \, \, \, \, \, \, \, \, \, \, r_{{\cal C}}
(i0) := r(i0) \exp\{4i \sum\limits_{l=n+1}^{N} \gamma_{l}\},&
\end{eqnarray*}
where $r(0) \! := \! (r(\lambda) \vert_{\lambda \in \Bbb R}) 
\vert_{\lambda=0}$, and $r(i0) \! := \! (r(\lambda) \vert_{
\lambda \in i \Bbb R}) \vert_{\lambda=0}$: the explicit 
expressions for ${\cal X}_{-,jk}^{\Sigma_{{\cal C},r}^{i}}
(\varsigma)$, $i \! \in \! \{1,2,3,4\}$, $j,k \! \in \! 
\{1,2\}$, are not written down here since they will not 
actually be needed. Since \cite{a7} $(r(\lambda) 
\vert_{\lambda \in \Bbb R}) \vert_{\lambda=0} 
\! = \! (r(\lambda) \vert_{\lambda \in i \Bbb R}) \vert
_{\lambda=0} \! = \! 0$, $\widehat{{\cal R}}_{{\cal C}}
^{(+)}(0) \! = \! \widehat{{\cal R}}_{{\cal C}}^{(-)}(0) 
\! = \! h_{1}^{+}(\lambda;\! \lambda_{0}) \! = \! 0$; 
hence, ${\cal X}_{ij}^{\Sigma_{{\cal C}}}(\lambda) \! = 
\! 0$, $i,j \! \in \! \{1,2\}$. To obtain the expression 
for ${\cal X}^{\Sigma_{r}}_{ij}(\lambda)$, $i,j \! \in \! 
\{1,2\}$, given above, use was made of the explicit 
representation for $\mu^{c}_{00}(\cdot;\! \lambda_{0}) \! 
:= \! (\underline{{\bf Id}}-C_{00}^{\natural}(\cdot;\! 
\lambda_{0}))^{-1} {\rm I}$ on $L^{\prime} \! \cup \! 
\overline{L^{\prime}}$ (recall the definition of $I_{1/2})$ 
in terms of parabolic-cylinder functions given in Sec.~7 of 
\cite{a7}. Now, substituting the expressions given above for 
${\cal X}^{\Sigma_{B,r}}_{\pm,ij} (\varsigma)$ and ${\cal X}
^{\Sigma_{A,r}}_{\pm,ij}(\varsigma)$, $i,j \! \in \! \{1,2
\}$, into the corresponding integrals for ${\cal X}^{\Sigma
_{r}}_{ij}(\lambda)$, $i,j \! \in \! \{1,2\}$, and using the 
following identities \cite{a19}, $\partial_{\varsigma} D_{a}
(\varsigma) \! = \! \frac{1}{2}(aD_{a-1}(\varsigma) \! - \! 
D_{a+1}(\varsigma))$, $\varsigma D_{a}(\varsigma) \! = \! D
_{a+1}(\varsigma) \! + \! a D_{a-1}(\varsigma)$, and $\vert 
\Gamma (i \nu) \vert^{2} \! = \! \pi / (\nu \sinh \pi \nu)$, 
as well as the following integral \cite{a19},
\begin{eqnarray*}
&\int_{0}^{\infty} \exp(-\frac{x^{2}}{4}) x^{a-1} D_{-b}(x) 
d x = \frac{\sqrt{\pi} \exp \{-\frac{1}{2}(a+b) \ln 2\} 
\Gamma(a)}{\Gamma(\frac{1}{2}(a+b)+\frac{1}{2})}, \, \, \, \, 
\, \, \, \, \, \Re(a) > 0,&
\end{eqnarray*}
one obtains the result stated in the Lemma. \hfill 
\rule{6.5pt}{6.5pt}
\begin{bbb}
As $\lambda \! \rightarrow \! \infty$, $\lambda \! \in \! \Bbb 
C \! \setminus \! (\widehat{\Gamma} \cup (\cup_{i=1}^{n} (\{\pm 
\lambda_{i}\} \! \cup \! \{\pm \overline{\lambda_{i}}\})))$, 
$\chi(\lambda)$ has the following asymptotic expansion,
\begin{eqnarray}
\chi(\lambda) & = & {\rm I} + \frac{1}{2 \lambda} \! \left( \! 
\left\{\overline{Q^{\chi}(x,t)} + 4 \sum\limits_{i=1}^{n} \! 
\left(\beta_{i} - \frac{\chi^{c}_{21}(\overline{\lambda_{i}})}
{\chi^{c}_{11}(\overline{\lambda_{i}})} \widehat{\delta}_{i} 
\right) \! \right\} \! \sigma_{-} \right. \nonumber \\
 & + & \left. \left\{Q^{\chi}(x,t) + 4 \sum\limits_{i=1}^{n} 
\left(\omega_{i}-\frac{\chi^{c}_{12}(\lambda_{i})}{\chi^{c}_{
22}(\lambda_{i})} \alpha_{i} \right) \! \right\} \! \sigma_{+} 
\! \right) + {\cal O}(\lambda^{-2}),
\end{eqnarray}
where $\lim\limits_{\lambda \rightarrow \infty} \! (\chi^{c}
(x,t;\! \lambda))_{12} \! := \! Q^{\chi}(x,t) / 2 \lambda$, 
$\{\alpha_{i},\omega_{i}\}_{i=1}^{n}$ satisfy the following 
non-degenerate system of $2n$ linear inhomogeneous algebraic 
equations,
\begin{eqnarray}
&\left[ \begin{array}{cccccc}
\left. \begin{array}{ccc} \cline{1-3}
\multicolumn{1}{|c}{} &   & \multicolumn{1}{c|}{} \\
\multicolumn{1}{|c}{} & \widehat{{\cal A}}^{+} & 
\multicolumn{1}{c|}{} \\
\multicolumn{1}{|c}{} &   & \multicolumn{1}{c|}{} \\ 
\cline{1-3}   
\end{array} \right. & \left. \begin{array}{ccc} 
\cline{1-3}
\multicolumn{1}{|c}{} &   & \multicolumn{1}{c|}{} \\
\multicolumn{1}{|c}{} & \widehat{{\cal B}}^{+} & 
\multicolumn{1}{c|}{} \\
\multicolumn{1}{|c}{} &   & \multicolumn{1}{c|}{} \\ 
\cline{1-3}   
\end{array} \right. \\
& \\
\left. \begin{array}{ccc} \cline{1-3}
\multicolumn{1}{|c}{} &   & \multicolumn{1}{c|}{} \\
\multicolumn{1}{|c}{} & \widehat{{\cal C}}^{+} & 
\multicolumn{1}{c|}{} \\
\multicolumn{1}{|c}{} &   & \multicolumn{1}{c|}{} \\ 
\cline{1-3}   
\end{array} \right. & \left. \begin{array}{ccc} 
\cline{1-3}
\multicolumn{1}{|c}{} &   & \multicolumn{1}{c|}{} \\
\multicolumn{1}{|c}{} & \widehat{{\cal D}}^{+} & 
\multicolumn{1}{c|}{} \\
\multicolumn{1}{|c}{} &   & \multicolumn{1}{c|}{} \\ 
\cline{1-3}   
\end{array} \right. 
\end{array} \right] 
\left[ \begin{array}{c}
          \alpha_{1} \\
          \alpha_{2} \\
          \vdots \\
          \alpha_{n} \\
          \omega_{1} \\
          \omega_{2} \\
          \vdots \\
          \omega_{n}
      \end{array} \right] = 
\left[ \begin{array}{c}
          g^{+}_{1} \chi^{c}_{12}(\lambda_{1}) \\
          g^{+}_{2} \chi^{c}_{12}(\lambda_{2}) \\
          \vdots \\
          g^{+}_{n} \chi^{c}_{12}(\lambda_{n}) \\
          \overline{g^{+}_{1}} \chi^{c}_{11} 
          (\overline{\lambda_{1}}) \\
          \overline{g^{+}_{2}} \chi^{c}_{11} 
          (\overline{\lambda_{2}}) \\
          \vdots \\
          \overline{g^{+}_{n}} \chi^{c}_{11} 
          (\overline{\lambda_{n}})
       \end{array} \right] \!,& 
\end{eqnarray}
where, for $i,j \! \in \! \{1,2,\ldots,n\}$, the $n \! \times \! 
n$ matrix blocks, $\widehat{{\cal A}}^{+}$, $\widehat{{\cal B}}^{
+}$, $\widehat{{\cal C}}^{+}$, and $\widehat{{\cal D}}^{+}$, are 
defined as follows,
\begin{eqnarray*}
&\widehat{{\cal A}}^{+}_{ij} := 
\left\{ \begin{array}{c}
\frac{ \lambda_{i} + g^{+}_{i} \chi^{c}_{
12}(\lambda_{i}) \chi^{c}_{22}(\lambda_{i}
) + \lambda_{i} g^{+}_{i} {\rm W} (\chi^{
c}_{12} (\lambda_{i}),\chi^{c}_{22} 
(\lambda_{i}))}{\lambda_{i} \chi^{c}_{22}
(\lambda_{i})}, \, \, \, \, \, \, \, i = j, \\
- \frac{2 g^{+}_{i} (- \lambda_{i} \chi^{
c}_{22} (\lambda_{i}) \chi^{c}_{12}
(\lambda_{j}) + \lambda_{j} \chi^{c}_{22}
(\lambda_{j}) \chi^{c}_{12}(\lambda_{i}) 
)}{\chi^{c}_{22}(\lambda_{j})(\lambda_{
i}^{2} - \lambda_{j}^{2})}, \, \, \, \, \, \, 
\, i \not= j,
\end{array} \right.& 
\end{eqnarray*}
\begin{eqnarray*}
&\widehat{{\cal B}}^{+}_{ij} := \left\{ \begin{array}{c}
-\frac{2 g^{+}_{i}(\lambda_{i} \chi^{c}_{22}(\lambda_{i}) 
\chi^{c}_{11}(\overline{\lambda_{i}})-\overline{\lambda_{i}} 
\chi^{c}_{21}(\overline{\lambda_{i}}) \chi^{c}_{12}(\lambda_{
i}))}{\chi^{c}_{11}(\overline{\lambda_{i}})(\lambda_{i}^{2}-
\overline{\lambda_{i}}^{2})}, \, \, \, \, \, \, \, i = j, \\
-\frac{2 g^{+}_{i}(\lambda_{i} \chi^{c}_{22}(\lambda_{i}) 
\chi^{c}_{11}(\overline{\lambda_{j}}) - \overline{\lambda_{j}
} \chi^{c}_{21}(\overline{\lambda_{j}}) \chi^{c}_{12}(
\lambda_{i}))}{\chi^{c}_{11}(\overline{\lambda_{j}})(
\lambda_{i}^{2}-\overline{\lambda_{j}}^{2})}, \, \, \, \, \, 
\, \, i \not= j,
\end{array} \right.&
\end{eqnarray*}
\begin{eqnarray*}
&\widehat{{\cal C}}^{+}_{ij} := \left\{ \begin{array}{c}
-\frac{2 \overline{g^{+}_{i}}(-\overline{\lambda_{i}} \chi^{
c}_{21} (\overline{\lambda_{i}}) \chi^{c}_{12}(\lambda_{i}) 
+ \lambda_{i} \chi^{c}_{22}(\lambda_{i}) \chi^{c}_{11}(
\overline{\lambda_{i}}))}{\chi^{c}_{22}(\lambda_{i})(
\overline{\lambda_{i}}^{2}-\lambda_{i}^{2})}, \, \, \, \, \, 
\, \, i = j, \\
- \frac{2 \overline{g^{+}_{i}}(-\overline{\lambda_{i}} \chi^{
c}_{21} (\overline{\lambda_{i}}) \chi^{c}_{12}(\lambda_{j}) 
+ \lambda_{j} \chi^{c}_{22}(\lambda_{j}) \chi^{c}_{11}(
\overline{\lambda_{i}}))}{\chi^{c}_{22}(\lambda_{j})(
\overline{\lambda_{i}}^{2} - \lambda_{j}^{2})}, \, \, \, \, 
\, \, \, i \not= j,
\end{array} \right.&
\end{eqnarray*}
\begin{eqnarray*}
&\widehat{{\cal D}}^{+}_{ij} := 
\left\{ \begin{array}{c}
\frac{ \overline{\lambda_{i}} - 
\overline{g^{+}_{i}} \chi^{c}_{21}(\overline{
\lambda_{i}}) \chi^{c}_{11}(\overline{
\lambda_{i}}) + \overline{\lambda_{i}} \, 
\overline{g^{+}_{i}} {\rm W} (\chi^{c}_{21}
(\overline{\lambda_{i}}),\chi^{c}_{11}(\overline{
\lambda_{i}}))}{\overline{\lambda_{i}} \chi^{c}_{
11}(\overline{\lambda_{i}})}, \, \, \, \, \, \, \,
i = j, \\
- \frac{2 \overline{g^{+}_{i}} (\overline{
\lambda_{i}} \chi^{c}_{21} (\overline{\lambda_{
i}}) \chi^{c}_{11}(\overline{\lambda_{j}}) - 
\overline{\lambda_{j}} \chi^{c}_{21}
(\overline{\lambda_{j}}) \chi^{c}_{11}
(\overline{\lambda_{i}}))}{\chi^{c}_{11}
(\overline{\lambda_{j}}) (\overline{\lambda_{
i}}^{2} - \overline{\lambda_{j}}^{2})}, \, \, \, 
\, \, \, \, i \not= j,
\end{array} \right. 
\end{eqnarray*}
$\{\beta_{i},\widehat{\delta_{i}}\}_{i=1}^{n}$ satisfy the 
following non-degenerate system of $2n$ linear inhomogeneous 
algebraic equations,
\begin{eqnarray}
&\left[ \begin{array}{cccccc}
\left. \begin{array}{ccc} \cline{1-3}
\multicolumn{1}{|c}{} &   & \multicolumn{1}{c|}{} \\
\multicolumn{1}{|c}{} & \widehat{{\cal E}}^{+} & 
\multicolumn{1}{c|}{} \\
\multicolumn{1}{|c}{} &   & \multicolumn{1}{c|}{} \\ 
\cline{1-3}   
\end{array} \right. & \left. \begin{array}{ccc} 
\cline{1-3}
\multicolumn{1}{|c}{} &   & \multicolumn{1}{c|}{} \\
\multicolumn{1}{|c}{} & \widehat{{\cal F}}^{+} & 
\multicolumn{1}{c|}{} \\
\multicolumn{1}{|c}{} &   & \multicolumn{1}{c|}{} \\ 
\cline{1-3}   
\end{array} \right. \\
& \\
\left. \begin{array}{ccc} \cline{1-3}
\multicolumn{1}{|c}{} &   & \multicolumn{1}{c|}{} \\
\multicolumn{1}{|c}{} & \widehat{{\cal G}}^{+} & 
\multicolumn{1}{c|}{} \\
\multicolumn{1}{|c}{} &   & \multicolumn{1}{c|}{} \\ 
\cline{1-3}   
\end{array} \right. & \left. \begin{array}{ccc} 
\cline{1-3}
\multicolumn{1}{|c}{} &   & \multicolumn{1}{c|}{} \\
\multicolumn{1}{|c}{} & \widehat{{\cal H}}^{+} & 
\multicolumn{1}{c|}{} \\
\multicolumn{1}{|c}{} &   & \multicolumn{1}{c|}{} \\ 
\cline{1-3}   
\end{array} \right. 
\end{array} \right] 
\left[ \begin{array}{c}
          \beta_{1} \\
          \beta_{2} \\
          \vdots \\
          \beta_{n} \\
          \widehat{\delta}_{1} \\
          \widehat{\delta}_{2} \\
          \vdots \\
          \widehat{\delta}_{n}
      \end{array} \right] = 
\left[ \begin{array}{c}
          g^{+}_{1} \chi^{c}_{22}(\lambda_{1}) \\
          g^{+}_{2} \chi^{c}_{22}(\lambda_{2}) \\
          \vdots \\
          g^{+}_{n} \chi^{c}_{22}(\lambda_{n}) \\
          \overline{g^{+}_{1}} \chi^{c}_{21} 
          (\overline{\lambda_{1}}) \\
          \overline{g^{+}_{2}} \chi^{c}_{21} 
          (\overline{\lambda_{2}}) \\
          \vdots \\
          \overline{g^{+}_{n}} \chi^{c}_{21} 
          (\overline{\lambda_{n}})
       \end{array} \right] \!,& 
\end{eqnarray}
where, for $i,j \! \in \! \{1,2,\ldots,n\}$, the $n \! \times 
\! n$ matrix blocks, $\widehat{{\cal E}}^{+}$, $\widehat{{\cal 
F}}^{+}$, $\widehat{{\cal G}}^{+}$, and $\widehat{{\cal H}}^{
+}$, are defined as follows,
\begin{eqnarray*}
&\widehat{{\cal E}}^{+}_{ij} := 
\left\{\begin{array}{c}
\frac{ \lambda_{i} - g^{+}_{i} 
\chi^{c}_{12}(\lambda_{i}) \chi^{c}_{22}(\lambda_{
i}) + \lambda_{i} g^{+}_{i} {\rm W} (\chi^{c}_{12} 
(\lambda_{i}),\chi^{c}_{22} (\lambda_{i}))}{
\lambda_{i} \chi^{c}_{22}(\lambda_{i})}, \, \, \, \, 
\, \, \, i = j, \\
\frac{2 g^{+}_{i} (\lambda_{j} \chi^{c}_{
12} (\lambda_{j}) \chi^{c}_{22}(\lambda_{
i}) - \lambda_{i} \chi^{c}_{12}(\lambda_{
i}) \chi^{c}_{22}(\lambda_{j}))}{\chi^{c}_{22}
(\lambda_{j}) (\lambda_{i}^{2} - \lambda_{j}^{2})}, 
\, \, \, \, \, \, \, i \not= j,
\end{array} \right.&
\end{eqnarray*}
\begin{eqnarray*}
&\widehat{{\cal F}}^{+}_{ij} := 
\left\{\begin{array}{c}
\frac{2 g^{+}_{i}(\lambda_{i} \chi^{c}_{12}(\lambda_{i}) 
\chi^{c}_{21}(\overline{\lambda_{i}})-\overline{\lambda_{
i}} \chi^{c}_{11}(\overline{\lambda_{i}}) \chi^{c}_{22}
(\lambda_{i}))}{\chi^{c}_{11}(\overline{\lambda_{i}})
(\lambda_{i}^{2}-\overline{\lambda_{i}}^{2})}, \, \, \, 
\, \, \, \, i = j, \\
\frac{2 g^{+}_{i}(\lambda_{i} \chi^{c}_{12}(\lambda_{i}) 
\chi^{c}_{21}(\overline{\lambda_{j}})-\overline{\lambda_{
j}} \chi^{c}_{11}(\overline{\lambda_{j}}) \chi^{c}_{22}(
\lambda_{i}))}{\chi^{c}_{11}(\overline{\lambda_{j}})(
\lambda_{i}^{2}-\overline{\lambda_{j}}^{2})}, \, \, \, \, 
\, \, \, i \not= j,
\end{array} \right.&
\end{eqnarray*}
\begin{eqnarray*}
&\widehat{{\cal G}}^{+}_{ij} := \left\{\begin{array}{c}
\frac{2 \overline{g^{+}_{i}}(\lambda_{i} \chi^{c}_{12}(
\lambda_{i}) \chi^{c}_{21}(\overline{\lambda_{i}})-\overline{
\lambda_{i}} \chi^{c}_{11}(\overline{\lambda_{i}}) \chi^{c}_{
22}(\lambda_{i}))}{\chi^{c}_{22}(\lambda_{i})(\overline{
\lambda_{i}}^{2}-\lambda_{i}^{2})}, \, \, \, \, \, \, \, 
i=j, \\
\frac{2 \overline{g^{+}_{i}}(\lambda_{j} \chi^{c}_{12}(
\lambda_{j}) \chi^{c}_{21}(\overline{\lambda_{i}})-\overline{
\lambda_{i}} \chi^{c}_{11}(\overline{\lambda_{i}}) \chi^{c}_{
22}(\lambda_{j}))}{\chi^{c}_{22}(\lambda_{j})(\overline{
\lambda_{i}}^{2}-\lambda_{j}^{2})}, \, \, \, \, \, \, \, 
i \not= j,
\end{array} \right.&
\end{eqnarray*}
\begin{eqnarray*}
&\widehat{{\cal H}}^{+}_{ij} := 
\left\{ \begin{array}{c}
\frac{\overline{\lambda_{i}} + \overline{g^{+}_{i}} 
\chi^{c}_{11} (\overline{\lambda_{i}}) 
\chi^{c}_{21} (\overline{\lambda_{i}}) 
- \overline{\lambda_{i}} \, \overline{g^{+}_{i}} 
{\rm W}(\chi^{c}_{11}(\overline{\lambda_{i}}),\chi^{
c}_{21}(\overline{\lambda_{i}}))}{\overline{\lambda_{
i}} \chi^{c}_{11}(\overline{\lambda_{i}})}, \, \, \, 
\, \, \, \, i = j, \\
\frac{2 \overline{g^{+}_{i}}(\overline{\lambda_{i}} 
\chi^{c}_{11}(\overline{\lambda_{i}}) 
\chi^{c}_{21}(\overline{\lambda_{j}}) - 
\overline{\lambda_{j}} \chi^{c}_{11}
(\overline{\lambda_{j}}) \chi^{c}_{21}
(\overline{\lambda_{i}}))}{\chi^{c}_{11}
(\overline{\lambda_{j}})(\overline{\lambda_{i}}^{2} 
- \overline{\lambda_{j}}^{2})}, \, \, \, \, \, \, 
\, i \not= j,
\end{array} \right.&
\end{eqnarray*}
with
\begin{eqnarray*}
&g^{+}_{j} := C_{j} e^{ 2 i \lambda_{j}^{2} x + 
4 i \lambda_{j}^{4} t} (\delta^{+}(\lambda_{j};\! 
\lambda_{0}))^{-2} \! \prod\limits_{l=n+1}^{N} \! 
\left(\! \frac{(\lambda_{j} - \lambda_{l})(\lambda_{
j} + \lambda_{l})}{(\lambda_{j} - \overline{\lambda_{
l}})(\lambda_{j} + \overline{\lambda_{l}})} \! 
\right)^{2}, \, \, \, \, \, \, \, \, \, 1 \leq j \leq n,&
\end{eqnarray*}
$\delta^{+}(\lambda_{k};\! \lambda_{0})$, $k \! \in \! \{1,2,
\ldots,n\}$, given in Theorem~2.1, Eq.~(18), and ${\rm W}(
\chi^{c}_{ij}(z),\chi^{c}_{i^{\prime} j^{\prime}}(z))$ is the 
Wronskian of $\chi^{c}_{ij}(\lambda)$ and $\chi^{c}_{i^{\prime} 
j^{\prime}}(\lambda)$ evaluated at $z$ $(i,j,i^{\prime},j^{
\prime} \! \in \! \{1,2\})$: ${\rm W}(\chi^{c}_{ij}(z),\chi^{c}
_{i^{\prime} j^{\prime}}(z)) \! := \! (\chi^{c}_{ij}(\lambda) 
\partial_{\lambda} \chi^{c}_{i^{\prime} j^{\prime}}(\lambda) 
\! - \! \chi^{c}_{i^{\prime} j^{\prime}}(\lambda) \partial_{
\lambda} \chi^{c}_{ij}(\lambda)) \vert_{\lambda=z}$.
\end{bbb}

{\em Proof.\/} For $1 \! \leq \! i \! \leq \! n$, set
\begin{eqnarray}
&{\rm res}(\chi(\lambda);\! \lambda_{i}) =  
\left(\! \begin{array}{cc}
\alpha_{i} & a_{i} \\
\beta_{i} & b_{i} 
\end{array} \! \right) \!, \, \, \, \, \, \, \, \, \, \, 
{\rm res}(\chi(\lambda);\! \overline{\lambda_{i}}) 
= \left(\! \begin{array}{cc}
c_{i} & \omega_{i} \\
d_{i} & \widehat{\delta}_{i} 
\end{array} \! \right) \!.& 
\end{eqnarray}
{}From Eqs.~(92), (93), and (97), and the polar (residue) 
conditions in Lemma~3.3, one gets a system of linear 
algebraic equations for $\{\alpha_{i},\beta_{i},a_{i},
b_{i},c_{i},d_{i},\omega_{i},\widehat{\delta}_{i}\}_{
i=1}^{n}$: {}from this system, one shows that, for $1 
\! \leq \! i \! \leq \! n$,
\begin{eqnarray}
\left. \begin{array}{l}
(\chi^{c}_{12}(\lambda_{i}) \alpha_{i} + 
\chi^{c}_{22}(\lambda_{i}) a_{i}) 
g^{+}_{i} = 0 \, \Rightarrow \, a_{i} = - \frac{ 
\chi^{c}_{12}(\lambda_{i})}{\chi^{c}_{22}(\lambda_{
i})} \alpha_{i}, \\
(\chi^{c}_{12}(\lambda_{i}) \beta_{i} + \chi^{c}_{
22}(\lambda_{i}) b_{i}) g^{+}_{i} = 0 \, \Rightarrow 
\, b_{i} = - \frac{\chi^{c}_{12}(\lambda_{i})}{
\chi^{c}_{22}(\lambda_{i})} \beta_{i}, \\ 
(\chi^{c}_{11}(\overline{\lambda_{i}}) 
c_{i} + \chi^{c}_{21}(\overline{\lambda_{
i}}) \omega_{i}) \overline{g^{+}_{i}} = 0 \, 
\Rightarrow \, c_{i} = - \frac{\chi^{c}_{
21}(\overline{\lambda_{i}})}{\chi^{c}_{11}
(\overline{\lambda_{i}})} \omega_{i}, \\ 
(\chi^{c}_{11}(\overline{\lambda_{i}}) 
d_{i} + \chi^{c}_{21}(\overline{\lambda_{
i}}) \widehat{\delta}_{i}) \overline{g^{+}_{i}} 
= 0 \, \Rightarrow \, d_{i} = - \frac{\chi^{c}_{
21}(\overline{\lambda_{i}})}{\chi^{c}_{11}
(\overline{\lambda_{i}})} \widehat{\delta}_{i}; 
\end{array} \right\} 
\end{eqnarray}
using Eqs.~(98), which show that the matrices $\{{\rm res}
(\chi(\lambda);\! \lambda_{i})\}_{i=1}^{n}$ and $\{{\rm 
res}(\chi(\lambda);\! \overline{\lambda_{i}})\}_{i=1}^{n}$, 
respectively, are degenerate, one simplifies the resulting 
system of linear algebraic equations for $\{\alpha_{i},
\beta_{i},a_{i},b_{i},c_{i},d_{i},\omega_{i},\widehat{
\delta}_{i}\}_{i=1}^{n}$ and obtains Eqs.~(95) and (96): the 
non-degeneracy of systems~(95) and (96) is a consequence of 
the unique solvability of the original RH problem (Lemma~2.2).
Substituting Eqs.~(97) into Eq.~(92) and defining $Q^{\chi}
(x,t)$ as in the Proposition, one obtains, {}from Eq.~(93), 
Lemma~4.1, and Eqs.~(98), Eq.~(94). \hfill \rule{6.5pt}{6.5pt}
\begin{fff}
As $t \! \rightarrow \! + \infty$ and $x \! \rightarrow \! 
-\infty$ such that $\lambda_{0} \! > \! M$ and $(x,t) \! 
\in \! \Omega_{n}$,
\begin{eqnarray*}
&Q(x,t) = \underline{Q}_{+}^{{\cal C}}(x,t) + 4 i 
\sum\limits_{j=1}^{n} \! \left(\omega_{j} - 
\frac{\chi^{c}_{12}(\lambda_{j})}{\chi^{c}_{
22}(\lambda_{j})} \alpha_{j} \right) + {\cal 
O} \! \left(C(\lambda_{0}) \exp \{- a b t\} 
\right),&
\end{eqnarray*}
where $\underline{Q}_{+}^{{\cal C}}(x,t) \! := \! i Q^{\chi}
(x,t)$, $a$ and $b$ are given in Lemma~3.3, and $C(\lambda_{
0}) \! \in \! {\cal L}^{\infty}(\Bbb R_{>M};\! \Bbb C)$.
\end{fff}

{\em Proof.\/} Since, {}from Lemma~3.2, $Q(x,t) \! = \! 2 i \! 
\lim\limits_{\lambda \rightarrow \infty} \! (\lambda m^{\sharp}
(x,t;\! \lambda))_{12}$, the result follows {}from Lemma~3.3 
and Proposition~4.2. \hfill \rule{6.5pt}{6.5pt}
\begin{bbb}
As $t \! \rightarrow \! + \infty$ and $x \! \rightarrow 
\! - \infty$ such that $\lambda_{0} \! > \! M$ and $(x,
t) \! \in \! \Omega_{n}$,
\begin{eqnarray}
&Q(x,t) = Q_{+}^{{\cal C}}(x,t) + 4 i \! \left(\omega_{n}- 
\frac{\chi^{c}_{12}(\lambda_{n})}{\chi^{c}_{22}(\lambda_{
n})} \alpha_{n} \right) + {\cal O} \! \left(\frac{C_{1}(
\lambda_{0}) \ln t}{t} \right) + {\cal O}(C_{2}(\lambda_{0}) 
e^{-a_{0}b_{0}t}),&
\end{eqnarray}
where
\begin{eqnarray}
&\alpha_{n} = \frac{\widehat{a}_{12} \overline{g^{+}_{n}} 
\chi^{c}_{11}(\overline{\lambda_{n}}) + \widehat{a}_{22} g^{
+}_{n} \chi^{c}_{12}(\lambda_{n})}{(\widehat{a}_{11} 
\widehat{a}_{22} - \widehat{a}_{12} \widehat{a}_{21})},& \\
&\omega_{n} = \frac{\widehat{a}_{11} \overline{g^{+}_{n}} 
\chi^{c}_{11}(\overline{\lambda_{n}}) + \widehat{a}_{21} g^{
+}_{n} \chi^{c}_{12}(\lambda_{n})}{(\widehat{a}_{11} 
\widehat{a}_{22} - \widehat{a}_{12} \widehat{a}_{21})},& \\
&\widehat{a}_{11} := \frac{\lambda_{n} + g^{+}_{
n} \chi^{c}_{12}(\lambda_{n}) \chi^{c}_{22}(\lambda_{
n}) + \lambda_{n} g^{+}_{n} {\rm W}(\chi^{c}_{12}
(\lambda_{n}),\chi^{c}_{22}(\lambda_{n}))}{\lambda_{n} 
\chi^{c}_{22}(\lambda_{n})},& \\
&\widehat{a}_{12} := \frac{2 g^{+}_{n} (\lambda_{
n} \chi^{c}_{22}(\lambda_{n}) \chi^{c}_{11}(\overline{
\lambda_{n}}) - \overline{\lambda_{n}} \chi^{c}_{21}
(\overline{\lambda_{n}}) \chi^{c}_{12}(\lambda_{n}))
}{\chi^{c}_{11}(\overline{\lambda_{n}})(\lambda_{n}^{
2} - \overline{\lambda_{n}}^{2})},& \\
&\widehat{a}_{21} := \frac{2 \overline{g^{+}_{n}} 
(\lambda_{n} \chi^{c}_{22}(\lambda_{n}) \chi^{c}_{11} 
(\overline{\lambda_{n}}) - \overline{\lambda_{n}} 
\chi^{c}_{21}(\overline{\lambda_{n}}) \chi^{c}_{12}
(\lambda_{n}))}{\chi^{c}_{22}(\lambda_{n})(\overline{
\lambda_{n}}^{2} - \lambda_{n}^{2})},& \\
&\widehat{a}_{22} := \frac{\overline{\lambda_{n}} 
- \overline{g^{+}_{n}} \chi^{c}_{21}(\overline{\lambda_{
n}}) \chi^{c}_{11}(\overline{\lambda_{n}}) + \overline{
\lambda_{n}} \, \overline{g^{+}_{n}} {\rm W}(\chi^{c}_{
21}(\overline{\lambda_{n}}),\chi^{c}_{11}(\overline{
\lambda_{n}}))}{\overline{\lambda_{n}} \chi^{c}_{11}
(\overline{\lambda_{n}})},&
\end{eqnarray}
$Q_{+}^{{\cal C}}(x,t)$ is given in Theorem~2.1, Eqs.~(20)--(22) 
and (24), $a_{0} \! := \! \min (a,8 \min \{\eta_{l}\}_{l=1}
^{n-1})$ $(> 0)$, $b_{0} \! := \! \min (b,\min \{\vert \xi_{
n} \! - \! \xi_{l} \vert\}_{l=1}^{n-1})$, $C_{1}(\lambda_{0}) \! 
\in \! {\cal S}(\Bbb R_{> M};\! \Bbb C)$, and $C_{2}(\lambda_{0}
) \! \in \! {\cal L}^{\infty}(\Bbb R_{> M};\! \Bbb C)$.
\end{bbb}

{\em Proof.\/} Solving Eqs.~(95) as $t \! \to \! +\infty$ and 
$x \! \to \! -\infty$ such that $\lambda_{0} \! > \! M$ and 
$(x,t) \! \in \! \Omega_{n}$ for $\{\alpha_{i}\}_{i=1}^{n}$ and 
$\{\omega_{i}\}_{i=1}^{n}$ via Cramer's rule, one shows that
\begin{eqnarray}
&\alpha_{i}, \, \, \, \omega_{i} \sim {\cal O} (\exp \{- a^{
\flat} \min\limits_{1 \leq l \leq n-1} \vert \xi_{n} - \xi_{l} 
\vert t\}), \, \, \, \, \, \, \, \, \, 1 \leq i \leq n-1,&
\end{eqnarray}
where $a^{\flat} \! := \! 8 \min \{\eta_{l}\}_{l=1}^{
n-1}$ $(> 0)$, and $\alpha_{n}$ and $\omega_{n}$ are given 
by Eqs.~(100) and (101): the result now follows {}from 
Corollary~4.1 and the estimates in Eq.~(106). \hfill 
\rule{6.5pt}{6.5pt}
\begin{bbb}
As $t \! \rightarrow \! + \infty$ and $x \! \rightarrow 
\! - \infty$ such that $\lambda_{0} \! > \! M$ and $(x,
t) \! \in \! \Omega_{n}$,
\begin{eqnarray*}
&Q(x,t) = Q^{+}_{{\rm as}} (x,t) + {\cal O} \! \left(
\frac{C(\lambda_{0}) \ln t}{t} \right) \!,&
\end{eqnarray*}
where $Q^{+}_{{\rm as}}(x,t)$ is given in Theorem~2.1,
Eqs.~(14)--(29), and $C(\lambda_{0}) \! \in \! {\cal 
S}(\Bbb R_{> M};\! \Bbb C)$.
\end{bbb}

{\em Proof.\/} Substitute $\chi^{c}_{ij}(\cdot)$, $i,j \! 
\in \! \{1,2\}$, {}from Lemma~4.1 into Eqs.~(99)--(105) and 
neglect exponentially small terms. \hfill \rule{6.5pt}{6.5pt}
\section{Asymptotic Evaluation of $((\Psi^{-1}(x,t;
0))_{11})^{2}$}
In this section, the phase integral, $((\Psi^{-1}(x,t;0))
_{11})^{2}$, which appears in the gauge transformation 
(Proposition~2.3, Eq.~(9)) is evaluated asymptotically as 
$t \! \to \! +\infty$ $(x/t \! \sim \! {\cal O}(1))$.
\begin{ccc}
As $t \! \rightarrow \! + \infty$ and $x \! \rightarrow 
\! - \infty$ such that $\lambda_{0} \! > \! M$ and $(x,
t) \! \in \! \Omega_{n}$,
\begin{eqnarray*}
&((\Psi^{-1}(x,t;0))_{11})^{2} = \exp\{2 \ln(\chi^{c}_{22}
(0))\} \exp\{\frac{2 i}{\pi} (\int_{0}^{\lambda_{0}} \frac{
\ln (1 - \vert r(\varrho) \vert^{2})}{\varrho} d \varrho - 
\int_{0}^{\infty} \frac{\ln(1 + \vert r(i\varrho) \vert^{2}
)}{\varrho} d \varrho)\}& \\
&\, \, \, \, \, \, \, \, \, \, \, \, \, \, \, \, \, \, \, 
\, \, \, \, \, \, \, \, \, \, \, \, \, \, \, \, \, \, \, 
\, \, \, \, \, \, \, \times \, \exp\{-4i \! 
\sum\limits_{l=n+1}^{N} \! \gamma_{l}\} \exp\{2 \ln (1-
\sum\limits_{i=1}^{n}(\frac{2 b_{i}}{\lambda_{i}}+\frac{2 
\widehat{\delta}_{i}}{\overline{\lambda_{i}}}))\} + {\cal 
O}(C(\lambda_{0}) e^{-abt}),&
\end{eqnarray*}
where $b_{i} \! = \! - \frac{\chi^{c}_{12}(\lambda_{i})}{
\chi^{c}_{22}(\lambda_{i})} \beta_{i}$, $1 \! \leq \! i \! 
\leq \! n$, $\{\beta_{i},\widehat{\delta}_{i}\}_{i=1}^{n}$ 
satisfy Eqs.~(96), $a$ and $b$ are given in Lemma~3.3, and 
$C(\lambda_{0}) \! \in \! {\cal L}^{\infty}(\Bbb R_{> M};\! 
\Bbb C)$.
\end{ccc}

{\em Proof.\/} {}From Lemma~2.1, the proof of Lemma~3.1, 
Proposition~3.2 (Eqs.~(85)), Lemma 3.2 (Eqs.~(86)), and 
Lemma~3.3, one gets that
\begin{eqnarray*}
&\Psi(x,t;0) = \chi(0) (\delta(0))^{\sigma_{3}} \! 
\prod\limits_{l=n+1}^{N} \! (d_{l_{+}} (0))^{\sigma_{
3}} + {\cal O}(C_{1}(\lambda_{0}) \exp\{-abt\}),&
\end{eqnarray*}
where $C_{1}(\lambda_{0}) \! \in \! {\cal L}^{\infty}(\Bbb 
R_{>M};\! M_{2}(\Bbb C))$. {}From Propositions~2.1--2.3, the 
parametrization for the discrete eigenvalues (Sec.~2), 
Lemma~3.2 (Eqs.~(87)), Proposition~4.1 (Eqs.~(92) and (93)), 
the proof of Proposition~4.2 (Eqs.~(97) and (98)), and the 
$\sigma_{1}$ and $\sigma_{3}$ symmetry reductions for 
$\chi(\lambda)$, one shows that,
\begin{eqnarray*}
&\Psi^{-1}(x,t;0) \! = \! (\widehat{h}(0))^{\sigma_{3}} 
\left( \begin{array}{cc}
\! \! 1 - \sum\limits_{i=1}^{n}(\frac{2 b_{i}}{
\lambda_{i}} + \frac{2 \widehat{\delta}_{i}}{
\overline{\lambda_{i}}}) & \! \! 0 \\
\! \! 0 & \! \! 1 - \sum\limits_{i=1}^{n} (\frac{2 
\alpha_{i}}{\lambda_{i}} + \frac{2 c_{i}}{\overline{
\lambda_{i}}})
\end{array} \! \! \right) \! + {\cal O}(C_{2}(\lambda_{0})
e^{-abt}),&
\end{eqnarray*}
where $\widehat{h}(0) \! := \! \chi^{c}_{22}(0)
(\delta(0))^{-1} \exp \{- 2 i \! \sum_{l=n+1}^{N} \! 
\gamma_{l}\}$, $b_{i}$, $1 \! \leq \! i \! \leq \! n$, 
and $\{\beta_{i},\widehat{\delta}_{i}\}_{i=1}^{n}$ are 
as given in the Lemma, $c_{i} \! = \! - \frac{\chi^{
c}_{21}(\overline{\lambda_{i}})}{\chi^{c}_{11}
(\overline{\lambda_{i}})} \omega_{i}$, $1 \! \leq \! 
i \! \leq \! n$, $\{\alpha_{i},\omega_{i}\}_{i=1}^{n}$ 
are defined by system~(95), and $C_{2}(\lambda_{0}) \! 
\in \! {\cal L}^{\infty}(\Bbb R_{>M};\! M_{2}(\Bbb C))$: 
using the expression for $\delta^{+}(\lambda;\! \lambda
_{0})$ given in Proposition~3.1 (and Remark~3.1), one 
obtains the result stated in the Lemma. \hfill 
\rule{6.5pt}{6.5pt}

In order to estimate $(\chi^{c}_{22}(0))^{2}$, the following 
proposition and lemma are necessary:
\begin{bbb}
Define $Q^{\sharp}(x,t) \! := \! 2 i \! \lim\limits_{
\lambda \to \infty}(\lambda \chi^{c}(x,t;\! \lambda))_{12}$.
Then
\begin{eqnarray*}
&(\vert \vert Q^{\sharp}(\cdot,t) \vert \vert_{{\cal L}^{
2}(\Bbb R;\Bbb C)})^{2} = \frac{2}{\pi} (\int_{0}^{\infty} 
\frac{\ln (1 + \vert r(i \varrho) \vert^{2})}{\varrho} d 
\varrho - \int_{0}^{\infty} \frac{\ln (1 - \vert r(\varrho) 
\vert^{2})}{\varrho} d \varrho),& \\
&(\chi^{c}_{22}(0))^{2} = (\delta^{+}(0;\! \lambda_{0}))^{2} 
\exp \{i \int_{+\infty}^{x} \vert Q^{\sharp}(\varrho,t) 
\vert^{2} d \varrho\}.&
\end{eqnarray*}
\end{bbb}

{\em Proof.\/} Follows {}from the definition of $\chi^{c}
(\lambda)$ given in Proposition~4.1, Proposition~2.2, and 
Proposition~8.1 in \cite{a7}. \hfill \rule{6.5pt}{6.5pt}
\begin{ccc}[{\rm \cite{a22}}]
As $t \! \to \! +\infty$ and $x \! \to \! -\infty$ such that 
$\lambda_{0} \! > \! M$,
\begin{eqnarray*}
&Q^{\sharp}(x,t) = \frac{u_{1,1,0}^{+}(\lambda_{0})
e^{i(4 \lambda_{0}^{4} t - \nu(\lambda_{0}) \ln \! t)}}{
\sqrt{t}} + \frac{u_{-1,2,0}^{+}(\lambda_{0})}{t} + {\cal 
O} \! \left(\frac{C(\lambda_{0})(\ln t)^{2}}{t^{3/2}} 
\right) \!,&
\end{eqnarray*}
where 
\begin{eqnarray*}
&u_{1,1,0}^{+}(\lambda_{0}) = \sqrt{\frac{\nu(\lambda_{
0})}{2 \lambda_{0}^{2}}} \exp \{i \theta^{+}(\lambda_{0})
\},& \\
&\theta^{+}(\lambda_{0}) \! = \! \phi^{+}(\lambda_{0}) \! -
\! \frac{3\pi}{4} \! + \! \arg \Gamma(i \nu(\lambda_{0})) \! 
+ \arg r(\lambda_{0}) \! - \! 3 \nu(\lambda_{0}) \ln 2 \! 
+ \! 2 \! \sum\limits_{l=n+1}^{N} \! \! \! \arg \! \left(\! 
\frac{(\lambda_{0} - \overline{\lambda_{l}})(\lambda_{0} + 
\overline{\lambda_{l}})}{(\lambda_{0} - \lambda_{l})
(\lambda_{0} + \lambda_{l})} \! \right) \!,& \\
&u_{-1,2,0}^{+}(\lambda_{0}) \! = \! -\frac{i}{8 \pi \lambda
_{0}^{2}} \! \left(\! \left. \frac{d(r(\varrho) \vert_{
\varrho \in \Bbb R})}{d\varrho} \right\vert_{\varrho=0} \! 
- \! \left. \frac{d(r(\varrho) \vert_{\varrho \in i \Bbb R})
}{d\varrho} \right\vert_{\varrho=0} \right) \! \exp \! 
\left\{\! i \! \left(4 \sum\limits_{l=n+1}^{N} \gamma_{l} + 
2 \vartheta^{+}(\lambda_{0}) \right) \! \right\} \!,& \\
&\vartheta^{+}(\lambda_{0})=-\int_{0}^{\lambda_{0}} \frac{
\ln(1 - \vert r(\varrho) \vert^{2})}{\varrho} \frac{d
\varrho}{\pi} + \int_{0}^{\infty} \frac{\ln(1+\vert r(i 
\varrho) \vert^{2})}{\varrho} \frac{d\varrho}{\pi},&
\end{eqnarray*}
$\phi^{+}(\cdot)$ is given in Theorem~2.1, Eq.~(22), and 
$C(\lambda_{0}) \! \in \! {\cal S}(\Bbb R_{>M};\! \Bbb 
C)$.
\end{ccc}

{\em Comment to Proof.\/} Up to the leading $({\cal O}(t^{
-1}))$ term, the asymptotic expansion was proved in \cite{a7}. 
The ${\cal O}(t^{-1})$ term constitutes the leading-order 
contribution {}from the first-order stationary phase point at 
$\lambda \! = \! 0$: the complete proof of this asymptotic 
expansion can be found in \cite{a22}.
\hfill \rule{6.5pt}{6.5pt}
\begin{bbb}
As $t \! \to \! +\infty$ and $x \! \to \! -\infty$ such that 
$\lambda_{0} \! > \! M$,
\begin{eqnarray*}
(\chi^{c}_{22}(0))^{2} & = & \exp \! \left\{\! i \! \left(
\! \sqrt{\frac{2}{t}} \int_{\lambda_{0}}^{\infty} \! \frac{
\sqrt{\nu(\mu)}}{\mu^{2}} \! \left(R^{+}_{i}(0) \cos(
\kappa^{+}(\mu;t)) \! - \! R^{+}_{r}(0) \sin(\kappa^{+}(\mu;
t)) \right) \! \frac{d\mu}{\pi} \right) \! \right\} \\
 & + & {\cal O} \! \left(\frac{C(\lambda_{0})(\ln t)^{
2}}{\lambda_{0}^{2} t} \right) \!,
\end{eqnarray*}
where $R^{+}_{i}(0) \! = \! \Im \{R^{+}(0)\}$, $R^{+}_{r}
(0) \! = \! \Re \{R^{+}(0)\}$, $R^{+}(0) \! := \! \left(
\left. \frac{d(r(\varrho) \vert_{\varrho \in \Bbb R})}{
d\varrho} \right\vert_{\varrho=0} \! - \! \left. \frac{d(r(
\varrho) \vert_{\varrho \in i \Bbb R})}{d\varrho} \right\vert
_{\varrho=0} \right) \linebreak[4] \cdot \exp \{4i\sum_{l=n+1
}^{N} \gamma_{l}\}$, $\kappa^{+}(\lambda_{0};t) \! := \! 
4 \lambda_{0}^{4} t \! - \! \nu(\lambda_{0}) \ln t \! + \! 
\theta^{+}(\lambda_{0}) \! - \! 2 \vartheta^{+}(\lambda_{0})$, 
and $C(\lambda_{0}) \! \in \! {\cal S}(\Bbb R_{>M};\! \Bbb C)$.
\end{bbb}

{\em Proof.\/} Writing $\int_{+\infty}^{x} \! \vert Q^{\sharp}
(\varrho,t) \vert^{2} d \varrho \! = \! - (\vert \vert Q^{
\sharp}(\cdot,t) \vert \vert_{{\cal L}^{2}(\Bbb R;\Bbb C)}
)^{2} \! + \! \int_{-\infty}^{x} \! \vert Q^{\sharp}(\varrho,
t) \vert^{2} d \varrho$, using the expressions for $(\vert 
\vert Q^{\sharp}(\cdot,t) \vert \vert_{{\cal L}^{2}(\Bbb R;
\Bbb C)})^{2}$ and $(\chi^{c}_{22}(0))^{2}$ given in 
Proposition~5.1, the asymptotic expansion for $Q^{\sharp}(x,
t)$ given in Lemma~5.2, the following inequalities, $\vert 
\! \exp \{(\cdot)\} \! - \! 1 \vert \! \leq \! \vert (\cdot) 
\vert \sup_{s \in [0,1]} \vert \! \exp \{s(\cdot)\} 
\vert$ and $0 \! < \! \nu(\lambda_{0}) \! \leq \! \nu_{{\rm 
max}} \! \equiv \! - \frac{1}{2 \pi} \ln(1 \! - \! \sup_{
\lambda \in \Bbb R} \vert r(\lambda) \vert^{2}) \! < \! 
\infty$, and the fact that $r(\lambda) \! \in \! {\cal S}
(\widehat{\Gamma};\! \Bbb C)$, one obtains the result stated 
in the Proposition. \hfill \rule{6.5pt}{6.5pt}
\begin{ccc}
As $t \! \to \! +\infty$ and $x \! \to \! -\infty$ such that 
$\lambda_{0} \! > \! M$ and $(x,t) \! \in \! \Omega_{n}$,
\begin{eqnarray*}
&((\Psi^{-1}(x,t;0))_{11})^{2} = \exp \{i \arg q_{{\rm as}
}^{+}(x,t)\} + {\cal O} \! \left(\frac{C(\lambda_{0})(\ln 
t)^{2}}{t} \right) \!,&
\end{eqnarray*}
where $\arg q_{{\rm as}}^{+}(x,t)$ is given in Theorem~2.2, 
Eqs.~(48)--(51), and $C(\lambda_{0}) \! \in \! {\cal S}(\Bbb 
R_{> M};\! \Bbb C)$.
\end{ccc}

{\em Proof.\/} According to Lemma~5.1, in order to evaluate 
$((\Psi^{-1}(x,t;0))_{11})^{2}$, estimates for $\exp\{2\ln(
\chi^{c}_{22}(0))\}$ and $\{b_{i},\widehat{\delta}_{i}\}_{i
=1}^{n}$ are required: the estimation for $\exp\{2\ln(\chi^{
c}_{22}(0))\}$ is given in Proposition~5.2; hence, it remains 
to estimate $\{b_{i},\widehat{\delta}_{i}\}_{i=1}^{n}$. Solving 
system~(96) as $t \! \to \! +\infty$ and $x \! \to \! -\infty$ 
such that $\lambda_{0} \! > \! M$ and $(x,t) \! \in \! \Omega_{
n}$ for $\{\beta_{i}\}_{i=1}^{n}$ and  $\{\widehat{\delta}_{i}
\}_{i=1}^{n}$ via Cramer's rule, one shows that, $\beta_{i}$, 
$\widehat{\delta}_{i} \! \sim \! {\cal O}(\exp \{-a^{\flat} \! 
\min\limits_{1 \leq i \leq n-1} \! \vert \xi_{n} \! - \! \xi_{
i} \vert t\})$, $1 \! \leq \! i \! \leq \! n \! - \! 1$, and
\begin{eqnarray*}
&\beta_{n}=\frac{\beta_{n}^{+}}{(\widehat{{\cal E}}_{nn}^{+} 
\widehat{{\cal H}}_{nn}^{+}-\widehat{{\cal F}}_{nn}^{+} 
\widehat{{\cal G}}_{nn}^{+})}, \, \, \, \, \, \, \, \, \, 
\, \, \, \widehat{\delta}_{n}=\frac{\widehat{\delta}_{n}^{+}}
{(\widehat{{\cal E}}_{nn}^{+} \widehat{{\cal H}}_{nn}^{+} - 
\widehat{{\cal F}}_{nn}^{+} \widehat{{\cal G}}_{nn}^{+})},&
\end{eqnarray*}
where
\begin{eqnarray*}
&\beta_{n}^{+} \! := \! \frac{g_{n}^{+} \chi^{c}_{22}
(\lambda_{n})}{\chi^{c}_{11}(\overline{\lambda_{n}})} \! + \! 
\frac{\vert g_{n}^{+} \vert^{2} \chi^{c}_{21}(\overline{
\lambda_{n}}) \chi^{c}_{22}(\lambda_{n})}{\overline{\lambda
_{n}}} \! - \! \frac{\vert g_{n}^{+} \vert^{2} \chi^{c}_{22}
(\lambda_{n}) W(\chi^{c}_{11}(\overline{\lambda_{n}}),\chi^{
c}_{21}(\overline{\lambda_{n}}))}{\chi^{c}_{11}(\overline{
\lambda_{n}})}& \\
&\! \! \! \! \! \! \! \! \! \! \! \! \! \! \! \! \! \! \! 
\! \! \! \! \! \! \! \! \! \! \! \! \! \! \! \! \! \! \! 
\! \! \! \! \! \! \! \! \! \! \! \! \! \! \! \! \! \! \! 
\! \! \! \! \! \! \! \! \! \! \! \! \! \! \! \! \! \! \! 
\! \! \! \! \! \! \! \! \! \! + \, 
\frac{2 \overline{\lambda_{n}} \vert g_{n}^{+} \vert^{2} 
\chi^{c}_{22}(\lambda_{n}) \chi^{c}_{21}(\overline{\lambda_{
n}})}{(\lambda_{n}^{2} - \overline{\lambda_{n}}^{2})},& \\
&\widehat{{\cal E}}_{nn}^{+} \widehat{{\cal H}}_{nn}^{+} - 
\widehat{{\cal F}}_{nn}^{+} \widehat{{\cal G}}_{nn}^{+} 
:= \frac{1}{\chi^{c}_{22}(\lambda_{n}) \chi^{c}_{11}
(\overline{\lambda_{n}})} + \frac{\overline{g_{n}^{+}} 
W(\chi^{c}_{21}(\overline{\lambda_{n}}),\chi^{c}_{11}(
\overline{\lambda_{n}}))}{\chi^{c}_{22}(\lambda_{n}) \chi^{
c}_{11}(\overline{\lambda_{n}})}+\frac{g_{n}^{+} W(\chi^{c}
_{12}(\lambda_{n}),\chi^{c}_{22}(\lambda_{n}))}{\chi^{c}_{
11}(\overline{\lambda_{n}}) \chi^{c}_{22}(\lambda_{n})}& \\
&\, \, \, \, \, \, \, \, \, \, \, \, \, \, \, \, \, \, 
\, \, \, \, \, \, + \, \, 
\frac{\overline{g_{n}^{+}} \chi^{c}_{21}(\overline{\lambda_{n}
})}{\overline{\lambda_{n}} \chi^{c}_{22}(\lambda_{n})}-\frac{
g_{n}^{+} \chi^{c}_{12}(\lambda_{n})}{\lambda_{n} \chi^{c}_{11} 
(\overline{\lambda_{n}})}+\frac{(2\overline{\lambda_{n}})^{2} 
\vert g_{n}^{+} \vert^{2} \chi^{c}_{22}(\lambda_{n}) \chi^{c}
_{11}(\overline{\lambda_{n}})}{(\lambda_{n}^{2}-\overline{
\lambda_{n}}^{2})^{2}},& \\
&\widehat{\delta}_{n}^{+} := \frac{\overline{g_{n
}^{+}} \chi^{c}_{21}(\overline{\lambda_{n}})}{\chi^{c}_{22}
(\lambda_{n})} - \frac{2 \overline{\lambda_{n}} \vert g_{
n}^{+} \vert^{2} \chi^{c}_{11}(\overline{\lambda_{n}}) 
\chi^{c}_{22}(\lambda_{n})}{(\lambda_{n}^{2}-\overline{
\lambda_{n}}^{2})}.&
\end{eqnarray*}
Substituting the expressions for $\chi^{c}_{ij}(\cdot)$, $i,j \! 
\in \! \{1,2\}$, given in Lemma~4.1 into the above equations for 
$\beta_{n}^{+}$, $\widehat{\delta}_{n}^{+}$, and $\widehat{{\cal 
E}}_{nn}^{+} \widehat{{\cal H}}_{nn}^{+} \! - \! \widehat{{\cal 
F}}_{nn}^{+} \widehat{{\cal G}}_{nn}^{+}$, and recalling that 
(Eqs.~(98)) $b_{i} \! = \! - \frac{\chi^{c}_{12}(\lambda_{i})}{
\chi^{c}_{22}(\lambda_{i})} \beta_{i}$, $1 \! \leq \! i \! \leq 
\! n$, one obtains, as a result of Lemma~5.1, keeping only 
${\cal O}(1)$ and ${\cal O}(t^{-1/2})$ terms, the result stated 
in the Lemma. \hfill \rule{6.5pt}{6.5pt}
\begin{fff}
As $t \! \to \! + \infty$ and $x \! \to \! - \infty$ such 
that $\lambda_{0} \! > \! M$ and $(x,t) \! \in \! \Omega_{
n}$,
\begin{eqnarray*}
&q(x,t) = Q_{{\rm as}}^{+}(x,t) \exp\{i \arg q_{{\rm as}}^{
+}(x,t)\} + {\cal O} \! \left(\frac{C(\lambda_{0}) (\ln t)^{
2}}{t} \right) \!,&
\end{eqnarray*}
where $Q^{+}_{{\rm as}}(x,t)$ is given in Theorem~2.1, 
Eqs.~(14)--(29), $\arg q_{{\rm as}}^{+}(x,t)$ is given 
in Theorem~2.2, Eqs.~(48)--(51), and $C(\lambda_{0}) \! 
\in \! {\cal S}(\Bbb R_{>M};\! \Bbb C)$.
\end{fff} 

{\em Proof.\/} Consequence of Proposition~2.3 and Lemma~5.3. 
\hfill \rule{6.5pt}{6.5pt}
\begin{fff}
As $t \! \to \! + \infty$ and $x \! \to \! + \infty$ such 
that $\widehat{\lambda}_{0} \! := \! \sqrt{\frac{1}{2} 
(\frac{x}{t} \! - \! \frac{1}{s})} \! > \! M$, $\frac{x}{t} 
\! > \! \frac{1}{s}$, $s \! \in \! \Bbb R_{>0}$, and $(x,t) 
\! \in \! \widetilde{\Omega}_{n}$,
\begin{eqnarray*}
&u(x,t) = v_{{\rm as}}^{+} (x,t) w_{{\rm as}}^{+} (x,t) + 
{\cal O} \! \left(\frac{C(\widehat{\lambda}_{0})(\ln t)^{2}
}{t} \right) \!,&
\end{eqnarray*}
where $v_{{\rm as}}^{+}(x,t)$ and $w_{{\rm as}}^{+}(x,t)$ 
are given in Theorem~2.3, Eqs.~(58)--(70), and $C(\widehat{
\lambda}_{0}) \! \in \! {\cal S}(\Bbb R_{>M};\linebreak[4] 
\Bbb C)$.
\end{fff}

{\em Proof.\/} Consequence of Proposition~2.4 and 
Corollary~5.1. \hfill \rule{6.5pt}{6.5pt}
\section{Asymptotics as $t \! \rightarrow \! - \infty$}
In this section, the asymptotic paradigm presented in 
Secs.~3--5 is reworked for the case when $t \! \to \! -\infty$: 
since the proofs of all obtained results are analogous, they 
will be omitted. This section is divided into three parts: (1) 
in Subsection~6.1, extended and model RH problems are formulated 
as $t \! \to \! -\infty$; (2) in Subsection~6.2, the model RH 
problem formulated in (1) above is solved asymptotically as $t 
\! \to \! -\infty$ for the Schwartz class of non-reflectionless 
generic potentials; and (3) in Subsection~6.3, the phase integral, 
$((\Psi^{-1}(x,t;0))_{11})^{2}$, is evaluated asymptotically as 
$t \! \to \! -\infty$.
\subsection{Extended and Model RH Problems}
\begin{bbbb}
In the solitonless sector $({\cal Z}_{d} \! \equiv \! 
\emptyset)$, as $t \! \rightarrow \! - \infty$ and 
$x \! \rightarrow \! + \infty$ such that $\lambda_{0} 
\! > \! M$,
\begin{eqnarray*}
&m(x,t;\! \lambda) = \widetilde{\Delta}(\lambda) + {\cal O} 
\! \left(\frac{C(\lambda_{0})}{\sqrt{-t}} \right) \!,&
\end{eqnarray*}
where $\widetilde{\Delta}(\lambda) \! := \! (\delta^{-}
(\lambda;\! \lambda_{0}))^{\sigma_{3}}$,
\begin{eqnarray*}
&\delta^{-}(\lambda;\! \lambda_{0}) = ((\lambda - 
\lambda_{0})(\lambda + \lambda_{0}))^{- i \nu} \exp 
\{\sum\limits_{l \in \{\pm\}} \widetilde{\rho}_{l} 
(\lambda)\},& \\
&\widetilde{\rho}_{\pm}(\lambda) = - \frac{1}{2 \pi 
i} \int_{\pm \lambda_{0}}^{\pm \infty} \ln(\varsigma 
- \lambda) d \ln (1 - \vert r(\varsigma) \vert^{2}),& 
\end{eqnarray*}
$\nu \! := \! \nu(\lambda_{0})$ is given by Eq.~(21), 
$\vert \vert (\delta^{-}(\cdot;\! \lambda_{0}))^{\pm 1} 
\vert \vert_{{\cal L}^{\infty}(\Bbb C;\Bbb C)} \! := \! 
\sup_{\lambda \in \Bbb C} \vert (\delta^{-}(\lambda;\! 
\lambda_{0}))^{\pm 1} \vert \! < \! \infty$, $\left(
\overline{\delta^{-}(\pm \overline{\lambda};\! \lambda_{
0})} \right)^{-1} \! = \! \delta^{-}(\lambda;\! \lambda_{
0})$, the principal branch of the logarithmic function is 
taken, $\ln (\mu \! - \! \lambda) \! := \! \ln \! 
\vert \mu \! - \! \lambda \vert \! + \! i \arg (\mu \! - 
\! \lambda)$, $\arg (\mu \! - \! \lambda) \! \in \! (-\pi,
\pi)$, and $C(\lambda_{0}) \! \in \! {\cal S}(\Bbb R_{>M};
\! M_{2}(\Bbb C))$.
\end{bbbb}
\begin{eeee}
{\rm Hereafter, all explicit $x,t$ dependences are 
suppressed, except where absolutely necessary, and 
$\delta^{-}(\lambda;\! \lambda_{0}) \! := \! 
\widetilde{\delta}(\lambda)$.\/}
\end{eeee}
\begin{cccc}
There exists a unique solution $\widetilde{m}^{\widetilde{
\Delta}}(\lambda) := m(\lambda)(\widetilde{\Delta}
(\lambda))^{-1} \colon \Bbb C \setminus ({\cal Z}_{d} \cup 
\widehat{\Gamma}) \rightarrow {\rm SL\/}(2,\! \Bbb C)$ 
of the following RH problem,
\begin{enumerate}
\item $\widetilde{m}^{\widetilde{\Delta}}(\lambda)$ is 
meromorphic $\forall \, \lambda \! \in \! \Bbb C \! 
\setminus \! \widehat{\Gamma},$
\item 
\begin{eqnarray*}
&\widetilde{m}^{\widetilde{\Delta}}_{+} (\lambda) = 
\widetilde{m}^{\widetilde{\Delta}}_{-} (\lambda) 
\widetilde{v}^{\widetilde{\Delta}} (\lambda), 
\, \, \, \, \, \, \, \lambda \in \widehat{\Gamma},& 
\end{eqnarray*}
where 
\begin{eqnarray*}
&\widetilde{v}^{\widetilde{\Delta}} (\lambda) = e^{-i 
\theta(\lambda) {\rm ad} (\sigma_{3})} 
\! \left( \begin{array}{cc}
\! \! (1 - r(\lambda) \overline{r(\overline{\lambda})}) 
\widetilde{\delta}_{-}(\lambda) (\widetilde{\delta}_{+}
(\lambda))^{-1} & \! \! r(\lambda) \widetilde{\delta}_{
-} (\lambda) \widetilde{\delta}_{+}(\lambda) \\
\! \! - \overline{r(\overline{\lambda})}(\widetilde{
\delta}_{-}(\lambda))^{-1}(\widetilde{\delta}_{+}
(\lambda))^{-1} & \! \! (\widetilde{\delta}_{-}(\lambda)
)^{-1} \widetilde{\delta}_{+}(\lambda)
\end{array} \! \right) \!,&
\end{eqnarray*}
\item $\widetilde{m}^{\widetilde{\Delta}}(\lambda)$ 
has simple poles at $\{\pm \lambda_{i},\pm \overline{
\lambda_{i}}\}_{i=1}^{N}$ with $(1 \! \leq \! i \! 
\leq \! N)$
\begin{eqnarray*}
&{\rm res}(\widetilde{m}^{\widetilde{\Delta}}(\lambda);\! 
\lambda_{i}) = \lim\limits_{\lambda \rightarrow 
\lambda_{i}} \! \widetilde{m}^{\widetilde{\Delta}}
(\lambda) v_{i}(\widetilde{\delta}(\lambda_{i}))^{-2} 
\sigma_{-},& \\
&{\rm res}(\widetilde{m}^{\widetilde{\Delta}}(\lambda);\! 
-\lambda_{i}) = - \sigma_{3} {\rm res} (\widetilde{
m}^{\widetilde{\Delta}}(\lambda);\! \lambda_{i}) \sigma_{
3},& \\
&{\rm res}(\widetilde{m}^{\widetilde{\Delta}}(\lambda);\! 
\overline{\lambda_{i}} \,) = \lim\limits_{\lambda 
\rightarrow \overline{\lambda_{i}}} \! \widetilde{m}^{
\widetilde{\Delta}}(\lambda) \overline{v_{i}}
(\widetilde{\delta}(\overline{\lambda_{i}}))^{2} 
\sigma_{+},& \\
&{\rm res} (\widetilde{m}^{\widetilde{\Delta}}(\lambda);\! 
-\overline{\lambda_{i}} \,) = - \sigma_{3} {\rm res} 
(\widetilde{m}^{\widetilde{\Delta}}(\lambda);\! 
\overline{\lambda_{i}} \,) \sigma_{3},& 
\end{eqnarray*}
\item as $\lambda \! \rightarrow \! \infty$, $\lambda 
\! \in \! \Bbb C \! \setminus \! ({\cal Z}_{d} \! \cup 
\! \widehat{\Gamma})$,
\begin{eqnarray*}
&\widetilde{m}^{\widetilde{\Delta}} (\lambda) = {\rm I} 
+ {\cal O}(\lambda^{-1});& 
\end{eqnarray*}
\end{enumerate}
moreover, $Q(x,t) \! = \! 2 i \! \lim\limits_{\lambda \to \infty} 
\! (\lambda \widetilde{m}^{\widetilde{\Delta}}(x,t;\! \lambda))_{
12}$ is equal to $Q(x,t)$ in Lemma~2.2, Eq.~(11).
\end{cccc}
\begin{bbbb}
Introduce arbitrarily small, clockwise- and 
counter-clockwise-oriented, mutually disjoint (and 
disjoint with respect to $\widehat{\Gamma})$ circles 
$\widetilde{K}_{j}^{\pm}$ and $\widetilde{L}_{j}^{
\pm}$, $1 \! \leq \! j \! \leq \! n \! - \! 1$, around 
the eigenvalues $\{\pm \lambda_{j}\}_{j=1}^{n-1}$ and 
$\{\pm \overline{\lambda_{j}}\}_{j=1}^{n-1}$, 
respectively, and define
\begin{eqnarray*}
&\widetilde{m}^{\flat} (\lambda) := 
\left\{ \begin{array}{l} 
\widetilde{m}^{\widetilde{\Delta}}(\lambda), 
\, \, \, \, \, \, \, \lambda \in \Bbb C \setminus 
(\widehat{\Gamma} \cup (\bigcup\limits_{i=1}^{n-1} 
(\widetilde{K}_{i}^{\pm} \cup \widetilde{L}_{i}^{
\pm}))), \\
\widetilde{m}^{\widetilde{\Delta}}(\lambda) \! \left( 
{\rm I} - \frac{v_{i}(\widetilde{\delta}(\pm \lambda_{
i}))^{-2}}{(\lambda \mp \lambda_{i})} \sigma_{-} 
\right) \!, \, \, \, \, \, \, \, \lambda \in {\rm int} 
\widetilde{K}_{i}^{\pm}, \, \, \, \, \, 1 \leq i \leq 
n-1, \\
\widetilde{m}^{\widetilde{\Delta}} (\lambda) \! \left( 
{\rm I} + \frac{\overline{v_{i}}(\widetilde{\delta}(\pm 
\overline{\lambda_{i}}))^{2}}{(\lambda \mp \overline{
\lambda_{i}})} \sigma_{+} \right) \!, \, \, \, \, \, \, 
\, \, \, \lambda \in {\rm int} \widetilde{L}_{i}^{\pm}, 
\, \, \, \, \, \, \, 1 \leq i \leq n-1. 
\end{array} \right.& 
\end{eqnarray*}
Then $\widetilde{m}^{\flat}(\lambda)$ solves a RH problem on 
$(\sigma_{\pounds} \! \setminus \! \cup_{i=1}^{n-1} (\{\pm 
\lambda_{i}\} \! \cup \! \{\pm \overline{\lambda_{i}}\})) \! 
\cup \! (\cup_{i=1}^{n-1} (\widetilde{K}_{i}^{\pm} \! \cup \! 
\widetilde{L}_{i}^{\pm}))$ with the same jumps as $\widetilde{
m}^{\widetilde{\Delta}}(\lambda)$ on $\widehat{\Gamma}$, 
$\widetilde{m}_{+}^{\flat}(\lambda) \! = \! \widetilde{m}_{-}
^{\flat}(\lambda) \widetilde{v}^{\widetilde{\Delta}}(\lambda)$, 
and
\begin{eqnarray*}
&\widetilde{m}_{+}^{\flat}(\lambda) =
\left\{ \begin{array}{l} 
\widetilde{m}_{-}^{\flat}(\lambda) \! \left({\rm 
I} + \frac{v_{i}(\widetilde{\delta}(\pm \lambda_{
i}))^{-2}}{(\lambda \mp \lambda_{i})} \sigma_{-} 
\right) \!, \, \, \, \, \, \, \, \lambda \in \widetilde{
K}_{i}^{\pm}, \, \, \, \, \, \, 1 \leq i \leq n-1, \\
\widetilde{m}_{-}^{\flat}(\lambda) \! \left({\rm 
I} + \frac{\overline{v_{i}}(\widetilde{\delta} 
(\pm \overline{\lambda_{i}}))^{2}}{(\lambda \mp 
\overline{\lambda_{i}}) } \sigma_{+} \right) \!, \,
\, \, \, \, \, \, \, \, \lambda \in \widetilde{L}_{
i}^{\pm}, \, \, \, \, \, \, \, \, 1 \leq i \leq n-1.  
\end{array} \right.& 
\end{eqnarray*}
\end{bbbb}
\begin{eeee}
{\rm The superscripts $\pm$ on $\{\widetilde{K}_{i}^{\pm}\}_{
i=1}^{n-1}$ and $\{\widetilde{L}_{i}^{\pm}\}_{i=1}^{n-1}$, which 
are related with $\{\pm \lambda_{i}\}_{i=1}^{n-1}$ and $\{\pm 
\overline{\lambda_{i}}\}_{i=1}^{n-1}$, respectively, should 
\underline{{\bf not}} be confused with the subscripts $\pm$ 
appearing in the various RH problems in this and the next 
subsection, namely, $m_{\pm}(\lambda)$, $\widetilde{m}^{
\widetilde{\Delta}}_{\pm}(\lambda)$, $\widetilde{m}^{\flat}_{\pm}
(\lambda)$, $\widetilde{m}_{\pm}^{\sharp}(\lambda)$, $\widetilde{
\chi}_{\pm}(\lambda)$, $\widetilde{E}_{\pm}(\lambda)$, and 
$\widetilde{\chi}^{c}_{\pm}(\lambda)$.\/}
\end{eeee}
\begin{cccc}
Set
\begin{eqnarray*}
&\widetilde{m}^{\sharp}(\lambda) \! := \! 
\left\{ \begin{array}{l} 
\widetilde{m}^{\flat} (\lambda) \prod\limits_{l=1}^{n-1} 
(d_{l_{+}}(\lambda))^{- \sigma_{3}}, \, \, \, \, \, \, 
\, \lambda \in \Bbb C \setminus (\widehat{\Gamma} \cup 
(\bigcup\limits_{i=1}^{n-1} ( \widetilde{K}_{i}^{\pm} 
\cup \widetilde{L}_{i}^{\pm}))), \\
\widetilde{m}^{\flat}(\lambda) (\widetilde{J}_{
\widetilde{K}_{i}^{\pm}}(\lambda))^{-1} \prod\limits_{
l=1}^{n-1} (d_{l_{-}}(\lambda))^{- \sigma_{3}}, \, \, \, 
\, \, \, \, \lambda \in {\rm int} \widetilde{K}_{i}^{\pm}, 
\, \, \, \, \, \, 1 \leq i \leq n-1, \\
\widetilde{m}^{\flat} (\lambda)(\widetilde{J}_{\widetilde{
L}_{i}^{\pm}}(\lambda))^{-1} \prod\limits_{l=1}^{n-1} (d_{
l_{-}}(\lambda))^{-\sigma_{3}}, \, \, \, \, \, \, \, \, 
\lambda \in {\rm int} \widetilde{L}_{i}^{\pm}, \, \, \, 
\, \, \, \, 1 \leq i \leq n-1,
\end{array} \right.&
\end{eqnarray*}
where
\begin{eqnarray*}
&\left. \begin{array}{l}
d_{l_{+}} (\lambda) \! := \! \frac{(\lambda - 
\overline{\lambda_{l}}) (\lambda + \overline{
\lambda_{l}})}{(\lambda - \lambda_{l})(\lambda + 
\lambda_{l})}, \, \, \, \, \, \, \, \lambda \in \Bbb 
C \setminus (\bigcup\limits_{i=1}^{n-1} (\widetilde{
K}_{i}^{\pm} \cup \widetilde{L}_{i}^{\pm})), 
\, \, \, \, \, \, 1 \leq l \leq n-1, \\
d_{l_{-}}(\lambda) \! := \! \left\{ \begin{array}{l}
\frac{(\lambda - \overline{\lambda_{l}}) (\lambda + 
\overline{\lambda_{l}})}{(\lambda \pm \lambda_{l})}, 
\, \, \, \, \, \, \, \lambda \in \bigcup\limits_{i=1}^{n-1} 
{\rm int} \widetilde{K}_{i}^{\pm}, \, \, \, \, \, \, 1 
\leq l \leq n-1, \\
\frac{(\lambda \pm \overline{\lambda_{l}})}{(\lambda - 
\lambda_{l})(\lambda + \lambda_{l})}, \, \, \, \, \, \, 
\, \, \lambda \in \bigcup\limits_{i=1}^{n-1} {\rm int} 
\widetilde{L}_{i}^{\pm}, \, \, \, \, \, \, \, 1 \leq l 
\leq n-1,
\end{array} \right. 
\end{array} \right.& 
\end{eqnarray*}
and the ${\rm SL\/}(2,\! \Bbb C)$-valued, holomorphic in ${\rm 
int} \widetilde{K}_{i}^{\pm}$ and ${\rm int} \widetilde{L}_{
i}^{\pm}$, respectively, functions $\widetilde{J}_{\widetilde{
K}_{i}^{\pm}}(\lambda)$ and $\widetilde{J}_{\widetilde{L}_{i}
^{\pm}}(\lambda)$, $1 \! \leq \! i \! \leq \! n \! - \! 1$,
are given by
\begin{eqnarray*}
&\widetilde{J}_{\widetilde{K}_{i}^{\pm}}(\lambda) 
\! = \! \left( \begin{array}{cc}
\frac{\prod\limits_{l=1 \atop \not= i}^{n-1} \frac{
d_{l_{-}}^{-1} (\lambda)}{d_{l_{+}}^{-1}(\lambda)} \, 
\, - \, \, \frac{v_{i}(\widetilde{\delta}(\pm \lambda
_{i}))^{-2} \widetilde{C}_{i}^{\sharp}}{(d_{i_{-}}(
\lambda))^{2}} \prod\limits_{l=1 \atop \not=i}^{n-1} 
\frac{d_{l_{-}}^{-1}(\lambda)}{d_{l_{+}}(\lambda)}}{(
\lambda \mp \lambda_{i})} & \, \, \, \, \frac{\widetilde{
C}_{i}^{\sharp}}{(d_{i_{-}}(\lambda))^{2}} \prod\limits_{
l=1 \atop \not= i}^{n-1} \frac{d_{l_{-}}^{-1}(\lambda)}{
d_{l_{+}}(\lambda)} \\
- v_{i}(\widetilde{\delta}(\pm \lambda_{i}))^{-2} 
\prod\limits_{l=1 \atop \not= i}^{n-1} \frac{d_{l_{-}}
(\lambda)}{d_{l_{+}}(\lambda)} & \, \, \, \, (\lambda \mp 
\lambda_{i}) \prod\limits_{l = 1 \atop \not= i}^{n-1} 
\frac{d_{l_{-}}(\lambda)}{d_{l_{+}}(\lambda)}
\end{array} \right) \!,& \\
&\widetilde{J}_{\widetilde{L}_{i}^{\pm}}(\lambda) 
\! = \! \left( \begin{array}{cc}
(\lambda \mp \overline{\lambda_{i}}) \prod\limits_{
l=1 \atop \not= i}^{n-1} \frac{d_{l_{-}}^{-1}(\lambda)
}{d_{l_{+}}^{-1}(\lambda)} & \, \, \, \, \overline{v_{i}} 
(\widetilde{\delta}(\pm \overline{\lambda_{i}}))^{2} 
\prod\limits_{l=1 \atop \not= i}^{n-1} \frac{d_{l_{-}
}^{-1} (\lambda)}{d_{l_{+}}^{-1}(\lambda)} \\
- \frac{\overline{\widetilde{C}_{i}^{\sharp}}}{(d_{
i_{-}}(\lambda))^{-2}} \prod\limits_{l=1 \atop \not= 
i}^{n-1} \frac{d_{l_{-}}(\lambda)}{d_{l_{+}}^{-1} 
(\lambda)} & \, \, \, \, \frac{\prod\limits_{l=1 \atop 
\not=i}^{n-1} \frac{d_{l_{-}}(\lambda)}{d_{l_{+}}(\lambda)} 
\, \, - \, \, \frac{\overline{v_{i}}(\widetilde{\delta}
(\pm \overline{\lambda_{i}}))^{2} \overline{\widetilde{
C}_{i}^{\sharp}}}{(d_{i_{-}}(\lambda))^{-2}} \prod\limits
_{l=1 \atop \not=i}^{n-1} \frac{d_{l_{-}}(\lambda)}{d_{l_{
+}}^{-1}(\lambda)}}{(\lambda \mp \overline{\lambda_{i}})} 
\end{array} \right) \!,&
\end{eqnarray*}
with
\begin{eqnarray*}
&\widetilde{C}_{i}^{\sharp} \! = \! (v_{i})^{-1}(\widetilde{
\delta}(\pm \lambda_{i}))^{2}(d_{i_{-}}(\pm \lambda_{i
}))^{2} \prod\limits_{l=1 \atop \not= i}^{n-1} (d_{l_{
+}}(\pm \lambda_{i}))^{2}, \, \, \, \, \, \, \, \, 1 \leq 
i \leq n-1.&
\end{eqnarray*}
Then $\widetilde{m}^{\sharp}(\lambda) \colon \Bbb C \! 
\setminus \! (({\cal Z}_{d} \! \setminus \! 
\cup_{i=1}^{n-1} (\{\pm \lambda_{i}\} \! \cup \! 
\{\pm \overline{\lambda_{i}}\})) \! \cup \! 
(\widehat{\Gamma} \! \cup \! (\cup_{i=1}^{n-1} 
(\widetilde{K}_{i}^{\pm} \! \cup \! \widetilde{L}_{
i}^{\pm})))) \! \rightarrow \! {\rm SL\/}(2,\! \Bbb 
C)$ solves the following, extended RH problem on 
$(\sigma_{\pounds} \setminus \cup_{i=1}^{n-1}(\{\pm 
\lambda_{i}\} \! \cup \! \{\pm \overline{\lambda_{i}}
\})) \! \cup \! (\cup_{i=1}^{n-1}(\widetilde{K}_{i}^{
\pm} \! \cup \! \widetilde{L}_{i}^{\pm}))$,
\begin{eqnarray*}
&\widetilde{m}_{+}^{\sharp}(\lambda) = \widetilde{
m}_{-}^{\sharp}(\lambda) e^{- i \theta(\lambda) {\rm 
ad}(\sigma_{3})} \widetilde{v}^{\sharp}(\lambda),&
\end{eqnarray*}
where 
\begin{eqnarray*}
&\widetilde{v}^{\sharp}(\lambda) \vert_{\widehat{
\Gamma}} \! = \! \left( \begin{array}{cc}
\! \! \! (1 - r(\lambda) \overline{r(\overline{
\lambda})}) \frac{\widetilde{\delta}_{-}(\lambda)}{ 
\widetilde{\delta}_{+}(\lambda)} & \, \, \frac{r(\lambda)
}{(\widetilde{\delta}_{-}(\lambda) \widetilde{\delta}_{
+}(\lambda))^{-1}} \prod\limits_{l=1}^{n-1} \! \left( 
\! \frac{(\lambda - \overline{\lambda_{l}}) (\lambda 
+ \overline{\lambda_{l}})}{(\lambda - \lambda_{l})(
\lambda + \lambda_{l})} \! \right)^{2} \\
\! \! \! - \frac{ \overline{r(\overline{\lambda})}}{ 
\widetilde{\delta}_{-}(\lambda) \widetilde{\delta}_{
+}(\lambda)} \prod\limits_{l=1}^{n-1} \! \left( \! 
\frac{(\lambda - \overline{\lambda_{l}})(\lambda + 
\overline{\lambda_{l}})}{(\lambda - \lambda_{l})
(\lambda + \lambda_{l})} \! \right)^{-2} & \, \, \frac{
\widetilde{\delta}_{+}(\lambda)}{\widetilde{
\delta}_{-}(\lambda)} \end{array} \right) \!,& \\
&\widetilde{v}^{\sharp}(\lambda) \! = \! 
\left\{\begin{array}{l}
{\rm I} + \frac{(v_{i})^{-1}(\widetilde{\delta}(\pm 
\lambda_{i}))^{2}}{(\lambda \mp \lambda_{i})} \left(\! 
\frac{\lambda_{i}^{2} - \overline{\lambda_{i}}^{2}}{2 
\lambda_{i}} \! \right)^{2} \prod\limits_{l=1 \atop 
\not= i}^{n-1} \left(\! \frac{\overline{\lambda_{l}}^{
2} - \lambda_{i}^{2}}{\lambda_{l}^{2} - \lambda_{i}^{2
}} \! \right)^{2} \! \sigma_{+}, \, \, \, \, \, \, \, 
\, \, \lambda \in \bigcup\limits_{i=1}^{n-1} \widetilde{
K}_{i}^{\pm}, \\
{\rm I} + \frac{(\overline{v_{i}})^{-1}(\widetilde{\delta}
(\pm \overline{\lambda_{i}}))^{-2}}{(\lambda \mp \overline{
\lambda_{i}})} \left(\! \frac{\lambda_{i}^{2} - \overline{
\lambda_{i}}^{2}}{2 \overline{\lambda_{i}}} \! \right)^{2} 
\prod\limits_{l=1 \atop \not= i}^{n-1} \left(\! \frac{ 
\lambda_{l}^{2} - \overline{\lambda_{i}}^{2}}{\overline{
\lambda_{l}}^{2} - \overline{\lambda_{i}}^{2}} \! \right)^{
2} \! \sigma_{-}, \, \, \, \, \, \, \, \lambda \in 
\bigcup\limits_{i=1}^{n-1} \widetilde{L}_{i}^{\pm},
\end{array} \right.&
\end{eqnarray*}
with polar (residue) conditions, 
\begin{eqnarray*}
&{\rm res}(\widetilde{m}^{\sharp}(\lambda);\! \lambda_{i}) 
= \lim\limits_{\lambda \rightarrow \lambda_{i}} \! 
\widetilde{m}^{\sharp} (\lambda) v_{i} (\widetilde{\delta}
(\lambda_{i}))^{-2} \prod\limits_{l=1}^{n-1} \! \left(\! 
\frac{(\lambda_{i} - \lambda_{l})(\lambda_{i} + \lambda_{l}
)}{(\lambda_{i} - \overline{\lambda_{l}}) (\lambda_{i} + 
\overline{\lambda_{l}})} \! \right)^{2} \! \! \sigma_{-}, \, 
\, \, \, \, \, \, n \leq i \leq N,& \\
&{\rm res}(\widetilde{m}^{\sharp}(\lambda);\! - \lambda_{i}) 
= -\sigma_{3} \mathstrut {\rm res}(\widetilde{m}^{\sharp}
(\lambda);\! \lambda_{i}) \mathstrut \sigma_{3}, \, \, \, \, 
\, \, \, n \leq i \leq N,& \\
&{\rm res}(\widetilde{m}^{\sharp}(\lambda);\! \overline{
\lambda_{i}} \,) = \lim\limits_{\lambda \rightarrow 
\overline{\lambda_{i}}} \! \widetilde{m}^{\sharp}
(\lambda) \overline{v_{i}} (\widetilde{\delta}(\overline{
\lambda_{i}}))^{2} \prod\limits_{l=1}^{n-1} \! \left(\! 
\frac{(\overline{\lambda_{i}} - \overline{\lambda_{l}})(
\overline{\lambda_{i}} + \overline{\lambda_{l}})}{(
\overline{\lambda_{i}} - \lambda_{l})(\overline{\lambda_{
i}} + \lambda_{l})} \! \right)^{2} \! \! \sigma_{+}, \, \, 
\, \, \, \, \, n \leq i \leq N,& \\
&{\rm res}(\widetilde{m}^{\sharp}(\lambda);\! - \overline{
\lambda_{i}} \,) = - \sigma_{3} \mathstrut {\rm res}(
\widetilde{m}^{\sharp} (\lambda);\! \overline{\lambda_{i}
} \,) \mathstrut \sigma_{3}, \, \, \, \, \, \, \, n \leq i 
\leq N,&
\end{eqnarray*}
and, as $\lambda \! \rightarrow \! \infty$, $\lambda \! \in 
\! \Bbb C \! \setminus \! (({\cal Z}_{d} \! \setminus \! 
\cup_{i=1}^{n-1} (\{\pm \lambda_{i}\} \! \cup \! \{\pm 
\overline{\lambda_{i}}\})) \! \cup \! (\widehat{\Gamma} \! 
\cup \! (\cup_{i=1}^{n-1} (\widetilde{K}_{i}^{\pm} \! \cup 
\! \widetilde{L}_{i}^{\pm}))))$,
\begin{eqnarray*}
&\widetilde{m}^{\sharp}(\lambda)={\rm I} + {\cal O}
(\lambda^{-1});&
\end{eqnarray*}
moreover, $Q(x,t) \! = \! 2 i \! \lim\limits_{\lambda \to 
\infty} \! (\lambda \widetilde{m}^{\sharp}(x,t;\! \lambda)
)_{12}$ is equal to $Q(x,t)$ in Lemma~2.2, Eq.~(11).
\end{cccc}
\begin{cccc}
Let $\widetilde{\chi}(\lambda)$ solve the following RH 
problem on $\sigma_{\pounds} \! \setminus \! \cup_{i=1
}^{n-1} (\{\pm \lambda_{i}\} \! \cup \! \{\pm \overline{
\lambda_{i}}\})$,
\begin{eqnarray*}
\widetilde{\chi}_{+}(\lambda) = \widetilde{\chi}_{-}
(\lambda) e^{- i \theta(\lambda){\rm ad}(\sigma_{3})}
\widetilde{v}^{\sharp}(\lambda) \vert_{\widehat{\Gamma
}}, \, \, \, \, \, \, \, \lambda \in \widehat{\Gamma}, 
\end{eqnarray*}
with polar (residue) conditions,
\begin{eqnarray*}
&{\rm res}(\widetilde{\chi}(\lambda);\! \lambda_{i}) = 
\lim\limits_{\lambda \rightarrow \lambda_{i}} \! 
\widetilde{\chi}(\lambda) v_{i} (\widetilde{\delta}
(\lambda_{i}))^{-2} \prod\limits_{l=1}^{n-1} \! 
\left(\! \frac{(\lambda_{i} - \lambda_{l})(\lambda_{i} 
+ \lambda_{l})}{(\lambda_{i} - \overline{\lambda_{l}}) 
(\lambda_{i} + \overline{\lambda_{l}})} \! \right)^{2} 
\! \! \sigma_{-}, \, \, \, \, \, \, \, n \leq i \leq N,& 
\\
&{\rm res}(\widetilde{\chi}(\lambda);\! -\lambda_{i}) = 
- \sigma_{3} \mathstrut {\rm res}(\widetilde{\chi} 
(\lambda);\! \lambda_{i}) \mathstrut \sigma_{3}, \, \, \, 
\, \, \, \, n \leq i \leq N,& \\
&{\rm res}(\widetilde{\chi}(\lambda);\! \overline{\lambda_{i}
} \,) = \lim\limits_{\lambda \rightarrow \overline{\lambda_{
i}}} \widetilde{\chi}(\lambda) \overline{v_{i}}(\widetilde{
\delta}(\overline{\lambda_{i}}))^{2} \prod\limits_{l=1}^{n-1} 
\! \left(\! \frac{(\overline{\lambda_{i}} - \overline{
\lambda_{l}})(\overline{\lambda_{i}} + \overline{
\lambda_{l}})}{(\overline{\lambda_{i}} - \lambda_{
l})(\overline{\lambda_{i}} + \lambda_{l})} \! \right)^{2} 
\! \! \sigma_{+}, \, \, \, \, \, \, \, n \leq i \leq N,& \\
&{\rm res}(\widetilde{\chi}(\lambda);\! -\overline{\lambda_{
i}} \,) = - \sigma_{3} \mathstrut {\rm res} (\widetilde{\chi}
(\lambda);\! \overline{\lambda_{i}} \,) \mathstrut \sigma_{3}, 
\, \, \, \, \, \, \, n \leq i \leq N,& 
\end{eqnarray*}
and, as $\lambda \! \rightarrow \! \infty$, $\lambda \! \in \! 
\Bbb C \! \setminus \! (\cup_{i=n}^{N} (\{\pm \lambda_{i}\} \! 
\cup \! \{\pm \overline{\lambda_{i}}\}) \! \cup \! \widehat{
\Gamma})$,
\begin{eqnarray*}
&\widetilde{\chi}(\lambda) = {\rm I} + {\cal O}
(\lambda^{-1}).&
\end{eqnarray*}
Then as $t \! \rightarrow \! - \infty$ and $x \! \rightarrow \! 
+ \infty$ such that $\lambda_{0} \! > \! M$ and $(x,t) \! \in 
\! \Omega_{n}$, the function $\widetilde{E}(\lambda) \! := \! 
\widetilde{m}^{\sharp}(\lambda) (\widetilde{\chi}(\lambda))^{
-1}$ has the following asymptotics,
\begin{eqnarray*}
&\widetilde{E}(\lambda) = {\rm I} + {\cal O}
(\widetilde{F}(\lambda;\! \lambda_{0}) \exp \{
\widetilde{a} \widetilde{b} t\}),&
\end{eqnarray*}
where $\vert \vert \widetilde{F}(\cdot;\! \lambda_{0}) 
\vert \vert_{{\cal L}^{\infty}(\Bbb C;M_{2}(\Bbb C))} 
\! < \! \infty$, $\vert \vert \widetilde{F}(\lambda;
\cdot) \vert \vert_{{\cal L}^{\infty}(\Bbb R_{> M};M_{
2}(\Bbb C))} \! < \! \infty$, $\widetilde{F}(\lambda;
\! \lambda_{0}) \! \sim \! {\cal O} \! \left(\frac{C
(\lambda_{0})}{\lambda} \right)$ as $\lambda \! 
\rightarrow \! \infty$ with $C(\lambda_{0}) \! \in \! 
{\cal L}^{\infty}(\Bbb R_{> M};\! M_{2}(\Bbb C))$, 
$\widetilde{a} \! := \! 8 \min \{\eta_{i}\}_{i=1}^{
n-1} \, (> \! 0)$, and $\widetilde{b} \! := \! \min 
\{\vert \xi_{n} \! - \! \xi_{i} \vert \}_{i=1}^{n-1}$.
\end{cccc}
\subsection{Asymptotic Solution for $\widetilde{\chi}
(\lambda)$}
\begin{bbbb}
The solution of the model RH problem formulated in 
Lemma~6.1.3, $\widetilde{\chi}(\lambda) \colon \Bbb 
C \! \setminus \! (\widehat{\Gamma} \! \cup \! (\cup_{i
=n}^{N} (\{\pm \lambda_{i}\} \! \cup \! \{\pm \overline{
\lambda_{i}}\}))) \! \rightarrow \! {\rm SL\/}(2,\! \Bbb 
C)$, has the following representation,
\begin{eqnarray}
\widetilde{\chi}(\lambda) = \widetilde{\chi}_{d}(\lambda)+
\int_{\raise-0.5ex\hbox{$\scriptstyle{}\widehat{\Gamma}$}} 
\frac{\widetilde{\chi}_{-}(\varrho)(\widetilde{v}^{\sharp}
(\varrho) \vert_{\widehat{\Gamma}}-{\rm I})}{(\varrho-
\lambda)} \frac{d \varrho}{2 \pi i},
\end{eqnarray}
where
\begin{eqnarray}
&\widetilde{\chi}_{d}(\lambda) = {\rm I} + 
\sum\limits_{i=n}^{N} \! \left(\frac{{\rm res}
(\widetilde{\chi}(\lambda); \lambda_{i})}{(\lambda 
- \lambda_{i})} - \frac{\sigma_{3} \mathstrut {\rm 
res}(\widetilde{\chi}(\lambda);\lambda_{i}) \mathstrut 
\sigma_{3}}{(\lambda + \lambda_{i})} + \frac{{\rm res} 
(\widetilde{\chi}(\lambda);\overline{\lambda_{i}})}{
(\lambda - \overline{\lambda_{i}})} - \frac{\sigma_{3} 
\mathstrut {\rm res} (\widetilde{\chi} (\lambda);
\overline{\lambda_{i}}) \mathstrut \sigma_{3}}{
(\lambda + \overline{\lambda_{i}})} \right) \!.& 
\end{eqnarray}
The solution of Eq.~(107) can be written as the following
ordered product,
\begin{eqnarray*}
&\widetilde{\chi}(\lambda) = \widetilde{\chi}_{d}
(\lambda) \widetilde{\chi}^{c}(\lambda),&
\end{eqnarray*}
where $\widetilde{\chi}_{d}(\lambda)$ is given by 
Eq.~(108), and $\widetilde{\chi}^{c}(\lambda)$ solves
the following RH problem: (1) $\widetilde{\chi}^{c}
(\lambda)$ is piecewise holomorphic $\forall \, \lambda
\! \in \! \Bbb C \! \setminus \! \widehat{\Gamma}$; (2)
$\widetilde{\chi}^{c}_{+}(\lambda) \! = \! \widetilde{
\chi}^{c}_{-}(\lambda) \exp \{- i \theta(\lambda){\rm 
ad}(\sigma_{3})\}(\widetilde{v}^{\sharp}(\lambda) 
\vert_{\widehat{\Gamma}})$, $\lambda \! \in \! \widehat{
\Gamma}$; and (3) as $\lambda \! \rightarrow \! \infty$, 
$\lambda \! \in \! \Bbb C \! \setminus \! \widehat{
\Gamma}$, $\widetilde{\chi}^{c}(\lambda) \! = \! {\rm 
I} \! + \! {\cal O}(\lambda^{-1})$.
\end{bbbb}
\begin{cccc}
Let $\widetilde{\epsilon}_{0}$ denote an arbitrarily fixed, 
sufficiently small positive real number. For $\widetilde{
\aleph} \! \in \! \{0,\pm \lambda_{0}\}$, set $\widetilde{
{\cal N}}(\widetilde{\aleph};\! \widetilde{\epsilon}_{0}) 
\! := \! \{\mathstrut \lambda; \, \vert \lambda \! - \! 
\widetilde{\aleph} \vert \! \leq \! \widetilde{\epsilon}_{
0}\}$. Then as $t \! \to \! -\infty$ and $x \! \to \! 
+\infty$ such that $\lambda_{0} \! > \! M$ and $\lambda \! 
\in \! \Bbb C \! \setminus \! \cup_{\widetilde{\aleph} \in 
\{0,\pm \lambda_{0}\}} \widetilde{{\cal N}}(\widetilde{
\aleph};\! \widetilde{\epsilon}_{0})$, $\widetilde{\chi}^{
c}(\lambda)$ has the following asymptotic expansion,
\begin{eqnarray*}
\widetilde{\chi}^{c}(\lambda) & = & {\rm I} + \frac{1}{
4} \sqrt{- \frac{\nu(\lambda_{0})}{2 \lambda_{0}^{2} t}} 
\! \left(\frac{1}{\lambda - \lambda_{0}} + \frac{1}{
\lambda + \lambda_{0}} \right) \! \left( \exp \{- i
(\phi^{-}(\lambda_{0}) + \widehat{\Phi}^{-}(\lambda_{0};
t))\} \sigma_{-} \right. \\
 & + & \left. \exp \{i (\phi^{-}(\lambda_{0}) + \widehat{
\Phi}^{-}(\lambda_{0};t))\} \sigma_{+} \right) + {\cal O}
\! \left(\frac{\widetilde{G}(\lambda;\! \lambda_{0}) \ln 
\! \vert t \vert}{t} \right) \!,
\end{eqnarray*}
where $\nu(\lambda_{0})$, $\phi^{-}(\lambda_{0})$, and 
$\widehat{\Phi}^{-}(\lambda_{0};t)$ are given in 
Theorem~2.1, Eqs.~(21), (23), and (24), $\vert \vert 
\widetilde{G}(\cdot;\! \lambda_{0}) \vert \vert_{{\cal 
L}^{\infty}(\Bbb C \setminus \cup_{\widetilde{\aleph} 
\in \{0,\pm \lambda_{0}\}} \widetilde{{\cal N}}
(\widetilde{\aleph};\widetilde{\epsilon}_{0});M_{2}
(\Bbb C))} \! < \! \infty$, $\widetilde{G}(\lambda;\! 
\cdot) \! \in \! {\cal S}(\Bbb R_{>M};\! M_{2}(\Bbb C))$, 
$\widetilde{G}(\lambda;\! \lambda_{0}) \! \sim \! {\cal 
O} \! \left(\frac{C(\lambda_{0})}{\lambda} \right)$ as 
$\lambda \! \rightarrow \! \infty$ with $C(\lambda_{0}) 
\! \in \! {\cal S}(\Bbb R_{>M};\! M_{2}(\Bbb C))$, and 
satisfies the following involutions, $\widetilde{\chi
}^{c}(-\lambda) \! = \! \sigma_{3} \widetilde{\chi}^{c}
(\lambda) \sigma_{3}$ and $\widetilde{\chi}^{c}(\lambda) 
\! = \! \sigma_{1} \overline{\widetilde{\chi}^{c}
(\overline{\lambda})} \sigma_{1}$.
\end{cccc}
\begin{bbbb} 
For $n \! \leq \! i \! \leq \! N$, set
\begin{eqnarray*}
&{\rm res}(\widetilde{\chi}(\lambda);\! \lambda_{i})
= \left( \! \begin{array}{cc}
\alpha_{i}^{-} & a_{i}^{-} \\
\beta_{i}^{-} & b_{i}^{-} 
\end{array} \! \right), \, \, \, \, \, \, \, \, \, 
{\rm res}(\widetilde{\chi}(\lambda);\! \overline{
\lambda_{i}}) = \left( \! \begin{array}{cc}
c_{i}^{-} & \omega_{i}^{-} \\
d_{i}^{-} & \widehat{\delta}_{i}^{-} 
\end{array} \! \right) \!.&
\end{eqnarray*}
Then as $\lambda \! \rightarrow \! \infty$, $\lambda 
\! \in \! \Bbb C \! \setminus \! (\widehat{\Gamma} 
\! \cup \! (\cup_{i=n}^{N} (\{\pm \lambda_{i}\} \! 
\cup \! \{\pm \overline{\lambda_{i}}\})))$, 
$\widetilde{\chi}(\lambda)$ has the following asymptotic 
expansion,
\begin{eqnarray*}
\widetilde{\chi}(\lambda) & = & {\rm I} + \frac{1}{2 \lambda} 
\left( \! \left\{\overline{Q^{\widetilde{\chi}}(x,t)} + 4 
\sum\limits_{i=n}^{N} \! \left(\beta_{i}^{-} - \frac{
\widetilde{\chi}^{c}_{21}(\overline{\lambda_{i}})}{\widetilde{
\chi}^{c}_{11}(\overline{\lambda_{i}})} \widehat{\delta}_{i}^{
-} \right) \! \right\} \sigma_{-} \right. \\
 & + & \left. \left\{Q^{\widetilde{\chi}}(x,t) + 4 \sum\limits
_{i=n}^{N} \! \left( \omega_{i}^{-} - \frac{\widetilde{\chi}^{
c}_{12}(\lambda_{i})}{\widetilde{\chi}^{c}_{22}(\lambda_{i})} 
\alpha_{i}^{-} \right) \! \right\} \sigma_{+} \! \right) + 
{\cal O}(\lambda^{-2}),
\end{eqnarray*}
where $\lim\limits_{\lambda \rightarrow \infty}(\widetilde{\chi}
^{c}(x,t;\! \lambda))_{12} \! := \! Q^{\widetilde{\chi}}(x,t) / 
2 \lambda$, $\{\alpha_{i}^{-},\omega_{i}^{-}\}_{i=n}^{N}$ 
satisfy the following non-degenerate system of $2(N \! - \! 
n \! + \! 1)$ linear inhomogeneous algebraic equations,
\begin{eqnarray*}
&\left[ \begin{array}{cccccc}
\left. \begin{array}{ccc} \cline{1-3}
\multicolumn{1}{|c}{} &   & \multicolumn{1}{c|}{} \\
\multicolumn{1}{|c}{} & \widehat{{\cal A}}^{-} & 
\multicolumn{1}{c|}{} \\
\multicolumn{1}{|c}{} &   & \multicolumn{1}{c|}{} \\ 
\cline{1-3}   
\end{array} \right. & \left. \begin{array}{ccc} 
\cline{1-3}
\multicolumn{1}{|c}{} &   & \multicolumn{1}{c|}{} \\
\multicolumn{1}{|c}{} & \widehat{{\cal B}}^{-} & 
\multicolumn{1}{c|}{} \\
\multicolumn{1}{|c}{} &   & \multicolumn{1}{c|}{} \\ 
\cline{1-3}   
\end{array} \right. \\
& \\
\left. \begin{array}{ccc} \cline{1-3}
\multicolumn{1}{|c}{} &   & \multicolumn{1}{c|}{} \\
\multicolumn{1}{|c}{} & \widehat{{\cal C}}^{-} & 
\multicolumn{1}{c|}{} \\
\multicolumn{1}{|c}{} &   & \multicolumn{1}{c|}{} \\ 
\cline{1-3}   
\end{array} \right. & \left. \begin{array}{ccc} 
\cline{1-3}
\multicolumn{1}{|c}{} &   & \multicolumn{1}{c|}{} \\
\multicolumn{1}{|c}{} & \widehat{{\cal D}}^{-} & 
\multicolumn{1}{c|}{} \\
\multicolumn{1}{|c}{} &   & \multicolumn{1}{c|}{} \\ 
\cline{1-3}   
\end{array} \right. 
\end{array} \right] 
\left[ \begin{array}{c}
          \alpha_{n}^{-} \\
          \alpha_{n+1}^{-} \\
          \vdots \\
          \alpha_{N}^{-} \\
          \omega_{n}^{-} \\
          \omega_{n+1}^{-} \\
          \vdots \\
          \omega_{N}^{-}
      \end{array} \right] = 
\left[ \begin{array}{c}
          g^{-}_{n} \widetilde{\chi}^{c}_{12} 
          (\lambda_{n}) \\
          g^{-}_{n+1} \widetilde{\chi}^{c}_{12} 
          (\lambda_{n+1}) \\
          \vdots \\
          g^{-}_{N} \widetilde{\chi}^{c}_{12} 
          (\lambda_{N}) \\
          \overline{g^{-}_{n}} \widetilde{
          \chi}^{c}_{11}(\overline{\lambda_{n}}) \\
          \overline{g^{-}_{n+1}} \widetilde{
          \chi}^{c}_{11}(\overline{\lambda_{n+1}}) \\
          \vdots \\
          \overline{g^{-}_{N}} \widetilde{
\chi}^{c}_{11}(\overline{\lambda_{N}})
       \end{array} \right] \!,& 
\end{eqnarray*}
where, for $i,j \! \in \! \{n,n \! + \! 1,\ldots,N\}$, the 
$(N \! - \! n \! + \! 1) \! \times \! (N \! - \! n \! + \! 
1)$ matrix blocks, $\widehat{{\cal A}}^{-}$, $\widehat{{\cal 
B}}^{-}$, $\widehat{{\cal C}}^{-}$, and $\widehat{{\cal D}}
^{-}$, are defined as follows,
\begin{eqnarray*}
\widehat{{\cal A}}^{-}_{ij} := 
\left\{ \begin{array}{c}
\frac{ \lambda_{i} + g^{-}_{i} 
\widetilde{\chi}^{c}_{12}(\lambda_{i}) \widetilde{
\chi}^{c}_{22}(\lambda_{i}) + \lambda_{i} g^{-}_{i} 
{\rm W}( \widetilde{\chi}^{c}_{12}(\lambda_{i}),
\widetilde{\chi}^{c}_{22}(\lambda_{i}))}{\lambda_{i} 
\widetilde{\chi}^{c}_{22}(\lambda_{i})}, \, \, \, \, 
\, \, \, i = j, \\
- \frac{2 g^{-}_{i}(- \lambda_{i} \widetilde{\chi}^{
c}_{22}(\lambda_{i}) \widetilde{\chi}^{c}_{12}(\lambda_{
j}) + \lambda_{j} \widetilde{\chi}^{c}_{22}(\lambda_{j}) 
\widetilde{\chi}^{c}_{12}(\lambda_{i}))}{\widetilde{
\chi}^{c}_{22}(\lambda_{j})(\lambda_{i}^{2} - \lambda_{
j}^{2})}, \, \, \, \, \, \, \, i \not= j,
\end{array} \right. 
\end{eqnarray*}
\begin{eqnarray*}
\widehat{{\cal B}}^{-}_{ij} := 
\left\{ \begin{array}{c}
- \frac{ 2 g^{-}_{i} ( \lambda_{i} \widetilde{\chi}^{
c}_{22}(\lambda_{i}) \widetilde{\chi}^{c}_{11}(\overline{
\lambda_{i}}) - \overline{\lambda_{i}} \widetilde{\chi}^{
c}_{21}(\overline{\lambda_{i}}) \widetilde{\chi}^{c}_{12}
(\lambda_{i}))}{\widetilde{\chi}^{c}_{11}(\overline{
\lambda_{i}}) (\lambda_{i}^{2} - \overline{\lambda_{i}}^{
2})}, \, \, \, \, \, \, \, i = j, \\
- \frac{ 2 g^{-}_{i} (\lambda_{i} \widetilde{\chi}^{c}_{
22}(\lambda_{i}) \widetilde{\chi}^{c}_{11}(\overline{
\lambda_{j}}) - \overline{\lambda_{j}} \widetilde{\chi}^{
c}_{21}(\overline{\lambda_{j}}) \widetilde{\chi}^{c}_{12}
(\lambda_{i}))}{\widetilde{\chi}^{c}_{11}(\overline{
\lambda_{j}}) (\lambda_{i}^{2} - \overline{\lambda_{j}}^{
2})}, \, \, \, \, \, \, \, i \not= j,
\end{array} \right. 
\end{eqnarray*}
\begin{eqnarray*}
\widehat{{\cal C}}^{-}_{ij} := 
\left\{ \begin{array}{c}
- \frac{ 2 \overline{g^{-}_{i}} (\lambda_{i} 
\widetilde{\chi}^{c}_{22} (\lambda_{i}) \widetilde{
\chi}^{c}_{11}(\overline{\lambda_{i}}) - \overline{
\lambda_{i}} \widetilde{\chi}^{c}_{21} (\overline{
\lambda_{i}}) \widetilde{\chi}^{c}_{12}(\lambda_{
i}))}{\widetilde{\chi}^{c}_{22}(\lambda_{i}) 
(\overline{\lambda_{i}}^{2} - \lambda_{i}^{2})}, 
\, \, \, \, \, \, \, i = j, \\
- \frac{2 \overline{g^{-}_{i}} (- \overline{\lambda_{
i}} \widetilde{\chi}^{c}_{21} (\overline{\lambda_{i}}) 
\widetilde{\chi}^{c}_{12}(\lambda_{j}) + \lambda_{j} 
\widetilde{\chi}^{c}_{22}(\lambda_{j}) \widetilde{
\chi}^{c}_{11}(\overline{\lambda_{i}}))}{\widetilde{
\chi}^{c}_{22}(\lambda_{j}) (\overline{\lambda_{i}}^{
2} - \lambda_{j}^{2})}, \, \, \, \, \, \, \, 
i \not= j,
\end{array} \right. 
\end{eqnarray*}
\begin{eqnarray*}
\widehat{{\cal D}}^{-}_{ij} := 
\left\{ \begin{array}{c}
\frac{ \overline{\lambda_{i}} - 
\overline{g^{-}_{i}} \widetilde{\chi}^{c}_{21}
(\overline{\lambda_{i}}) \widetilde{\chi}^{c}_{11}
(\overline{\lambda_{i}}) + \overline{\lambda_{i}} 
\, \overline{g^{-}_{i}} {\rm W} ( \widetilde{\chi}^{
c}_{21}(\overline{\lambda_{i}}), \widetilde{\chi}^{
c}_{11}(\overline{\lambda_{i}}))}{\overline{\lambda_{
i}} \widetilde{\chi}^{c}_{11}(\overline{\lambda_{i}})}, 
\, \, \, \, \, \, \, i = j, \\
- \frac{2 \overline{g^{-}_{i}} (\overline{\lambda_{i}} 
\widetilde{\chi}^{c}_{21}(\overline{\lambda_{i}}) 
\widetilde{\chi}^{c}_{11} (\overline{\lambda_{j}}) - 
\overline{\lambda_{j}} \widetilde{\chi}^{c}_{21} 
(\overline{\lambda_{j}}) \widetilde{\chi}^{c}_{11}
(\overline{\lambda_{i}}))}{\widetilde{\chi}^{c}_{11}
(\overline{\lambda_{j}})(\overline{\lambda_{i}}^{2} 
- \overline{\lambda_{j}}^{2} ) }, \, \, \, \, \, \, 
\, i \not= j,
\end{array} \right.
\end{eqnarray*}
$\{\beta_{i}^{-},\widehat{\delta}_{i}^{-}\}_{i=n}^{N}$ satisfy 
the following non-degenerate system of $2(N \! - \! n \! + \! 
1)$ linear inhomogeneous algebraic equations,
\begin{eqnarray*}
&\left[ \begin{array}{cccccc}
\left. \begin{array}{ccc} \cline{1-3}
\multicolumn{1}{|c}{} &   & \multicolumn{1}{c|}{} \\
\multicolumn{1}{|c}{} & \widehat{{\cal E}}^{-} & 
\multicolumn{1}{c|}{} \\
\multicolumn{1}{|c}{} &   & \multicolumn{1}{c|}{} \\ 
\cline{1-3}   
\end{array} \right. & \left. \begin{array}{ccc} 
\cline{1-3}
\multicolumn{1}{|c}{} &   & \multicolumn{1}{c|}{} \\
\multicolumn{1}{|c}{} & \widehat{{\cal F}}^{-} & 
\multicolumn{1}{c|}{} \\
\multicolumn{1}{|c}{} &   & \multicolumn{1}{c|}{} \\ 
\cline{1-3}   
\end{array} \right. \\
& \\
\left. \begin{array}{ccc} \cline{1-3}
\multicolumn{1}{|c}{} &   & \multicolumn{1}{c|}{} \\
\multicolumn{1}{|c}{} & \widehat{{\cal G}}^{-} & 
\multicolumn{1}{c|}{} \\
\multicolumn{1}{|c}{} &   & \multicolumn{1}{c|}{} \\ 
\cline{1-3}   
\end{array} \right. & \left. \begin{array}{ccc} 
\cline{1-3}
\multicolumn{1}{|c}{} &   & \multicolumn{1}{c|}{} \\
\multicolumn{1}{|c}{} & \widehat{{\cal H}}^{-} & 
\multicolumn{1}{c|}{} \\
\multicolumn{1}{|c}{} &   & \multicolumn{1}{c|}{} \\ 
\cline{1-3}   
\end{array} \right. 
\end{array} \right] 
\left[ \begin{array}{c}
          \beta_{n}^{-} \\
          \beta_{n+1}^{-} \\
          \vdots \\
          \beta_{N}^{-} \\
          \widehat{\delta}^{-}_{n} \\
          \widehat{\delta}^{-}_{n+1} \\
          \vdots \\
          \widehat{\delta}^{-}_{N}
      \end{array} \right] = 
\left[ \begin{array}{c}
          g^{-}_{n} \widetilde{\chi}^{c}_{22} 
          (\lambda_{n}) \\
          g^{-}_{n+1} \widetilde{\chi}^{c}_{22} 
          (\lambda_{n+1}) \\
          \vdots \\
          g^{-}_{N} \widetilde{\chi}^{c}_{22} 
          (\lambda_{N}) \\
          \overline{g^{-}_{n}} \widetilde{
          \chi}^{c}_{21}(\overline{\lambda_{n}}) \\
          \overline{g^{-}_{n+1}} \widetilde{\chi}^{
          c}_{21}(\overline{\lambda_{n+1}}) \\
          \vdots \\
          \overline{g^{-}_{N}} \widetilde{
\chi}^{c}_{21}(\overline{\lambda_{N}})
       \end{array} \right] \!,& 
\end{eqnarray*}
where, for $i,j \! \in \! \{n,n \! + \! 1,\ldots,N\}$, the 
$(N \! - \! n \! + \! 1) \! \times \! (N \! - \! n \! + \! 
1)$ matrix blocks, $\widehat{{\cal E}}^{-}$, $\widehat{{\cal 
F}}^{-}$, $\widehat{{\cal G}}^{-}$, and $\widehat{{\cal H}}
^{-}$, are defined as follows,
\begin{eqnarray*}
\widehat{{\cal E}}^{-}_{ij} := 
\left\{ \begin{array}{c}
\frac{ \lambda_{i} - g^{-}_{i} 
\widetilde{\chi}^{c}_{12}(\lambda_{i}) \widetilde{
\chi}^{c}_{22}(\lambda_{i}) + \lambda_{i} g^{-}_{i} 
{\rm W}(\widetilde{\chi}^{c}_{12}(\lambda_{i}),
\widetilde{\chi}^{c}_{22} (\lambda_{i}))}{\lambda_{i} 
\widetilde{\chi}^{c}_{22}(\lambda_{i})}, \, \, \, \, 
\, \, \, i = j, \\
\frac{2 g^{-}_{i} (\lambda_{j} \widetilde{\chi}^{c}_{
12}(\lambda_{j}) \widetilde{\chi}^{c}_{22}(\lambda_{i}) 
- \lambda_{i} \widetilde{\chi}^{c}_{12}(\lambda_{i}) 
\widetilde{\chi}^{c}_{22}(\lambda_{j}))}{\widetilde{
\chi}^{c}_{22}(\lambda_{j})(\lambda_{i}^{2} - \lambda_{
j}^{2})}, \, \, \, \, \, \, \, i \not= j,
\end{array} \right. 
\end{eqnarray*}
\begin{eqnarray*}
\widehat{{\cal F}}^{-}_{ij} := 
\left\{ \begin{array}{c}
\frac{2 g^{-}_{i} (\lambda_{i} \widetilde{\chi}^{c}_{12} 
(\lambda_{i}) \widetilde{\chi}^{c}_{21}(\overline{
\lambda_{i}}) - \overline{\lambda_{i}} \widetilde{\chi}^{
c}_{11}(\overline{\lambda_{i}}) \widetilde{\chi}^{c}_{22}
(\lambda_{i}))}{\widetilde{\chi}^{c}_{11}(\overline{
\lambda_{i}})(\lambda_{i}^{2} - \overline{\lambda_{i}}^{
2})}, \, \, \, \, \, \, \, i = j, \\
\frac{2 g^{-}_{i} (\lambda_{i} \widetilde{\chi}^{c}_{12} 
(\lambda_{i}) \widetilde{\chi}^{c}_{21}(\overline{
\lambda_{j}}) - \overline{\lambda_{j}} \widetilde{
\chi}^{c}_{11}(\overline{\lambda_{j}}) \widetilde{
\chi}^{c}_{22}(\lambda_{i}))}{\widetilde{\chi}^{c}_{
11}(\overline{\lambda_{j}}) (\lambda_{i}^{2} - 
\overline{\lambda_{j}}^{2})}, \, \, \, \, \, \, \, 
i \not= j,
\end{array} \right. 
\end{eqnarray*}
\begin{eqnarray*}
\widehat{{\cal G}}^{-}_{ij} := 
\left\{ \begin{array}{c}
\frac{2 \overline{g^{-}_{i}} (\lambda_{i} \widetilde{
\chi}^{c}_{12} (\lambda_{i}) \widetilde{\chi}^{c}_{21}
(\overline{\lambda_{i}}) - \overline{\lambda_{i}} 
\widetilde{\chi}^{c}_{11}(\overline{\lambda_{i}}) 
\widetilde{\chi}^{c}_{22}(\lambda_{i}))}{\widetilde{
\chi}^{c}_{22}(\lambda_{i})(\overline{\lambda_{i}}^{2}
- \lambda_{i}^{2})}, \, \, \, \, \, \, \, i = j, \\
\frac{2 \overline{g^{-}_{i}} (\lambda_{j} \widetilde{
\chi}^{c}_{12}(\lambda_{j}) \widetilde{\chi}^{c}_{21}
(\overline{\lambda_{i}}) - \overline{\lambda_{i}} 
\widetilde{\chi}^{c}_{11}(\overline{\lambda_{i}}) 
\widetilde{\chi}^{c}_{22} (\lambda_{j}))}{\widetilde{
\chi}^{c}_{22}(\lambda_{j})(\overline{\lambda_{i}}^{2} 
- \lambda_{j}^{2})}, \, \, \, \, \, \, \, i \not= j,
\end{array} \right. 
\end{eqnarray*}
\begin{eqnarray*}
\widehat{{\cal H}}^{-}_{ij} := 
\left\{ \begin{array}{c}
\frac{\overline{\lambda_{i}} + 
\overline{g^{-}_{i}} \widetilde{\chi}^{c}_{11} 
(\overline{\lambda_{i}}) \widetilde{\chi}^{c}_{21} 
(\overline{\lambda_{i}}) - \overline{\lambda_{i}} 
\, \overline{g^{-}_{i}} {\rm W}(\widetilde{\chi}^{
c}_{11}(\overline{\lambda_{i}}),\widetilde{\chi}^{
c}_{21}(\overline{\lambda_{i}}))}{\overline{\lambda_{
i}} \, \widetilde{\chi}^{c}_{11}(\overline{\lambda_{
i}})}, \, \, \, \, \, \, \, i = j, \\
\frac{2 \overline{g^{-}_{i}} (\overline{\lambda_{i}} 
\widetilde{\chi}^{c}_{11} (\overline{\lambda_{i}}) 
\widetilde{\chi}^{c}_{21}(\overline{\lambda_{j}}) - 
\overline{\lambda_{j}} \widetilde{\chi}^{c}_{11}
(\overline{\lambda_{j}}) \widetilde{\chi}^{c}_{21}
(\overline{\lambda_{i}}))}{\widetilde{\chi}^{c}_{11}
(\overline{\lambda_{j}}) (\overline{\lambda_{i}}^{2} 
- \overline{\lambda_{j}}^{2})}, \, \, \, \, \, \, \, 
i \not= j,
\end{array} \right. 
\end{eqnarray*}
with $a_{i}^{-} \! = \! - \frac{\widetilde{\chi}^{c}_{
12}(\lambda_{i})}{\widetilde{\chi}^{c}_{22}(\lambda_{
i})} \alpha_{i}^{-}$, $b_{i}^{-} \! = \! - \frac{ 
\widetilde{\chi}^{c}_{12}(\lambda_{i})}{\widetilde{
\chi}^{c}_{22}(\lambda_{i})} \beta_{i}^{-}$, $c_{i}^{-} 
\! = \! - \frac{ \widetilde{\chi}^{c}_{21}(\overline{
\lambda_{i}})}{\widetilde{\chi}^{c}_{11}(\overline{
\lambda_{i}})} \omega_{i}^{-}$, $d_{i}^{-} \! = \! - 
\frac{ \widetilde{\chi}^{c}_{21} (\overline{\lambda_{i}}
)}{\widetilde{\chi}^{c}_{11}(\overline{\lambda_{i}})}
\widehat{\delta}_{i}^{-}$, $n \! \leq \! i \! \leq \! N$, 
\begin{eqnarray*}
&g^{-}_{j} := C_{j} e^{2 i \lambda_{j}^{2} x + 4 i 
\lambda_{j}^{4} t} (\delta^{-}(\lambda_{j};\! \lambda_{
0}))^{-2} \prod\limits_{l=1}^{n-1} \! \left( \! \frac{
(\lambda_{j} - \lambda_{l})(\lambda_{j} + \lambda_{l})
}{(\lambda_{j} - \overline{\lambda_{l}})(\lambda_{j} + 
\overline{\lambda_{l}})} \! \right)^{2}, 
\, \, \, \, \, \, \, \, \, n \leq j \leq N, 
\end{eqnarray*}
$\delta^{-}(\lambda_{k};\! \lambda_{0})$, $k \! \in \! 
\{n,n \! + \! 1,\ldots,N\}$, given in Theorem~2.1, 
Eq.~(19), and ${\rm W}(\widetilde{\chi}^{c}_{ij}(z),
\widetilde{\chi}^{c}_{i^{\prime} j^{\prime}}(z))$ is the 
Wronskian of $\widetilde{\chi}^{c}_{ij}(\lambda)$ and 
$\widetilde{\chi}^{c}_{i^{\prime} j^{\prime}}(\lambda)$ 
evaluated at $z$ $(i,j,i^{\prime},j^{\prime} \! \in \! 
\{1,2\})$: ${\rm W}(\widetilde{\chi}^{c}_{ij}(z),
\widetilde{\chi}^{c}_{i^{\prime} j^{\prime}}(z)) \! := \! 
(\widetilde{\chi}^{c}_{ij}(\lambda) \partial_{\lambda} 
\widetilde{\chi}^{c}_{i^{\prime} j^{\prime}} (\lambda) \! 
- \! \widetilde{\chi}^{c}_{i^{\prime} j^{\prime}}(\lambda) 
\partial_{\lambda} \widetilde{\chi}^{c}_{ij}(\lambda)) 
\vert_{\lambda=z}$.
\end{bbbb}
\begin{ffff}
As $t \! \rightarrow \! - \infty$ and $x \! \rightarrow 
\! + \infty$ such that $\lambda_{0} \! > \! M$ and $(x,t) 
\! \in \! \Omega_{n}$,
\begin{eqnarray*}
&Q(x,t) = \underline{Q}_{-}^{{\cal C}}(x,t) + 4 i 
\sum\limits_{j=n}^{N} \! \left(\omega_{j}^{-} - \frac{
\widetilde{\chi}^{c}_{12}(\lambda_{j})}{\widetilde{\chi}^{
c}_{22}(\lambda_{j})} \alpha_{j}^{-} \right) + {\cal O}
(C(\lambda_{0}) \exp \{\widetilde{a} \widetilde{b} 
t\}),&
\end{eqnarray*}
where $\underline{Q}_{-}^{{\cal C}}(x,t) \! := \! i Q^{
\widetilde{\chi}}(x,t)$, $\widetilde{a}$ and $\widetilde{b}$ 
are given in Lemma~6.1.3, and $C(\lambda_{0}) \! \in \! {\cal 
L}^{\infty}(\Bbb R_{>M};\! \Bbb C)$.
\end{ffff}
\begin{bbbb}
As $t \! \rightarrow \! - \infty$ and $x \! \rightarrow \! 
+ \infty$ such that $\lambda_{0} \! > \! M$ and $(x,t) \! 
\in \! \Omega_{n}$,
\begin{eqnarray*}
&Q(x,t) = Q_{-}^{{\cal C}}(x,t) + 4 i \! \left(\omega_{
n}^{-} - \frac{\widetilde{\chi}^{c}_{12}(\lambda_{n})}{
\widetilde{\chi}^{c}_{22}(\lambda_{n})} \alpha_{n}^{-} 
\right) + {\cal O} \! \left(\frac{C_{1}(\lambda_{0}) \ln 
\! \vert t \vert}{t} \right) + {\cal O}(C_{2}(\lambda_{0}) 
e^{\widetilde{a}_{0} \widetilde{b}_{0} t}),&
\end{eqnarray*}
where $\alpha_{i}^{-},\omega_{i}^{-} \! \sim \! {\cal O}
(\exp \{ \widetilde{a}^{\flat} \min\limits_{n+1 \leq l 
\leq N} \vert \xi_{n} \! - \! \xi_{l} \vert t\})$, $n \! 
+ \! 1 \! \leq \! i \! \leq \! N$, $\widetilde{a}^{\flat}
\! := \! 8 \min \{\eta_{l}\}_{l=n+1}^{N}$ $(> \! 0)$,
\begin{eqnarray*}
&\alpha_{n}^{-} = \frac{\widehat{a}_{12}^{-} \overline{
g^{-}_{n}} \widetilde{\chi}^{c}_{11}(\overline{\lambda_{
n}}) + \widehat{a}_{22}^{-} g^{-}_{n} \widetilde{\chi}^{
c}_{12}(\lambda_{n})}{(\widehat{a}_{11}^{-} \widehat{a}_{
22}^{-} - \widehat{a}_{12}^{-} \widehat{a}_{21}^{-})},& \\ 
&\omega_{n}^{-} = \frac{\widehat{a}_{11}^{-} \overline{
g^{-}_{n}} \widetilde{\chi}^{c}_{11}(\overline{\lambda_{
n}}) + \widehat{a}_{21}^{-} g^{-}_{n} \widetilde{\chi}^{
c}_{12}(\lambda_{n})}{(\widehat{a}_{11}^{-} \widehat{a}_{
22}^{-} - \widehat{a}_{12}^{-} \widehat{a}_{21}^{-})},& \\
&\widehat{a}_{11}^{-} := \frac{\lambda_{n}+g^{-}_{n} 
\widetilde{\chi}^{c}_{12}(\lambda_{n}) \widetilde{\chi}^{
c}_{22}(\lambda_{n}) + \lambda_{n} g^{-}_{n} {\rm W}(
\widetilde{\chi}^{c}_{12}(\lambda_{n}),\widetilde{\chi}^{
c}_{22}(\lambda_{n}))}{\lambda_{n} \widetilde{\chi}^{c}_{
22}(\lambda_{n})},& \\
&\widehat{a}_{12}^{-} := \frac{2 g^{-}_{n} (\lambda_{n} 
\widetilde{\chi}^{c}_{22}(\lambda_{n}) \widetilde{\chi}^{c}_{
11} (\overline{\lambda_{n}}) - \overline{\lambda_{n}} 
\widetilde{\chi}^{c}_{21}(\overline{\lambda_{n}}) \widetilde{
\chi}^{c}_{12}(\lambda_{n}))}{\widetilde{\chi}^{c}_{11}(
\overline{\lambda_{n}})(\lambda_{n}^{2} - \overline{\lambda_{
n}}^{2})},& \\
&\widehat{a}_{21}^{-} := \frac{ 2 \overline{g^{-}_{n}} 
(\lambda_{n} \widetilde{\chi}^{c}_{22}(\lambda_{n}) \widetilde{
\chi}^{c}_{11} (\overline{\lambda_{n}})-\overline{\lambda_{n}} 
\widetilde{\chi}^{c}_{21}(\overline{\lambda_{n}}) \widetilde{
\chi}^{c}_{12}(\lambda_{n}))}{\widetilde{\chi}^{c}_{22}(
\lambda_{n})(\overline{\lambda_{n}}^{2}-\lambda_{n}^{2})},& \\
&\widehat{a}_{22}^{-} := \frac{\overline{\lambda_{n}} - 
\overline{g^{-}_{n}} \widetilde{\chi}^{c}_{21}(\overline{
\lambda_{n}}) \widetilde{\chi}^{c}_{11}(\overline{\lambda_{n}}) 
+ \overline{\lambda_{n}} \, \overline{g^{-}_{n}} {\rm W}(
\widetilde{\chi}^{c}_{21}(\overline{\lambda_{n}}),\widetilde{
\chi}^{c}_{11}(\overline{\lambda_{n}}))}{\overline{\lambda_{n}}
\widetilde{\chi}^{c}_{11}(\overline{\lambda_{n}})},&
\end{eqnarray*}
$Q^{{\cal C}}_{-}(x,t)$ is given in Theorem~2.1, Eqs.~(20), 
(21), (23) and (24), $\widetilde{a}_{0}\! := \! \min (
\widetilde{a},\widetilde{a}^{\flat}) \, (> \! 0)$, $\widetilde{
b}_{0} \! := \! \min(\widetilde{b},\min \{\vert \xi_{n} \! 
- \! \xi_{l} \vert\}_{l=n+1}^{N})$, $C_{1}(\lambda_{0}) \! \in 
\! {\cal S}(\Bbb R_{>M};\! \Bbb C)$, and $C_{2}(\lambda_{0}) \! 
\in \! {\cal L}^{\infty}(\Bbb R_{>M};\! \Bbb C)$.
\end{bbbb}
\begin{bbbb}
As $t \! \rightarrow \! - \infty$ and $x \! \rightarrow 
\! + \infty$ such that $\lambda_{0} \! > \! M$ and $(x,
t) \! \in \! \Omega_{n}$,
\begin{eqnarray*}
&Q(x,t)=Q^{-}_{{\rm as}}(x,t) + {\cal O} \! \left(\frac{
C(\lambda_{0}) \ln \vert t \vert}{t} \right) \!,&
\end{eqnarray*}
where $Q^{-}_{{\rm as}}(x,t)$ is given in Theorem~2.1, 
Eqs.~(14)--(29), and $C(\lambda_{0}) \! \in \! {\cal 
S}(\Bbb R_{> M};\! \Bbb C)$.
\end{bbbb}
\subsection{Asymptotics of $((\Psi^{-1}(x,t;0))_{11})^{
2}$ as $t \! \rightarrow \! - \infty$}
\begin{bbbb}
Define $Q^{\natural}(x,t) \! := \! 2 i \! \lim\limits_{
\lambda \to \infty}(\lambda \widetilde{\chi}^{c}(x,t;\! \lambda
))_{12}$. Then
\begin{eqnarray*}
&(\widetilde{\chi}^{c}_{22}(0))^{2} = (\delta^{-}(0;\! \lambda
_{0}))^{2} \exp\{i \int_{+\infty}^{x} \vert Q^{\natural}
(\varrho,t) \vert^{2} d \varrho\}.&
\end{eqnarray*}
\end{bbbb}
\begin{cccc}[{\rm \cite{a22}}]
As $t \! \to \! -\infty$ and $x \! \to \! +\infty$ such that 
$\lambda_{0} \! > \! M$,
\begin{eqnarray*}
&Q^{\natural}(x,t) = \frac{u_{1,1,0}^{-}(\lambda_{0}) e^{i(4 
\lambda_{0}^{4} t + \nu(\lambda_{0}) \ln \! \vert t \vert)}}
{\sqrt{-t}} + \frac{u_{-1,2,0}^{-}(\lambda_{0})}{(-t)} + 
{\cal O} \! \left(\frac{C(\lambda_{0})(\ln \vert t \vert)^{2}
}{(-t)^{3/2}} \right) \!,&
\end{eqnarray*}
where 
\begin{eqnarray*}
&u_{1,1,0}^{-}(\lambda_{0}) = \sqrt{\frac{\nu(\lambda_{0})
}{2 \lambda_{0}^{2}}} \exp \{i \theta^{-}(\lambda_{0})\},& \\
&\theta^{-}(\lambda_{0}) \! = \! \phi^{-}(\lambda_{0}) \! 
+ \! \frac{3\pi}{4} \! - \! \arg \Gamma(i \nu(\lambda_{
0})) \! + \! \arg r(\lambda_{0}) \! + \! 3 \nu(\lambda_{0}) 
\ln 2 \! + \! 2 \! \sum\limits_{l=1}^{n-1} \! \! \arg 
\! \left(\! \frac{(\lambda_{0} - \overline{\lambda_{
l}})(\lambda_{0} + \overline{\lambda_{l}})}{(\lambda_{0} - 
\lambda_{l})(\lambda_{0} + \lambda_{l})} \! \right) \!,& \\
&u_{-1,2,0}^{-}(\lambda_{0}) \! = \! \frac{i}{8 \pi \lambda
_{0}^{2}} \! \left(\! \left. \frac{d(r(\varrho) \vert_{
\varrho \in \Bbb R})}{d\varrho} \right\vert_{\varrho=0} \! 
- \! \left. \frac{d(r(\varrho) \vert_{\varrho \in i \Bbb R}
)}{d\varrho} \right\vert_{\varrho=0} \right) \! \exp \! 
\left\{\! i \! \left(4 \sum\limits_{l=1}^{n-1} \gamma_{l}+
2 \vartheta^{-}(\lambda_{0}) \right) \! \right\} \!,& \\
&\vartheta^{-}(\lambda_{0}) = - \int_{\lambda_{0}}^{\infty} 
\frac{\ln(1 - \vert r(\varrho) \vert^{2})}{\varrho} \frac{d
\varrho}{\pi},&
\end{eqnarray*}
$\phi^{-}(\cdot)$ is given in Theorem~2.1, Eq.~(23), and 
$C(\lambda_{0}) \! \in \! {\cal S}(\Bbb R_{> M};\! \Bbb 
C)$.
\end{cccc}
\begin{bbbb}
As $t \! \to \! -\infty$ and $x \! \to \! +\infty$ such that 
$\lambda_{0} \! > \! M$,
\begin{eqnarray*}
(\widetilde{\chi}^{c}_{22}(0))^{2} & = & \exp \! \left\{\! 
i \! \left(\! \sqrt{\frac{2}{-t}} \int_{\lambda_{0}}^{
\infty} \! \frac{\sqrt{\nu(\mu)}}{\mu^{2}} \! \left(R^{-}_{
i}(0) \cos(\kappa^{-}(\mu;t)) \! - \! R^{-}_{r}(0) \sin(
\kappa^{-}(\mu;t)) \right) \! \frac{d\mu}{\pi} \right) \! 
\right\} \\
 & + & {\cal O} \! \left(\frac{C(\lambda_{0})(\ln \vert t 
\vert)^{2}}{\lambda_{0}^{2} t} \right) \!,
\end{eqnarray*}
where $R^{-}_{i}(0) \! = \! \Im \{R^{-}(0)\}$, $R^{-}_{
r}(0) \! = \! \Re \{R^{-}(0)\}$, $R^{-}(0) \! := \! 
\left(\left. \frac{d(r(\varrho) \vert_{\varrho \in \Bbb R})}{
d\varrho} \right\vert_{\varrho=0} \! - \! \left. \frac{d(r(
\varrho) \vert_{\varrho \in i \Bbb R})}{d\varrho} \right\vert
_{\varrho=0} \right) \linebreak[4] \cdot \exp \{4i \sum_{l=1}
^{n-1} \gamma_{l}\}$, $\kappa^{-}(\lambda_{0};t) \! := \! 4 
\lambda_{0}^{4} t \! + \! \nu(\lambda_{0}) \ln \! \vert t 
\vert \! + \! \theta^{-}(\lambda_{0}) \! - \! 2 \vartheta^{
-}(\lambda_{0})$, and $C(\lambda_{0}) \! \in \! {\cal S}(\Bbb 
R_{> M};\! \Bbb C)$.
\end{bbbb}
\begin{cccc}
As $t \! \to \! -\infty$ and $x \! \to \! +\infty$ such that 
$\lambda_{0} \! > \! M$ and $(x,t) \! \in \! \Omega_{n}$, 
\begin{eqnarray*}
&((\Psi^{-1}(x,t;0))_{11})^{2} = \exp \{2 \ln(\widetilde{
\chi}^{c}_{22}(0))\} \exp \{\frac{2i}{\pi} \! \int_{\lambda_{
0}}^{\infty} \frac{\ln (1 - \vert r(\varrho) \vert^{2})}{
\varrho} d \varrho\} \exp \{-4i \sum\limits_{l=1}^{n-1} 
\gamma_{l}\}& \\
&\, \, \, \, \, \, \, \, \, \, \, \, \, \, \, \, \, \, \, \, 
\, \, \, \, \, \, \, \, \, \, \, \, \times \, \exp \{2 \ln 
(1-\sum\limits_{i=n}^{N} (\frac{2 b_{i}^{-}}{\lambda_{i}} + 
\frac{2 \widehat{\delta}_{i}^{-}}{\overline{\lambda_{i}}}))\} 
+ {\cal O}(C(\lambda_{0})\exp \{\widetilde{a} \widetilde{b} 
t\}),&
\end{eqnarray*}
where $(\widetilde{\chi}^{c}_{22}(0))^{2}$ is given in 
Proposition~6.3.2, $b_{i}^{-} \! = \! - \frac{\widetilde{
\chi}^{c}_{12}(\lambda_{i})}{\widetilde{\chi}^{c}_{22}(
\lambda_{i})} \beta_{i}^{-}$, $n \! \leq \! i \! \leq \! N$, 
$\beta^{-}_{j}, \widehat{\delta}_{j}^{-} \! \sim \! {\cal 
O}(\exp \{\widetilde{a}^{\flat} \min\limits_{n+1 \leq j \leq 
N} \vert \xi_{n} \! - \! \xi_{j} \vert t\})$, $n \! + \! 1 
\! \leq \! j \! \leq \! N$,
\begin{eqnarray*}
&\beta_{n}^{-}=\frac{\beta_{n}^{{\cal N},-}}{(\widehat{{\cal 
E}}_{nn}^{-} \widehat{{\cal H}}_{nn}^{-}-\widehat{{\cal F}}_{
nn}^{-} \widehat{{\cal G}}_{nn}^{-})}, \, \, \, \, \, \, \, \,
\, \, \widehat{\delta}_{n}^{-}=\frac{\widehat{\delta}_{n}^{{\cal 
N},-}}{(\widehat{{\cal E}}_{nn}^{-} \widehat{{\cal H}}_{nn}^{
-}-\widehat{{\cal F}}_{nn}^{-} \widehat{{\cal G}}_{nn}^{-})},&
\end{eqnarray*}
with
\begin{eqnarray*}
&\beta_{n}^{{\cal N},-} \! := \! \frac{g_{n}^{-} 
\widetilde{\chi}^{c}_{22}(\lambda_{n})}{\widetilde{\chi}^{
c}_{11}(\overline{\lambda_{n}})} \! + \! \frac{\vert g_{n}
^{-} \vert^{2} \widetilde{\chi}^{c}_{21}(\overline{\lambda_{
n}}) \widetilde{\chi}^{c}_{22}(\lambda_{n})}{\overline{
\lambda_{n}}} \! - \! \frac{\vert g_{n}^{-} \vert^{2} 
\widetilde{\chi}^{c}_{22}(\lambda_{n}) W(\widetilde{\chi}^{
c}_{11}(\overline{\lambda_{n}}),\widetilde{\chi}^{c}_{21}(
\overline{\lambda_{n}}))}{\widetilde{\chi}^{c}_{11}(
\overline{\lambda_{n}})}& \\
&\! \! \! \! \! \! \! \! \! \! \! \! \! \! \! \! \! \! \! 
\! \! \! \! \! \! \! \! \! \! \! \! \! \! \! \! \! \! \! 
\! \! \! \! \! \! \! \! \! \! \! \! \! \! \! \! \! \! \! 
\! \! \! \! \! \! \! \! \! \! \! \! \! \! \! \! \! \! \! 
\! \! \! \! + \, \frac{2 \overline{\lambda_{
n}} \vert g_{n}^{-} \vert^{2} \widetilde{\chi}^{c}_{22}(
\lambda_{n}) \widetilde{\chi}^{c}_{21}(\overline{\lambda_{
n}})}{(\lambda_{n}^{2} - \overline{\lambda_{n}}^{2})},& \\
&\widehat{{\cal E}}_{nn}^{-} \widehat{{\cal H}}_{nn}^{-}-
\widehat{{\cal F}}_{nn}^{-} \widehat{{\cal G}}_{nn}^{-} 
:= \frac{1}{\widetilde{\chi}^{c}_{22}(\lambda_{n}) 
\widetilde{\chi}^{c}_{11}(\overline{\lambda_{n}})} + \frac{
\overline{g_{n}^{-}} W(\widetilde{\chi}^{c}_{21}(\overline{
\lambda_{n}}),\widetilde{\chi}^{c}_{11}(\overline{\lambda_{
n}}))}{\widetilde{\chi}^{c}_{22}(\lambda_{n}) \widetilde{
\chi}^{c}_{11}(\overline{\lambda_{n}})}+\frac{g_{n}^{-} W
(\widetilde{\chi}^{c}_{12}(\lambda_{n}),\widetilde{\chi}^{
c}_{22}(\lambda_{n}))}{\widetilde{\chi}^{c}_{11}(\overline{
\lambda_{n}}) \widetilde{\chi}^{c}_{22}(\lambda_{n})}& \\
&\, \, \, \, \, \, \, \, \, \, \, \, \, \, \, \, \, \, 
\, \, \, \, \, \, + \, \, 
\frac{\overline{g_{n}^{-}} \widetilde{\chi}^{c}_{21}(
\overline{\lambda_{n}})}{\overline{\lambda_{n}} \widetilde{
\chi}^{c}_{22}(\lambda_{n})} - \frac{g_{n}^{-}\widetilde{
\chi}^{c}_{12}(\lambda_{n})}{\lambda_{n} \widetilde{\chi}^{
c}_{11}(\overline{\lambda_{n}})} + \frac{(2 \overline{
\lambda_{n}})^{2} \vert g_{n}^{-} \vert^{2} \widetilde{
\chi}^{c}_{22}(\lambda_{n}) \widetilde{\chi}^{c}_{11}
(\overline{\lambda_{n}})}{(\lambda_{n}^{2}-\overline{
\lambda_{n}}^{2})^{2}},& \\
&\widehat{\delta}_{n}^{{\cal N},-} := \frac{\overline{
g_{n}^{-}} \, \widetilde{\chi}^{c}_{21}(\overline{\lambda_{
n}})}{\widetilde{\chi}^{c}_{22}(\lambda_{n})} - \frac{2 
\overline{\lambda_{n}} \vert g_{n}^{-} \vert^{2} \widetilde{
\chi}^{c}_{11}(\overline{\lambda_{n}}) \widetilde{\chi}^{
c}_{22}(\lambda_{n})}{(\lambda_{n}^{2}-\overline{\lambda_{
n}}^{2})},&
\end{eqnarray*}
and $C(\lambda_{0}) \! \in \! {\cal L}^{\infty}(\Bbb R_{
>M};\! \Bbb C)$.
\end{cccc}
\begin{ffff}
As $t \! \rightarrow \! -\infty$ and $x \! \rightarrow \! 
+\infty$ such that $\lambda_{0} \! > \! M$ and $(x,t) \! 
\in \! \Omega_{n}$,
\begin{eqnarray*}
&q(x,t) = Q_{{\rm as}}^{-}(x,t) \exp\{i \arg q_{{\rm as}
}^{-}(x,t)\} + {\cal O} \! \left(\frac{C(\lambda_{0}) 
(\ln \vert t \vert)^{2}}{t} \right) \!,&
\end{eqnarray*}
where $Q_{{\rm as}}^{-}(x,t)$ is given in Theorem~2.1, 
Eqs.~(14)--(29), $\arg q_{{\rm as}}^{-}(x,t)$ is given 
in Theorem~2.2, Eqs.~(48)--(51), and $C(\lambda_{0}) \! 
\in \! {\cal S}(\Bbb R_{>M};\! \Bbb C)$.
\end{ffff}
\begin{ffff}
As $t \! \rightarrow \! -\infty$ and $x \! \rightarrow \! 
-\infty$ such that $\widehat{\lambda}_{0} \! := \! \sqrt{
\frac{1}{2}(\frac{x}{t} \! - \! \frac{1}{s})} \! > \! M$, 
$\frac{x}{t} \! > \! \frac{1}{s}$, $s \! \in \! \Bbb R_{>0}$, 
and $(x,t) \! \in \! \widetilde{\Omega}_{n}$,
\begin{eqnarray*}
&u(x,t) = v_{{\rm as}}^{-}(x,t) w_{{\rm as}}^{-}(x,t) + 
{\cal O} \! \left(\frac{C(\widehat{\lambda}_{0})(\ln \vert 
t \vert)^{2}}{t} \right) \!,&
\end{eqnarray*}
where $v_{{\rm as}}^{-}(x,t)$ and $w_{{\rm as}}^{-}(x,t)$ 
are given in Theorem~2.3, Eqs.~(58)--(70), and $C(\widehat{
\lambda}_{0}) \! \in \! {\cal S}(\Bbb R_{>M};\linebreak[4] 
\Bbb C)$.
\end{ffff}
\vspace{1.30cm}
\begin{flushleft}
{\LARGE {\bf Acknowledgements}}
\end{flushleft}
The authors are grateful to P.~A.~Deift for a copy of \cite{a20} 
prior to publication, A.~R.~Its for informative discussions, and 
V.~B.~Matveev for encouragement and support.


\clearpage
\begin{thebibliography}{100}
\bibitem{a1} J. R. Taylor, ed., {\em Optical Solitons - Theory 
and Experiment\/}, Cambridge Studies in Modern Optics, Vol.~10, 
CUP, Cambridge, 1992.\\
G. P. Agrawal, 2nd edn., {\em Nonlinear Fiber Optics\/}, Academic, 
San Diego, 1995.\\
A. Hasegawa and Y. Kodama, {\em Solitons in Optical 
Communications\/}, Oxford Series in Optical and Imaging Sciences, 
No.~7, OUP, Oxford, 1995.
\bibitem{a2} L. D. Faddeev and L. A. Takhtajan, {\em Hamiltonian 
Methods in the Theory of Solitons\/}, Springer-Verlag, Berlin, 
1987.
\bibitem{a3} M. Wadati, K. Konno, and Y. Ichikawa, {\em A 
Generalization of Inverse Scattering Method\/}, J. Phys. Soc. 
Japan, {\bf 46} (1979), pp.~1965--1966.\\
H. Eichhorn, {\em Application of the inverse scattering method 
to the generalised non-linear Schr\"{o}dinger equation\/}, 
Inverse Problems, {\bf 1} (1985), pp.~193--198.\\
I. Cherednik, {\em Basic Methods of Soliton Theory\/}, Advanced 
Series in Mathematical Physics, Vol.~25, World Scientific, 
Singapore, 1996.
\bibitem{a4} M. W. Chbat, P. R. Prucnal, M. N. Islam, C. E. 
Soccolich, and J. P. Gordon, {\em Long-range interference effects 
of soliton reshaping in optical fibers\/}, J. Opt. Soc. Am. B, 
{\bf 10} (1993), pp.~1386--1395.
\bibitem{a5} P. C. Schuur, {\em Asymptotic Analysis of Soliton 
Problems\/}, Lecture Notes in Mathematics, Vol.~1232, 
Springer-Verlag, Berlin, 1986.\\
R. F. Bikbaev, {\em On the asymptotics as $t \! \to \! \infty$ 
of the Cauchy problem solution for the Landau-Lifshitz equation\/}, 
Theor. Math. Phys., {\bf 77} (1988), pp.~1117--1123.\\
A. S. Fokas and A. R. Its, {\em Soliton Generation for 
Initial-Boundary-Value Problems\/}, Phys. Rev. Lett., {\bf 68} 
(1992), pp.~3117--3120.\\
E. Ya. Khruslov and V. P. Kotlyarov, {\em Soliton Asymptotics of 
Nondecreasing Solutions of Nonlinear Completely Integrable 
Evolution Equations\/}, in {\em Spectral Operator Theory and 
Related Topics\/}, V. A. Marchenko, ed., Advances in Soviet 
Mathematics, Vol.~19, AMS, Providence, 1994, pp.~129--180.
\bibitem{a6} V. A. Vysloukh and I. V. Cherednik, {\em 
Many-Soliton Components of Solutions of Nonlinear Schr\"{o}dinger 
Equation with Perturbing Term\/}, Theor. Math. Phys., {\bf 78} 
(1989), pp.~24--31.\\
Z. Chen and N. Huang, {\em Explicit N-soliton solution of the 
modified nonlinear Schr\"{o}dinger equation\/}, Phys. Rev. A, 
{\bf 41} (1990), pp.~4066--4069.
\bibitem{a7} A.~V. Kitaev and A.~H. Vartanian, {\em Leading-order 
temporal asymptotics of the modified nonlinear Schr\"{o}dinger
equation: solitonless sector\/}, Inverse Problems, {\bf 13} 
(1997), pp.~1311--1339.
\bibitem{a8} M. J. Ablowitz and H. Segur, {\em Solitons and the 
Inverse Scattering Transform\/}, SIAM, Philadelphia, 1981.\\
M. J. Ablowitz and P. A. Clarkson, {\em Solitons, Nonlinear 
Evolution Equations and Inverse Scattering\/}, LMS, No.~149, 
CUP, Cambridge, 1991.
\bibitem{a9} S. V. Manakov, {\em Nonlinear Fraunhoffer 
diffraction\/}, Sov. Phys. JETP, {\bf 38} (1974), 
pp.~693--696.\\
V. E. Zakharov and S. V. Manakov, {\em Asymptotic behaviour of 
nonlinear wave systems integrable by the inverse scattering 
method\/}, Sov. Phys. JETP, {\bf 44} (1976), pp.~106--112.\\
A. R. Its, {\em Asymptotics of solutions of the nonlinear 
Schr\"{o}dinger equation and isomonodromic deformations 
of systems of linear differential equations\/}, Sov. Math. 
Dokl., {\bf 24} (1981), pp.~452--456.
\bibitem{a10} S. P. Novikov, S. V. Manakov, L. P. Pitaevskii, 
and V. E. Zakharov, {\em Theory of Solitons: The Inverse 
Scattering Method\/}, Plenum, New York, 1984.
\bibitem{a11} V. E. Zakharov and A. B. Shabat, {\em Integration 
of the nonlinear equations of mathematical physics by the method 
of the inverse scattering transform. II\/}, Funct. Anal. Appl., 
{\bf 13} (1979), pp.~166--174.
\bibitem{a12} R. Beals, P. Deift, and C. Tomei, {\em Direct 
and Inverse Scattering on the Line\/}, Mathematical Surveys 
and Monographs, No.~28, AMS, Providence, 1988.
\bibitem{a13} R. Beals and R. R. Coifman, {\em Scattering and 
Inverse Scattering for First Order Systems\/}, Comm. Pure Appl. 
Math., {\bf 37} (1984), pp.~39--90.\\
R. Beals and R. R. Coifman, {\em Inverse Scattering and 
Evolution Equations\/}, Comm. Pure Appl. Math., {\bf 38} (1985), 
pp.~29--42.
\bibitem{a14} A. V. Kitaev, {\em Self-Similar Solutions of 
the Modified Nonlinear Schr\"{o}dinger Equation\/}, Theor. 
Math. Phys., {\bf 64} (1985), pp.~878--894.
\bibitem{a15} R. F. Bikbaev, private communication.
\bibitem{a16} D. J. Kaup and A. C. Newell, {\em An exact 
solution for a derivative nonlinear Schr\"{o}dinger 
equation\/}, J. Math. Phys., {\bf 19} (1979), pp.~798--801.
\bibitem{a17} V. S. Gerdzhikov, M. I. Ivanov, and P. P. 
Kulish, {\em Quadratic Bundle and Nonlinear Equations\/}, 
Theor. Math. Phys., {\bf 44} (1980), pp.~784--795.\\
J. Lee, {\em Global Solvability of the Derivative Nonlinear 
Schr\"{o}dinger Equation\/}, Trans. Amer. Math. Soc., {\bf 
314} (1989), pp.~107--118.
\bibitem{a18} X. Zhou, {\em The Riemann-Hilbert Problem and 
Inverse Scattering\/}, SIAM J. Math. Anal., {\bf 20} (1989), 
pp.~966--986.
\bibitem{a19} I. S. Gradshteyn and I. M. Ryzhik, 5th edn., 
{\em Tables of Integrals, Series, and Products\/}, A. Jeffrey, 
ed., Academic Press, San Diego, 1994.
\bibitem{a20} P. Deift, S. Kamvissis, T. Kriecherbauer, and 
X. Zhou, {\em The Toda Rarefaction Problem\/}, Comm. Pure 
Appl. Math., {\bf 49} (1996), pp.~35--83.
\bibitem{a21} P. A. Deift and X. Zhou, {\em Long-time 
Asymptotics for Integrable Systems. Higher Order Theory\/}, 
Commun. Math. Phys., {\bf 165} (1994), pp.~175--191.
\bibitem{a22} A. H. Vartanian, {\em Higher Order Asymptotics 
of the Modified Nonlinear Schr\"{o}dinger Equation\/}, to be 
submitted.
\end{thebibliography}
\end{document}